\pdfoutput=1
\newcommand*{\ATLASLATEXPATH}{latex/}
\documentclass[cernpreprint, texlive=2016, american]{\ATLASLATEXPATH atlasdoc}
\usepackage{\ATLASLATEXPATH atlaspackage}
\usepackage{\ATLASLATEXPATH atlasbiblatex}
\usepackage{\ATLASLATEXPATH atlasphysics}
\usepackage{ANA-TOPQ-2016-11-PAPER-defs}
\usepackage{multirow}
\usepackage{bigdelim}

\AtlasTitle{Measurement of the $t\bar{t}Z$ and $t\bar{t}W$ cross sections in
proton--proton collisions at $\sqrt{s}=13$ TeV with the ATLAS detector}
\AtlasAbstract{A measurement of the associated production of a top-quark pair
($t\bar{t}$) with a vector boson ($W$, $Z$) in proton--proton collisions at a
center-of-mass energy of 13 TeV is presented, using $36.1$~fb$^{-1}$ of
integrated luminosity collected by the ATLAS detector at the Large Hadron
Collider.  Events are selected in channels with two same- or opposite-sign
leptons (electrons or muons), three leptons or four leptons, and each channel
is further divided into multiple regions to maximize the sensitivity of the
measurement. The $t\bar{t}Z$ and $t\bar{t}W$ production cross sections are
simultaneously measured using a combined fit to all regions.  The best-fit
values of the production cross sections are $\sigma_{t\bar{t}Z} = 0.95 \pm
0.08_{\mathrm{stat.}} \pm 0.10_{\mathrm{syst.}}\,\mathrm{pb}$ and
$\sigma_{t\bar{t}W} = 0.87 \pm 0.13_{\mathrm{stat.}} \pm
0.14_{\mathrm{syst.}}\,\mathrm{pb}$ in agreement with the Standard Model
predictions. The measurement of the $t\bar{t}Z$ cross section is used to set
constraints on effective field theory operators which modify the $t\bar{t}Z$
vertex.}

\author{The ATLAS Collaboration}
\PreprintIdNumber{CERN-EP-2018-331}
\AtlasDate{\today}
\arXivId{1901.03584}
\AtlasDOI{10.1103/PhysRevD.99.072009}
\AtlasJournalRef{Phys.\ Rev.\ D 99 (2019) 072009}

\begin{document}
\maketitle

\section{Introduction}

Properties of the top quark have been explored by the Large Hadron Collider
(LHC) and previous collider experiments in great detail. The production cross
sections of top-quark pairs and single top quarks, as well as the top-quark
mass, spin correlations and $W$ boson helicity fractions have all been
measured. Other properties of the top quark are now becoming accessible, owing
to the large center-of-mass energy and luminosity at the LHC.  These include
its coupling to the Higgs boson and electroweak neutral-current couplings,
accessed by measurements of pair-produced top quarks in association with a
Higgs boson~\cite{HIGG-2017-02, CMS-HIG-17-018, CMS-HIG-17-035, HIGG-2018-13}
or a photon~\cite{TOPQ-2012-07, TOPQ-2015-21, CMS-TOP-14-008, TOPQ-2017-14}.

Measurements of top-quark pairs in association with a $Z$ or $W$ boson (\ttZ
and \ttW) provide a direct probe of the weak couplings of the top
quark~\cite{Dror:2015nkp, Buckley:2015lku, Bylund:2016phk}. These couplings may
be modified in the presence of physics beyond the Standard Model (BSM). Any
deviations from the SM predictions due to BSM effects can be parameterized in a
model-independent way using the framework of the Standard Model Effective Field
Theory (SMEFT)~\cite{Weinberg:1978kz, Buchmuller:1985jz, Leung:1984ni}. If no
deviations are observed, measurements of the \ttZ and \ttW production cross
sections, \sttZ and \sttW, can be used to set constraints on the weak couplings
of the top quark in the SMEFT context. The \ttZ and \ttW processes were
observed by ATLAS~\cite{TOPQ-2013-05, TOPQ-2015-22} and
CMS~\cite{CMS-TOP-14-021, CMS-TOP-17-005}, with measured cross sections
compatible with the SM prediction. At \SI{13}{\TeV}, ATLAS analysed
\SI{3.2}{\per \fb} of data using the same-sign dimuon, trilepton and
tetralepton channels, and measured $\sigma_{\ttZ} = 0.9 \pm 0.3$ pb and
$\sigma_{\ttW} = 1.5 \pm 0.8$ pb, while CMS analysed \SI{35.9}{\per \fb} and
measured $0.99^{+0.15}_{-0.13}$ pb and $0.77^{+0.18}_{-0.16}$ pb, respectively.

The production of \ttZ and \ttW is often an important background in searches
involving final states with multiple leptons and $b$-quarks. These processes
also constitute an important background in measurements of the associated
production of the Higgs boson with top quarks.
 
This paper presents measurements of the \ttZ and \ttW cross sections using
proton--proton ($pp$) collision data at a center-of mass energy $\sqrt{s} =
13\,\TeV$ corresponding to an integrated luminosity of \lumi, collected by the
ATLAS detector in 2015 and 2016.  The final states of top-quark pairs produced
in association with a $Z$ or a $W$ boson contain up to four isolated, prompt
leptons.\footnote{In this paper, lepton is used to denote electron or muon, and
prompt lepton is used to denote a lepton produced in a $Z$ or $W$ boson decay,
or in the decay of a $\tau$-lepton which arises from a $Z$ or $W$ boson decay.}
In this analysis, events with two opposite-sign (OS) or same-sign (SS) leptons,
three leptons or four leptons are considered.  The dominant backgrounds in
these four channels are $Z$+jets and \ttbar, events with non-prompt or
misidentified leptons, $WZ$, and $ZZ$ production, respectively.  An
interpretation of the \ttZ cross-section measurement in the SMEFT framework is
also performed.

\section{The ATLAS detector}
\label{s:atlas}

The ATLAS detector~\cite{PERF-2007-01} consists of three main subsystems: an
inner tracking system, electromagnetic (EM) and hadronic calorimeters, and a
muon spectrometer (MS).  The inner detector (ID) consists of a
high-gra\-nu\-la\-ri\-ty silicon pixel detector, including the insertable
B-layer~\cite{ATL-INDET-PUB-2015-001,Abbott:2018ikt}, which is the innermost
layer of the tracking system, and a silicon microstrip tracker, together
providing precision tracking in the pseudorapidity\footnote{ATLAS uses a
right-handed coordinate system with its origin at the nominal interaction point
(IP) in the center of the detector and the $z$-axis along the beam pipe. The
$x$-axis points from the IP to the center of the LHC ring, and the $y$-axis
points upward. Cylindrical coordinates $(r,\phi)$ are used in the transverse
plane, $\phi$ being the azimuthal angle around the $z$-axis. The pseudorapidity
is defined in terms of the polar angle $\theta$ as $\eta=-\ln\tan(\theta/2)$.}
range $|\eta|<2.5$, followed by a transition radiation tracker covering
$|\eta|<2.0$.  All these systems are immersed in a \SI{2}{T} magnetic field
provided by a thin superconducting solenoid.  The EM sampling calorimeter uses
lead and liquid argon (LAr) and is divided into barrel ($|\eta|<1.475$) and
endcap ($1.375<|\eta|<3.2$) regions.  Hadronic calorimetry is provided by a
steel/scintillator-tile calorimeter, segmented into three barrel structures, in
the range $|\eta|<1.7$, and by two copper/LAr hadronic endcap calorimeters that
cover the region $1.5<|\eta|<3.2$. The solid angle coverage is completed with
forward copper/LAr and tungsten/LAr calorimeter modules, optimized for EM and
hadronic measurements, respectively, covering the region $3.1<|\eta|<4.9$. The
muon spectrometer measures the deflection of muons in the range $|\eta|<2.7$
using multiple layers of high-precision tracking chambers located in toroidal
magnetic fields.  The field integral of the toroids ranges between \SI{2.0}{Tm}
and \SI{6.0}{Tm} for most of the detector.  The muon spectrometer is also
instrumented with separate trigger chambers covering $|\eta|<2.4$.  A two-level
trigger system~\cite{TRIG-2016-01}, using custom hardware followed by a
software-based trigger level, is used to reduce the event rate to an average of
around \SI{1}{kHz} for offline storage. 

\section{Data and simulated event samples}
\label{s:samples}

The data were collected with the ATLAS detector during 2015 and 2016 at a
proton--proton ($pp$) collision energy of \SI{13}{\TeV}.  The bunch spacing was
\SI{25}{ns} and the mean number of interactions per bunch crossing was 14 (25)
in 2015 (2016).  With strict data-quality requirements, the integrated
luminosity considered corresponds to \lumi~\cite{DAPR-2013-01,LUCID2}.

Monte Carlo (MC) simulation samples are used to model the expected signal and
background distributions in the different control, validation and signal
regions described below. The heavy-flavor hadron decays involving $b$- and
$c$-quarks, particularly important in this measurement, were modeled using
\EVTGEN~\cite{EvtGen} v1.2.0, except for those processes modeled using the
\SHERPA{}~\cite{Gleisberg:2008ta} generator, as described below. 
In all samples the top-quark mass
was set to $172.5\,\GeV,$ and the Higgs boson mass was set to $125\,\GeV$. The
response of the detector to stable\footnote{A particle is considered stable if
$c\tau \ge 1$~cm.} particles was emulated by a dedicated
simulation~\cite{SOFT-2010-01} based either fully on \GEANT~\cite{geant}, or on
a faster simulation~\cite{ATL-PHYS-PUB-2010-013} using a parameterized
calorimeter response and \GEANT for other detector systems. To account for
additional $pp$ interactions from the same and nearby bunch crossings (pileup),
minimum-bias interactions generated using \PYTHIA
v8.186~\cite{Sjostrand:2014zea}, referred to as \PYTHIA 8 in the following,
with the \tune{A2}~\cite{ATL-PHYS-PUB-2011-014} set of tuned MC parameters
(\tune{A2} tune) were superimposed on the hard-scattering events.  

Simulated events were corrected using per-event weights to describe the
distribution of the average number of interactions per proton bunch-crossing as
observed in data.  All samples were processed through the same reconstruction
software as used for the data.  Simulated events were corrected so that the
object identification, reconstruction and trigger efficiencies, energy scales
and energy resolutions match those determined from data control samples.

The associated production of a top-quark pair with one or two vector bosons was
generated at next-to-leading order (NLO) with \MGMCatNLO~\cite{Alwall:2014hca}
(referred to in the following as \MGAMC) version 2.3.2 interfaced to \PYTHIA 8.
The cross sections for the \ttZ and \ttW processes at \SI{13}{\TeV}, computed
including NLO QCD and electroweak corrections using \MGAMC, are \mbox{$\sttZ =
0.88\,\text{pb}$} and \mbox{$\sttW = 0.60\,\text{pb}$} with an uncertainty of
$\sim\!\!12\%$~\cite{Frixione:2015zaa, Alwall:2014hca, lhcxs}.  The uncertainty
is primarily due to higher-order QCD corrections, estimated by varying the
renormalization (\muR) and factorization (\muF) scales.  The $\gamma^{*}$
contribution and the \Zgamstar interference were included in the \ttZ samples,
with the dilepton invariant mass (\Mll) required to be above 5~\GeV. The
\pdf{NNPDF2\!.\!3NLO} parton distribution function (PDF) set~\cite{Ball:2015cx}
was used in the matrix-element (ME) computation. The
\tune{A14}~\cite{ATL-PHYS-PUB-2014-021} set of tuned MC parameters (\tune{A14}
tune) was used together with the \pdf{NNPDF2\!.\!3LO} PDF
set~\cite{Ball:2012cx} in the parton shower.

The $t$-channel production of a single top quark in association with a $Z$
boson (\tZ) was generated at leading order (LO) using \MGAMC\ v2.2.3 interfaced
to \PYTHIA v6.427~\cite{PythiaManual}, referred to as \PYTHIA 6 in the
following, with the \pdf{CTEQ6L1}~\cite{cteq6} PDF set and the
\tune{Perugia2012}~\cite{Skands:2010ak} set of tuned MC parameters at NLO in
QCD. The four-flavor scheme was used in the generation, and the sample was
normalized using the cross section computed at NLO in QCD using \MGAMC.

The production of a single top quark together with a $W$ and a $Z$ boson (\WtZ)
was generated with \MGAMC\ v2.3.3 using the \pdf{NNPDF3\!.\!0NLO} PDF
set~\cite{Ball:2015cx}. The generation was performed at NLO in QCD using the
five-flavor scheme. Diagrams containing a top-quark pair were removed to avoid
overlap with the \ttZ process. The parton shower was modeled by \PYTHIA 8 with
the \tune{A14} tune. The sample was normalized using the NLO cross section
obtained from the generator.

Events containing $Z$ or $W$ bosons with associated jets were simulated using
the \SHERPA{} 2.2.1 event generator. The matrix-element calculation was
performed using \COMIX~\cite{Gleisberg:2008fv} and
\OPENLOOPS~\cite{Cascioli:2011va} for up to two partons at NLO and four partons
at LO, and merged with the \SHERPA{} parton shower \cite{Schumann:2007mg}
according to the ME+PS@NLO prescription \cite{Hoeche:2012yf}. The
\pdf{NNPDF3\!.\!0NNLO} PDF set was used in conjunction with dedicated
parton-shower tuning developed by the \SHERPA{} authors. The $Z/W$ + jets
samples were normalized to next-to-next-to-leading-order (NNLO) QCD cross
sections for $Z/W$ production calculated by the FEWZ
program~\cite{Gavin:2010az}. 

Diboson processes with four charged leptons ($4\ell$), three charged leptons
and one neutrino ($\ell\ell\ell\nu$) or two charged leptons and two neutrinos
($\ell\ell\nu\nu$) were simulated using the \SHERPA 2.1.1 generator.  The matrix
elements included all diagrams with four electroweak vertices.  They were
calculated including up to three partons at LO, and the \pdf{CT10} PDF
set~\cite{Lai:2010vv} was used in conjunction with a dedicated parton-shower
tune developed by the \SHERPA authors. The invariant mass of any two
opposite-sign, same-flavor (OSSF) leptons was required to be greater
than \SI{5}{\GeV} in the generated events.

The production of three massive vector bosons with subsequent leptonic decays
of all three bosons was modeled at LO with the \SHERPA 2.1.1 generator and the
\pdf{CT10} PDF set. Up to two additional partons were included
in the matrix element at LO and the full NLO accuracy was used for the inclusive
process.  

Electroweak processes involving the vector-boson scattering (VBS) diagram and
producing two same-sign leptons,  two neutrinos and two partons were modeled
using \SHERPA 2.1.1 at LO accuracy and the \pdf{CT10} PDF set. Processes of
orders four and six in the electroweak coupling constant were considered, and up
to one additional parton was included in the matrix element. Other VBS processes
are found to be negligible in the analysis regions considered.

The \POWHEGBOX~\cite{Frixione:2007nw,Powheg, Powbox1, Powbox2} v2 generator
with the \pdf{NNPDF3.0NLO} PDF set was used for the generation of \ttbar
events.  The parton shower and the underlying event were simulated using
\PYTHIA 8 with the \pdf{NNPDF2.3LO} PDF set and the corresponding \tune{A14}
tune. The $h_{\mathrm{damp}}$ parameter, which controls the transverse momentum
of the first gluon emission beyond the Born configuration, was set to 1.5 times
the top-quark mass. The \ttbar samples were normalized to the NNLO
cross-section predictions, including soft-gluon resummation to
next-to-next-to-leading-logarithm order, as calculated with the
\tool{Top++2.0}~\cite{ref:xs6} program.

Electroweak $s$-channel and $t$-channel single-top-quark events, and $Wt$ final
states, were generated with \POWHEGBOX v1, and the parton shower modeled by
\PYTHIA v6.428. The \pdf{CT10} PDF set was used for $s$-channel production and
$Wt$ events, while for $t$-channel production the four-flavor scheme was used
for the NLO matrix element calculations together with the fixed four-flavor PDF
set \pdf{CT10f4}.  Diagram removal was employed to remove the overlap between
\ttbar and \Wt production~\cite{Re:2010bp}. The single-top-quark samples were
normalized to the cross sections computed at NLO reported in
Refs.~\cite{Aliev:2010zk, Kant:2014oha} for the $s$- and $t$-channels and at
NLO with next-to-next-to-leading-logarithm soft-gluon corrections for $Wt$
production~\cite{Kidonakis:2013zqa}.

Samples of \ttbar events produced in association with a Higgs boson (\ttH) were
generated using NLO matrix elements in \MGAMC\ with the \pdf{NNPDF3\!.\!0NLO}
PDF set and interfaced to \PYTHIA 8 for the modeling of the parton shower.
Higgs boson production via gluon--gluon fusion (ggF) and vector-boson fusion
(VBF) was generated using the \POWHEGBOX v2 generator with the \pdf{CT10} PDF
set.  The parton shower and underlying event were simulated using \PYTHIA 8
with the \pdf{CTEQ6L1} PDF set and \tune{AZNLO} tune.  Higgs boson production
with a vector boson was generated at LO using \PYTHIA 8 with the
\pdf{NNPDF2\!.\!3LO} PDF. All Higgs boson samples were normalized using
theoretical calculations presented in Ref.~\cite{lhcxs}.

The SM production of three and four top quarks was generated at LO with
\MGAMC+\PYTHIA 8, using the \tune{A14} tune together with the
\pdf{NNPDF2\!.\!3LO} PDF set. The samples were normalized using cross sections
computed at NLO~\cite{Barger:2010uw,Bevilacqua:2012em}.

The events with a $Z$ or $W$ boson in association with a photon were simulated
with up to three partons at LO using \SHERPA~2.1.1 (ME+PS mode) and the
\pdf{CT10} PDF set. They were normalized to the LO cross section obtained from
the generator.  The $t\bar{t}\gamma$ process was generated at LO with
\MGAMC+\PYTHIA 8, using the \tune{A14} tune together with the
\pdf{NNPDF2\!.\!3LO} PDF set and normalized to the NLO cross section. Events in
the \ttbar sample with radiated photons of high transverse momentum were vetoed
to avoid overlap with those from the $t\bar{t}\gamma$ sample.

\section{Object reconstruction}
\label{s:objects}

Electron candidates~\cite{PERF-2016-01} are reconstructed from energy deposits
(clusters) in the EM calorimeter that are associated with reconstructed tracks
in the ID.  Electrons are required to pass the `medium' likelihood
identification requirements described in Ref.~\cite{PERF-2016-01}. In the SS
dilepton channel, the `tight' likelihood requirement is used instead.  The
electrons are also required to have transverse momentum $\pt > 7\,\GeV$ and
$|\eta_\text{cluster}| < 2.47$, where $\eta_\text{cluster}$ is the
pseudorapidity of the calorimeter energy deposit associated with the electron
candidate.  Candidates in the EM calorimeter barrel/endcap transition region
$1.37 < |\eta_\text{cluster}| < 1.52$ are excluded.  

Muon candidates are reconstructed from a fit to track segments in the various
layers of the muon spectro\-meter, matched with tracks identified in the inner
detector.  Muons are required to have $\pt > 7\,\GeV$ and $\abseta < 2.5$ and
to pass the `medium' identification requirements defined in
Ref.~\cite{PERF-2015-10}. The `medium' criteria include requirements on the
numbers of hits in the ID and MS as well as a compatibility requirement between
momentum measurements in the ID and MS. They provide a high efficiency and
purity of selected muons. Electron candidates sharing a track with a muon
candidate are removed.  

To reduce the background due to non-prompt leptons from hadron decays, photon
conversions or jets misidentified as leptons (labeled as ``fake leptons''
throughout this paper), electron and muon candidates are required to be
isolated. In the OS dilepton and the tetralepton channels, as well as in those
trilepton regions that target the \ttZ process, the total sum of the track
transverse momenta in a surrounding cone of size $\Delta R_{\eta} \equiv
\sqrt{(\Delta\eta)^2 + (\Delta\phi)^2} = \min(10\,\GeV/\pT,r_{e,\mu})$,
excluding the track of the candidate, is required to be less than 6\% of the
candidate \pt, where $r_e = 0.2$ and $r_{\mu} = 0.3$.  In addition, the sum of
the cluster transverse energies in the calorimeter within a cone of size
$\Delta R_{\eta} = 0.2$ around any electron candidate, excluding energy
deposits of the candidate itself, is required to be less than 6\% of the
candidate \pt.

In the SS dilepton channel and those trilepton regions targeting the \ttW
process, where the fake-lepton background is particularly important, tighter
isolation requirements are imposed on candidate leptons. A multivariate
discriminant is built to distinguish prompt leptons from leptons arising from
heavy-hadron decays inside jets~\cite{HIGG-2017-02}. The discriminant uses
information from charged-particle tracks in a cone around the lepton candidate.
Jets are reconstructed from these tracks to obtain a track-jet, and the
discriminant is constructed from information such as the angular distance
between the lepton and the track-jet, the number of tracks in the track-jet and
the ratio of the lepton candidate \pt to the track-jet \pt. The rejection
factor obtained for leptons from $b$-hadron decays is about 20, while the
prompt-lepton efficiency is about 85\% (80\%) for $\pt \sim \SI{20}{\GeV}$ and
reaches a plateau of $\sim 98\%$ (96\%) for muons (electrons) at high \pt.
Simulated events are corrected to account for differences in the prompt-lepton
tagging discriminant between data and simulation. 

Another important background in the SS dilepton channel arises from electrons
with misidentified charge. To suppress this background, another multivariate
discriminant is used, which takes as inputs various track and cluster
properties of the electron candidates~\cite{HIGG-2017-02}. The discriminant
provides a 95\% efficiency for electrons with correct charge reconstruction
while achieving a rejection factor of $\sim 17$ for electrons with
misidentified charge that pass the `tight' likelihood identification
requirement. Correction factors are applied to selected electrons to match the
efficiency of the discriminant in simulation to that measured in data. 

For both the electrons and muons, the longitudinal impact parameter of the
associated track relative to the primary vertex,\footnote{A primary vertex
candidate is defined as a vertex with at least two associated tracks,
consistent with the beam collision region. The vertex candidate with the
largest sum of squared transverse momenta of its associated tracks is taken as
the primary vertex.} $z_{0}$, is required to satisfy $|z_0 \sin\theta|<0.5$ mm.
The significance of the transverse impact parameter $d_0$ is required to
satisfy $|d_0|/\sigma(d_0)<5$ for electrons and $|d_0|/\sigma(d_0)<3$ for
muons, where $\sigma(d_0)$ is the uncertainty in $d_0$.

Jets are reconstructed using the anti-$k_t$ algorithm~\cite{Cacciari:2008gp,
Cacciari:2005hq} with radius parameter $R = 0.4$, starting from topological
energy clusters in the calorimeters~\cite{PERF-2014-07}. The effect of pileup
on jet energies is accounted for by a jet-area-based
correction~\cite{Cacciari:2008gn} and the energy resolution of the jets is
improved by using global sequential corrections~\cite{PERF-2016-04}. Jets are
calibrated to the hadronic energy scale using energy- and
pseudorapidity-dependent calibration factors derived from data. Jets are
accepted if they fulfill the requirements $\pT > 25\,\GeV$ and $|\eta| < 2.5$.
To reduce the contribution from jets associated with pileup, jets with $\pT <
60\,\GeV$ and $|\eta| < 2.4$ are required to satisfy pileup rejection criteria
(JVT), based on a multivariate combination of track-based variables~\cite{PERF-2014-03}. 

Jets are tagged as likely to contain $b$-hadrons ($b$-tagged) with a
multivariate discriminant making use of the long lifetime, large decay
multiplicity, hard fragmentation and high mass of $b$-hadrons
\cite{PERF-2016-05}.  For the working point used in this analysis, the average
efficiency for correctly tagging a $b$-jet is approximately $77\%$, as
determined in simulated \ttbar events.  In simulation, the tagging algorithm
gives a rejection factor of $134$ against light-quark and gluon jets, and $6.2$
against charm-quark jets.  The $b$-tagging efficiency and mistagging rates in
simulation are corrected to reproduce those in data~\cite{PERF-2016-05}.

The missing transverse momentum $\mathbf{p}^\text{miss}_\text{T}$, with
magnitude \met, is a measure of the transverse momentum imbalance due to
particles escaping detection. It is computed~\cite{PERF-2014-04} as the
negative sum of the transverse momenta of all electrons, muons and jets and an
additional soft term. The soft term is constructed from all tracks that are
associated with the primary vertex but not with any lepton or jet.  In this
way, the \met is adjusted for the best calibration of the jets and the other
identified objects, while maintaining pileup independence in the soft
term~\cite{PERF-2014-04, PERF-2016-07}.

To prevent double-counting of electron energy deposits as jets, the closest jet
within $\Delta R_y = 0.2$ of a reconstructed electron is removed, where $\Delta
R_y \equiv \sqrt{(\Delta y)^2 + (\Delta\phi)^2}$ and $y$ is the rapidity of the
electron.  If the nearest jet surviving the above criterion is within $\Delta
R_y = 0.4$ of an electron, the electron is discarded to ensure that selected
electrons are sufficiently separated from nearby jet activity.  To reduce the
background from muons originating from heavy-flavor particle decays inside
jets, muons are removed if they are separated from the nearest jet by $\Delta
R_y < 0.4$.  However, if this jet has fewer than three associated tracks, the
muon is kept and the jet is removed instead; this ensures that high-energy
muons undergoing significant energy loss in the calorimeter are retained.

\section{Event selection and background estimation}
\label{s:selection}

Table~\ref{tab:intro-channels} lists the analysis channels and the targeted
decay modes of the \ttZ and \ttW processes.  Each channel is divided into
multiple analysis regions in order to enhance the sensitivity to the signal.
Simultaneous fits to the signal regions and dedicated control regions are
performed to extract the cross sections for \ttZ and \ttW production.

\begin{table}[htbp]
\centering
\caption{\label{tab:intro-channels} List of \ttW and \ttZ decay modes and
analysis channels targeting them. The symbols $b$ and $\nu$ denote a bottom
quark or antiquark and neutrino or antineutrino, respectively, with charge
conjugation implied.\spaceit}
\begin{tabular}{llll}
\toprule
Process & \ttbar decay & Boson decay & Channel\\
\midrule
\multirow{2}{*}{$\ttW$}
& $ (\ell^{\pm}\nu b) (q\bar{q} b) $ & $\ell^{\pm}\nu$ & SS dilepton\\
& $ (\ell^{\pm}\nu b) (\ell^{\mp}\nu b)$ & $\ell^{\pm}\nu$ & Trilepton\\
\midrule
\multirow{3}{*}{\ttZ}
& $(q\bar{q} b) (q\bar{q} b)$ & $\ell^{+}\ell^{-}$ & OS dilepton\\
& $(\ell^{\pm}\nu b) (q\bar{q} b)$ & $ \ell^{+}\ell^{-}$ & Trilepton\\
& $(\ell^{\pm}\nu b) (\ell^{\mp} \nu b)$ & $ \ell^{+}\ell^{-}$ & Tetralepton\\
\bottomrule
\end{tabular}
\end{table}

Only events collected using single-electron or single-muon triggers are
accepted.  Events are required to have at least one reconstructed primary
vertex.  In all regions considered, at least one reconstructed lepton with
$\pT > 27\,\GeV$ is required to match ($\Delta R_{\eta} < 0.15$) a lepton with
the same flavor reconstructed by the trigger algorithm.  Four channels are
defined: same-sign dilepton, opposite-sign dilepton, 
trilepton and tetralepton. 

The shapes of background distributions containing prompt leptons are modeled by
simulation. The  normalizations for the $WZ$ and $ZZ$ processes, as well as the
production of a $Z$ boson in association with heavy-flavor (HF) jets, are taken
from data control regions as defined in this section and which are included in
the fit discussed in Section~\ref{s:results}.  The yields in these data control
regions are extrapolated to the signal regions using simulation. Systematic
uncertainties in the extrapolation are taken into account in the overall
uncertainty in the background estimate.

The contribution from events containing an electron with misidentified charge
(referred to as ``charge-flip'' in the following) is estimated from data. The
charge-flip probability is extracted in events containing a pair of electrons
with \Mll\ close to the $Z$ boson mass. It is parameterized in $\pt$ and $\eta$
and is found to range from around 0.01\% to 2\%, increasing with $\pt$ and $\eta$,
for electrons passing the identification and isolation criteria applied in the
SS dilepton channel. The probability is extracted by maximizing a likelihood
which relates the number of events in which the two electrons have the same
charge to the total number of observed events. The background contribution from
events other than charge-flip electrons in the same-sign region is estimated
from a sideband of the \Mll\ distribution and subtracted before performing the
likelihood fit. 

The charge-flip background contribution in any SS dilepton region is estimated
by constructing a control region with identical requirements, but without 
any requirement on the lepton charge, and applying the appropriate
charge-flip probabilities.

Background sources involving one or more fake leptons are modeled using data
events from dedicated regions. For the SS dilepton (\SSLSR) and trilepton
channels the fake-lepton background is estimated using the matrix
method~\cite{TOPQ-2010-01}.  The matrix method makes use of events with the
same selection as the region of interest, but for which the electron
identification criteria are relaxed to the `loose' likelihood requirement of
Ref.~\cite{PERF-2016-01}, and neither electrons nor muons are required to be
isolated. These leptons are referred to as loose leptons, whereas leptons
satisfying the full set of identification and isolation criteria of
Section~\ref{s:objects} are referred to as tight leptons. The fake-lepton
background in any region of interest is obtained from the aforementioned events
using efficiencies for prompt and fake leptons to satisfy the tight criteria.

The lepton efficiencies are extracted in control regions with a likelihood fit,
by using the model of the matrix method, and assuming that the number of events
with two fake leptons is negligible.  The control regions are defined in
dilepton events, separately for events with exactly one $b$-tagged jet and
$\ge2\,b$-tagged jets. The prompt lepton efficiencies are measured in inclusive
OSSF events, while fake-lepton efficiencies are measured in events with
same-sign leptons. Both the prompt and fake-lepton efficiencies are
parameterized as a function of the lepton $\pt$. The measurement of fake-lepton
efficiencies is performed after subtracting the estimated contribution from
charge-flip events.  Contributions from processes with two prompt same-sign
leptons or one real lepton and a photon conversion (arising mainly from
$t\bar{t}\gamma$ production) are estimated from simulation and are also
subtracted. The prompt and fake-lepton efficiencies are extracted separately
for the regions targeting \ttZ and those targeting \ttW, due to the different
lepton isolation requirements applied in the two sets of regions.

In the \FLC, the matrix method is not used due to the small number of events in
data with four selected leptons. Instead, the contribution from backgrounds
containing fake leptons is estimated from simulation and corrected with scale
factors determined in control regions. The contributions from events containing
a photon conversion (denoted by $\gamma+X$) in the SS dilepton and trilepton
channels are estimated from simulation and scaled with these correction
factors, obtained separately for lepton type and origin.

\subsection{Opposite-sign dilepton analysis}
\label{s:selection2lOS}
The OS dilepton analysis targets the \ttZ process, where both top quarks decay
hadronically and the $Z$ boson decays to a pair of leptons (electrons or
muons).  Events are required to have exactly two OSSF leptons.  Events with
additional isolated leptons are rejected. The invariant mass of the lepton pair
is required to be in the $Z$ boson mass window, $|m_{\ell\ell} - m_{Z}| <
\SI{10}{\GeV}$. The leading (subleading) lepton is required to have a
transverse momentum of at least 30 (15)~\GeV.

The OS dilepton analysis is affected by large backgrounds from $Z$+jets or
\ttbar production, both characterized by the presence of two leptons.  In order
to improve the signal-to-background ratio and constrain these backgrounds from
data, three separate analysis regions are considered, depending on the number
of jets ($n_\text{jets}$) and number of $b$-tagged jets (\nbtags): \OSLSRB,
\OSLSRA and \OSLSRC. The signal region requirements are summarized in
Table~\ref{tab:SR2lOSSF_sel}. In signal region \OSLSRB, exactly five jets are
required, of which at least two must be $b$-tagged. In \OSLSRA (\OSLSRC), at
least six jets are required with exactly one (at least two) being $b$-tagged
jets. 

\begin{table}[htbp]
\begin{center} 
\caption{Summary of the event selection requirements in the OS dilepton signal
regions.  \spaceit}
\label{tab:SR2lOSSF_sel}
\begin{tabular}{@{}lrrr@{}}
 \toprule
Variable  & \OSLSRA &  \OSLSRB &  \OSLSRC\\
\midrule
Leptons                 &\multicolumn{3}{c}{$=2$, same flavor and opposite sign}\\
\Mll			&\multicolumn{3}{c}{$|\Mll - m_{Z}| <$ \SI{10}{\GeV}}\\
$\pt$ (leading lepton) &\multicolumn{3}{c}{$>$\SI{30}{\GeV}}\\
$\pt$ (subleading lepton) &\multicolumn{3}{c}{$>$\SI{15}{\GeV}}\\
\nbtags	              & 1 & $\ge2$ & $\ge2$  \\
$n_{\text{jets}}$     & $\ge6$ & 5 & $\ge6$\\
\bottomrule
\end{tabular}
\end{center}
\end{table}

In order to separate signal from background, boosted decision trees (BDTs) are
used. The BDTs are constructed and trained separately for each region against
all the contributing backgrounds, using as input 15, 14 and 17 variables for
\OSLSRA, \OSLSRB and \OSLSRC, respectively. Fourteen of the variables are
common to the three regions. The details of the variables used are given in
Table~\ref{tab:NN_inputv_vars} in the Appendix. In all three regions, the
variables with the largest discriminative power are found to be:

\begin{itemize}
\item the $\eta$ of the dilepton system,
\item the scalar sum of transverse momenta of all jets divided by the sum of
their energies,
\item the first Fox--Wolfram moment $H_1$~\cite{Fox:1978vu}.
\end{itemize}

Each of the signal regions is further divided into 19 equal-size bins of the
BDT distribution.  To avoid relying on simulation for the normalization of
$Z$+HF jet production, the $Z$+jets background is constrained by using events
with low values of the BDT discriminant. The simulated $Z$+jets background is
split into three components, $Z$+0HF, $Z$+1HF and $Z$+2HF, depending on the
number of reconstructed jets which are matched to a generator-level $b$- or
$c$-hadron (heavy-flavor, or HF jets). The normalization factors of the $Z$+1HF
and $Z$+2HF components of the $Z$+jets background are determined from the fit
to data, as described in Section~\ref{s:results}, while the normalization of
the $Z$+0HF component is taken from simulation.

\begin{figure}[htbp]
\centering
\subfloat[\label{fig:os_ttbar1}]{\includegraphics[width=\twofigwidth]{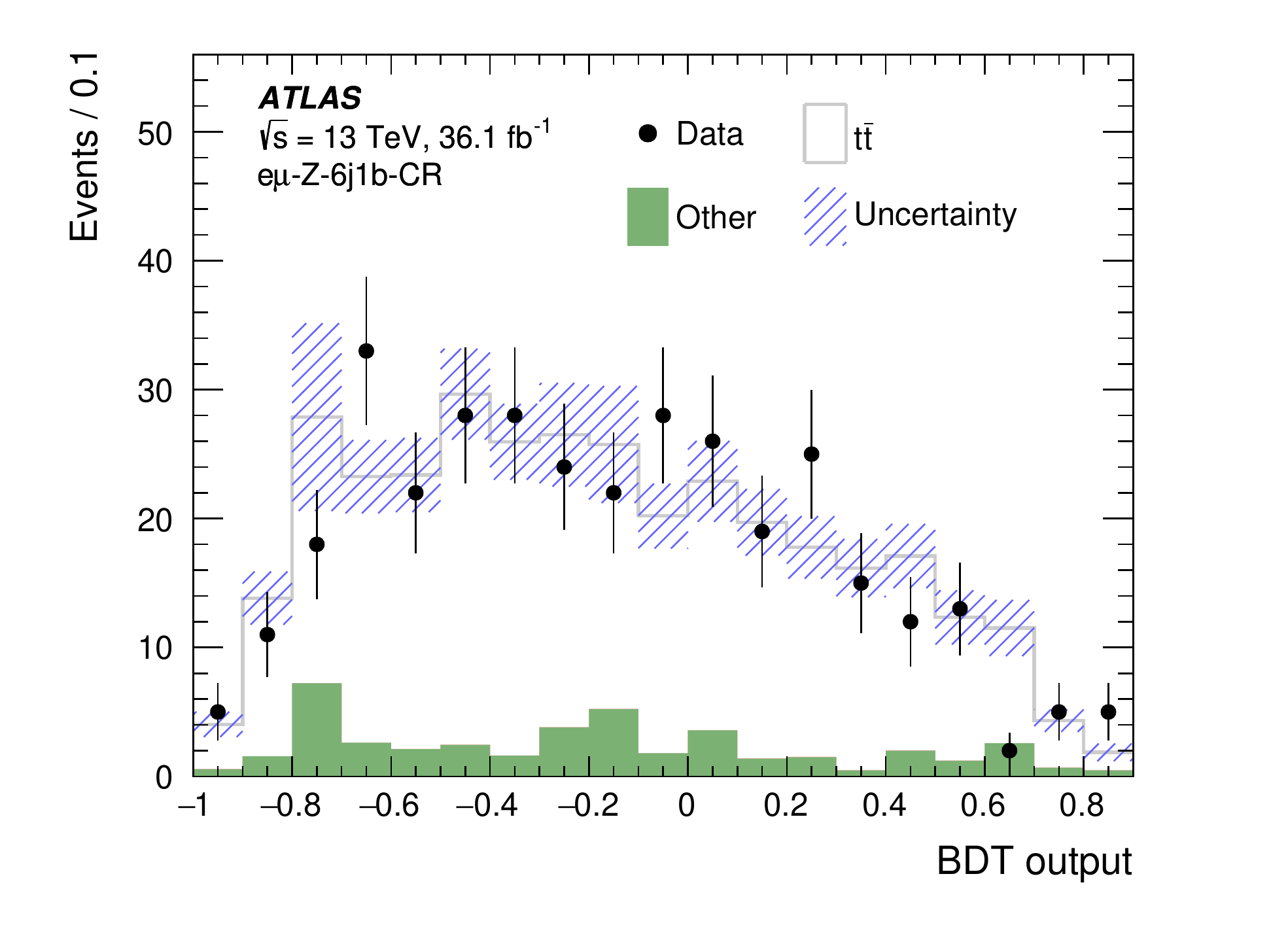}}
\subfloat[\label{fig:os_ttbar2}]{\includegraphics[width=\twofigwidth]{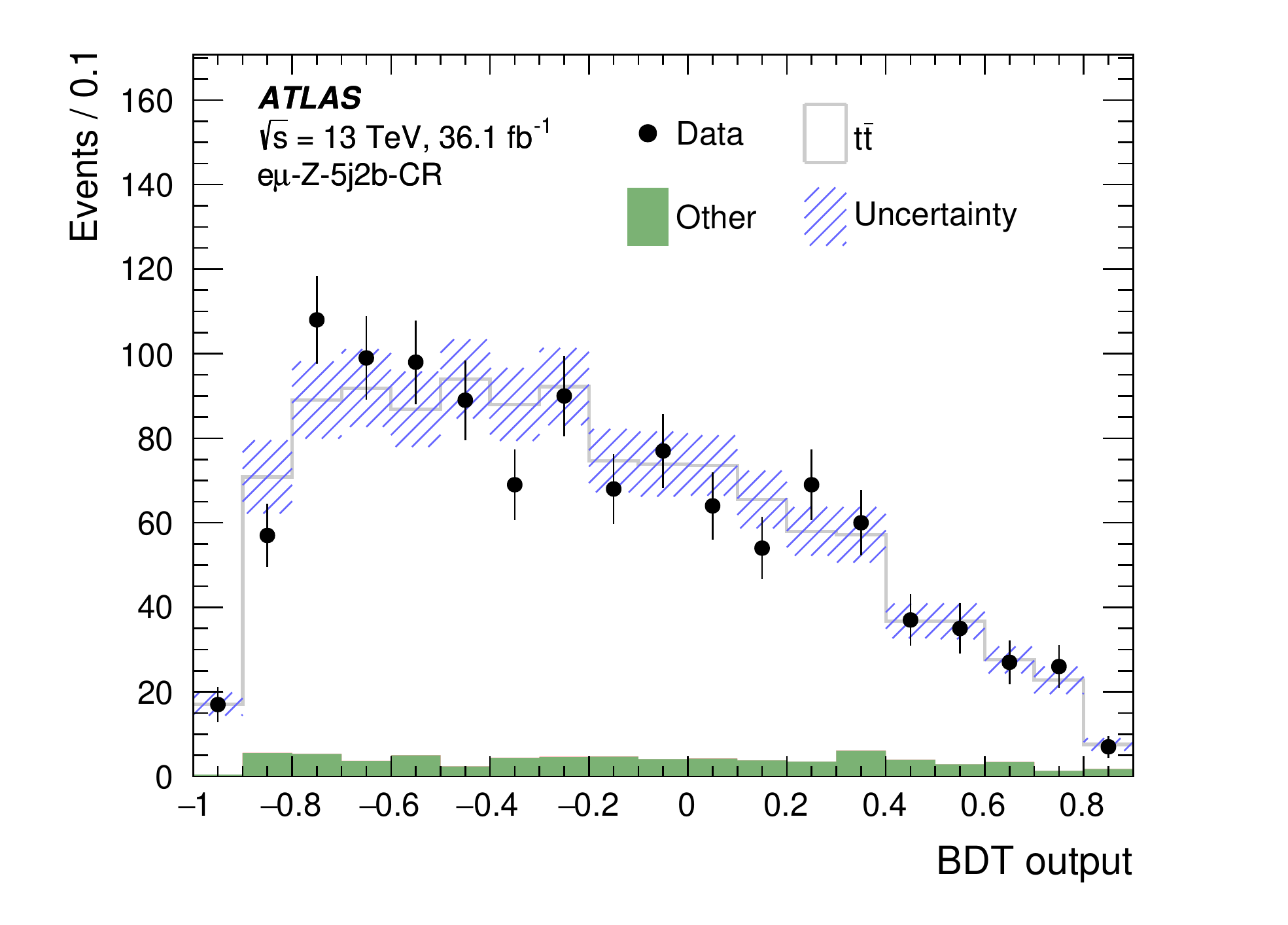}}\\
\subfloat[\label{fig:os_ttbar3}]{\includegraphics[width=\twofigwidth]{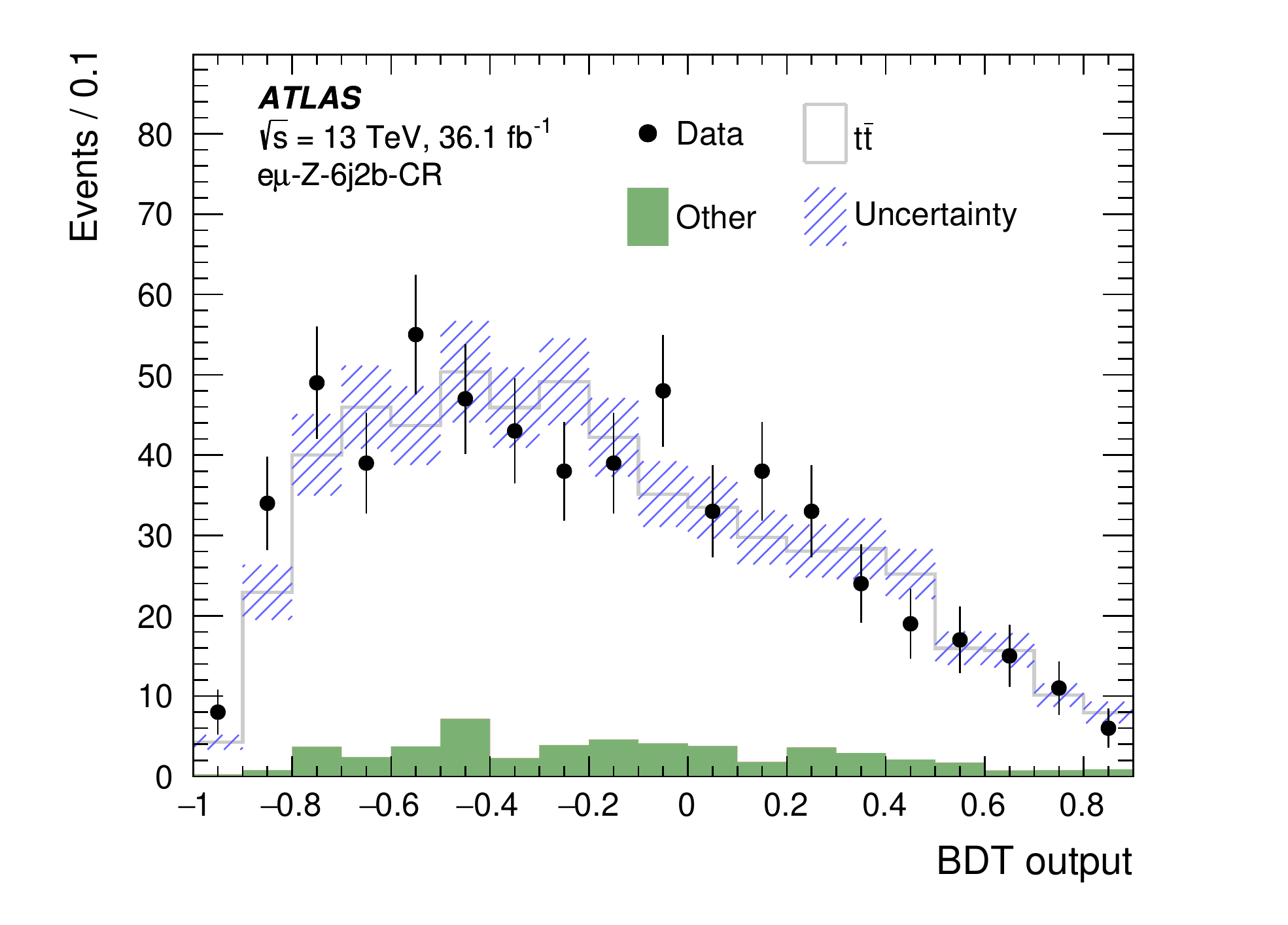}
} \caption{\label{fig:os_ttbar} The BDT distributions for the \ttbar control
regions \protect\subref{fig:os_ttbar1} \OSLSRA, \protect\subref{fig:os_ttbar2}
\OSLSRB, \protect\subref{fig:os_ttbar3} \OSLSRC. \hatch.  The `Other'
background contains SM processes with small cross sections producing two
opposite-sign prompt leptons, including the \ttZ process, whose contribution
is negligible. }
\end{figure}

A data-driven method is used to estimate the \ttbar background in the OS
dilepton signal regions. Control regions are defined which are identical to the
signal regions, except that the requirement of two leptons with the same flavor
and opposite sign is replaced by the requirement of two leptons with different
flavors and opposite sign. In this manner, three regions enriched in \ttbar
background are obtained.  The number of \ttbar events in each same-flavor
dilepton region is estimated from corresponding opposite-flavor regions,
corrected for non-\ttbar backgrounds and differences in contributions from
leptonic $\tau$-lepton decays.  This procedure is applied to each bin of the
distribution under consideration.  Figure~\ref{fig:os_ttbar} shows the BDT
distributions for the \ttbar control regions. Agreement between the data and
the expectation is observed.

\subsection{Same-sign dilepton analysis}

The SS dilepton signal regions target the \ttW process.  Events are required to
have two lepton candidates with the same sign and $\pt>27\,\GeV$.  The scalar
sum of the \pT of selected leptons and jets, \HT, is required to be above
$240\,\GeV$. Events containing additional loose leptons (with $\pT>7\,\GeV$)
are vetoed. Twelve signal regions are defined in total, categorized by the
number of $b$-tagged jets (one or $\ge2$) as well as the charge and the flavor
of the selected leptons. The signal regions are denoted by \eloneeltwo-SSp-$1b$,
\eloneeltwo-SSm-$1b$, \eloneeltwo-SSp-$2b$ and \eloneeltwo-SSm-$2b$, where 
``p'' or ``m'' indicates the charge of the selected
leptons. Considering separate signal regions for positive and negative charges
of the leptons increases the sensitivity of the analysis since \ttW events are
preferentially produced with positively charged $W$ bosons, while the
fake-lepton background and other processes such as \ttZ and $t\bar{t}H$ are
expected to be charge symmetric.

The event selection requirements in the SS dilepton regions are summarized in
Table~\ref{tab:SRs2l}. The presence of at least four jets and $\met>40\,\GeV$
is required in all signal regions except $2\mu$-SSp-$2b$ and $2\mu$-SSm-$2b$.
In these regions, the \met requirement is loosened to $\met>20\,\GeV$, and at
least two jets are required.  In the $2e$ and $2\mu$ signal regions, events
containing a pair of leptons whose invariant mass is within $10\,\GeV$ of the
$Z$ boson mass are vetoed. 

\begin{table}[htbp] 
\begin{center} 
\caption{\label{tab:SRs2l}Summary of the event selection requirements in the SS
dilepton signal regions.\spaceit}
\begin{tabular}{l c c c c} 
\toprule
Requirement &  \eloneeltwo-SS(p,m)-$1b$ & $2e$-SS(p,m)-$2b$ & $e\mu$-SS(p,m)-$2b$ & $2\mu$-SS(p,m)-$2b$\\
\midrule
\nbtags & =1 & $\geq2$ & $\geq2$ & $\geq2$\\ 
$\met$ & $>40$ \GeV & $>40$ \GeV  & $>40$ \GeV & $>20$ \GeV\\
$\HT$ & \multicolumn{4}{c}{$>240$ \GeV}\\ 
$\pt$ (leading lepton)  & \multicolumn{4}{c}{$>27$ \GeV}\\
$\pt$ (subleading lepton)  & \multicolumn{4}{c}{$>27$ \GeV}\\
$n_{\mathrm{jets}}$ &  $\ge 4$ & $\ge4$ & $\ge4$ & $\ge 2$ \\
$Z$ veto & \multicolumn{4}{c}{ $|m_{\ell\ell} - m_{Z}| > \SI{10}{\GeV}$ in the $2e$ and $2\mu$ regions}\\
\bottomrule
\end{tabular} 
\end{center} 
\end{table}

The control regions used to measure the fake-lepton efficiencies, as explained
at the beginning of this section, are defined to be orthogonal to the SS dilepton
signal regions: either the \met, $\HT$ or jet multiplicity requirements of the
signal regions are not satisfied.  The \ttW signal contribution in the control
regions where both leptons satisfy the ``tight'' criteria is found to be
non-negligible. To enhance the sensitivity of the analysis, the latter regions
are also included in the final fit used to measure the \ttW cross section, as
discussed in Section~\ref{s:results}. These six regions are further split
according to the charge of the leptons, and the resulting twelve regions are
denoted by \eloneeltwo-SS(p,m)-$(1,2)b$-CR, following the same notation as for
the signal regions defined above. In each control region, both leptons are
required to have $\pt>27\,\GeV$, and at least one (two) jets are required in
the $1b$ ($2b$) regions.  In addition, events containing a pair of leptons
whose invariant mass is compatible with the $Z$ boson mass are vetoed.  The
largest contamination from \ttW is found to be $25\%$, in the region
2$\mu$-SSp$2b$-CR.

\begin{figure}[htbp]
\centering
\subfloat[\label{fig:ssmm_val_a1}]{
  \includegraphics[width=\twofigwidth]{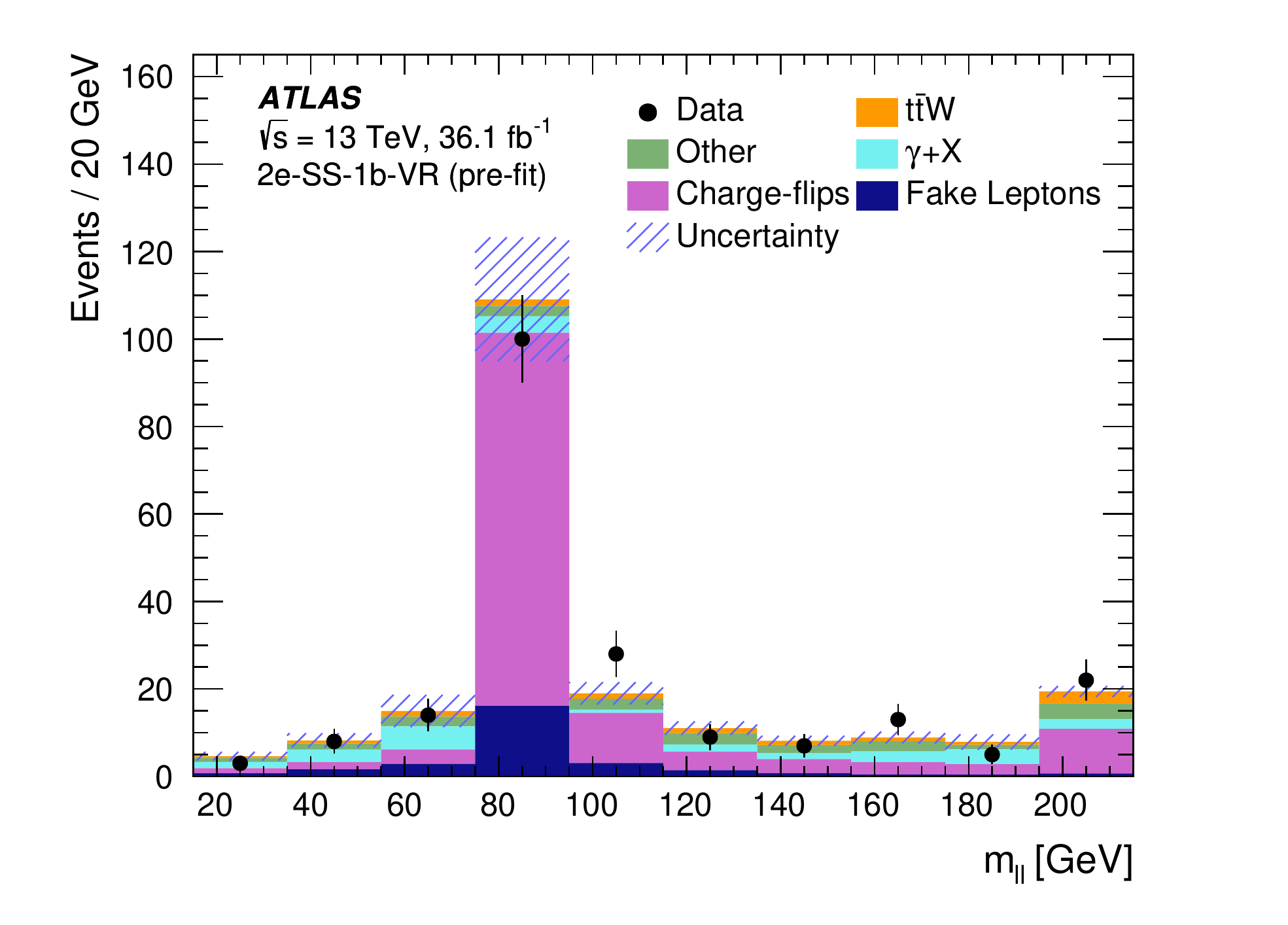}
}
\subfloat[\label{fig:ssmm_val_a2}]{
  \includegraphics[width=\twofigwidth]{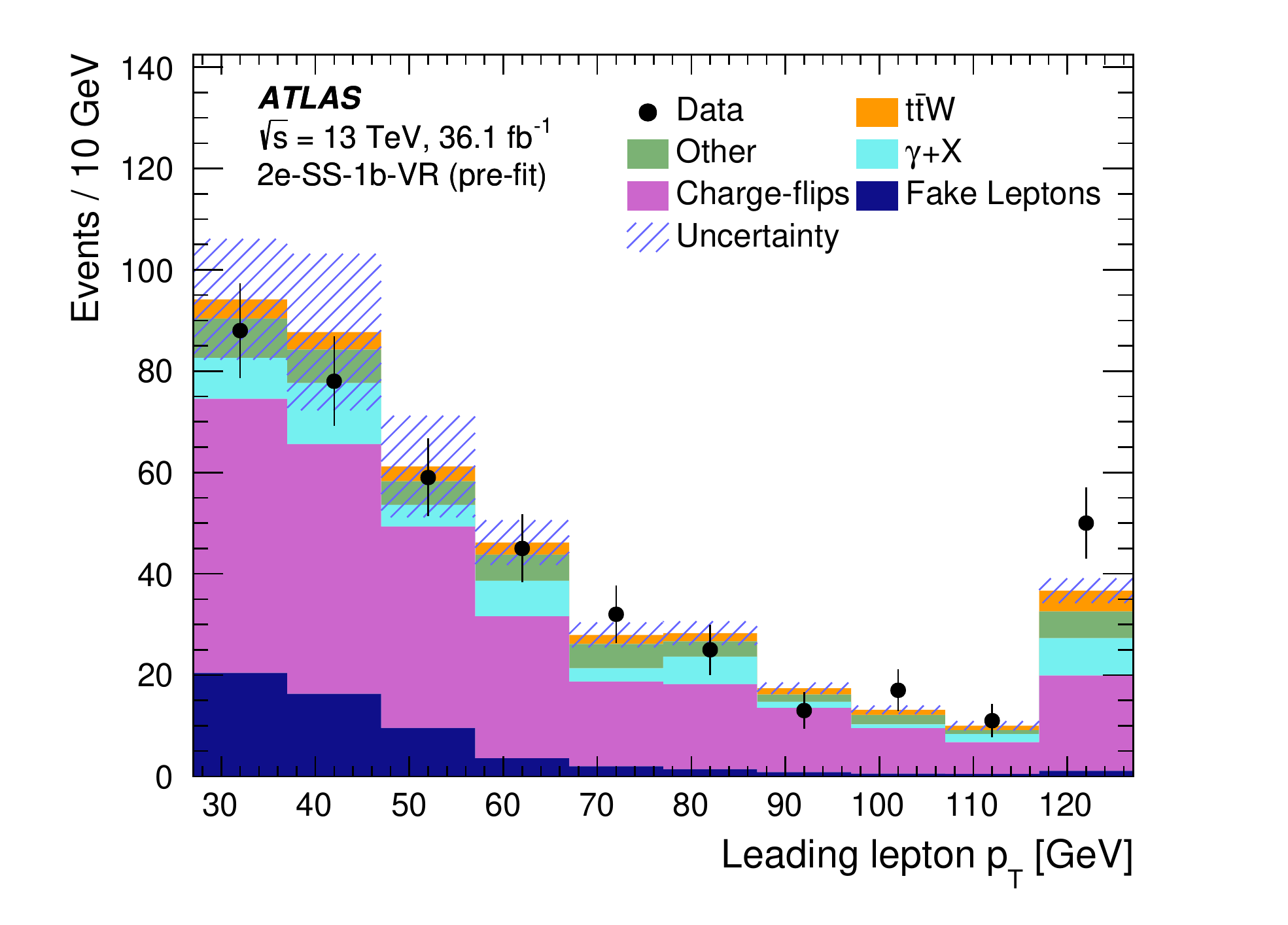}
}
\caption{Distributions in the 2$e$-SS-1b-VR validation region:
\protect\subref{fig:ssmm_val_a1} the invariant mass $m_{\ell\ell}$ of the
lepton pair and \protect\subref{fig:ssmm_val_a2} leading lepton transverse
momentum \pt. \hatch.  The `Other' background contains SM processes with small
cross sections producing two same-sign prompt leptons.  \oflo.}
\label{fig:ssmm_val_a} 
\end{figure}

The dominant background in the \SSLSR\ signal regions arises from events
containing fake leptons. Backgrounds from the production of prompt leptons with
correctly identified charge come primarily from $t\bar{t}H$ and $WZ$
production.  The charge-flip background is also significant in signal regions
with two electrons. In regions with two muons, this background is negligible as
the probability of misidentifying the charge of a muon in the relevant \pT
range is very small.  To validate the charge-flip background, a validation
region called 2$e$-SS-1b-VR is constructed similarly to the $2e$-SS signal
regions, except that the number of jets is required to be between one and
three, to ensure orthogonality with the signal regions. The requirement on \HT
is also removed, exactly one jet is required to be $b$-tagged, and the
invariant mass of the lepton pair is required to be greater than 15~\GeV. The
distributions of \Mll\ and the leading lepton \pt are shown in
Figure~\ref{fig:ssmm_val_a}, demonstrating good modeling of the charge-flip
background. 

\begin{figure}[htbp]
\centering
\subfloat[\label{fig:ssmm_val_b1}]{
  \includegraphics[width=\twofigwidth]{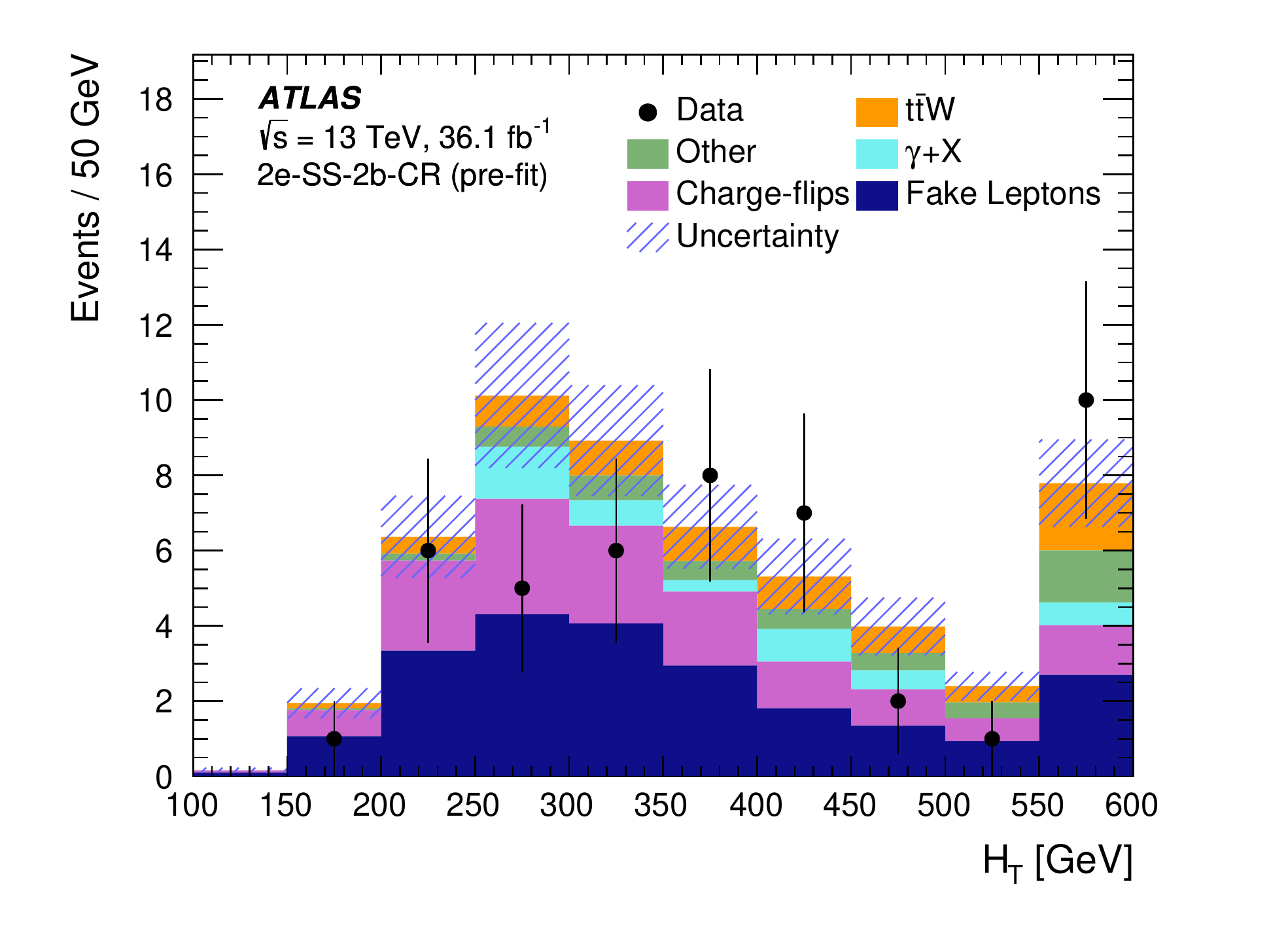}
}
\subfloat[\label{fig:ssmm_val_b2}]{
  \includegraphics[width=\twofigwidth]{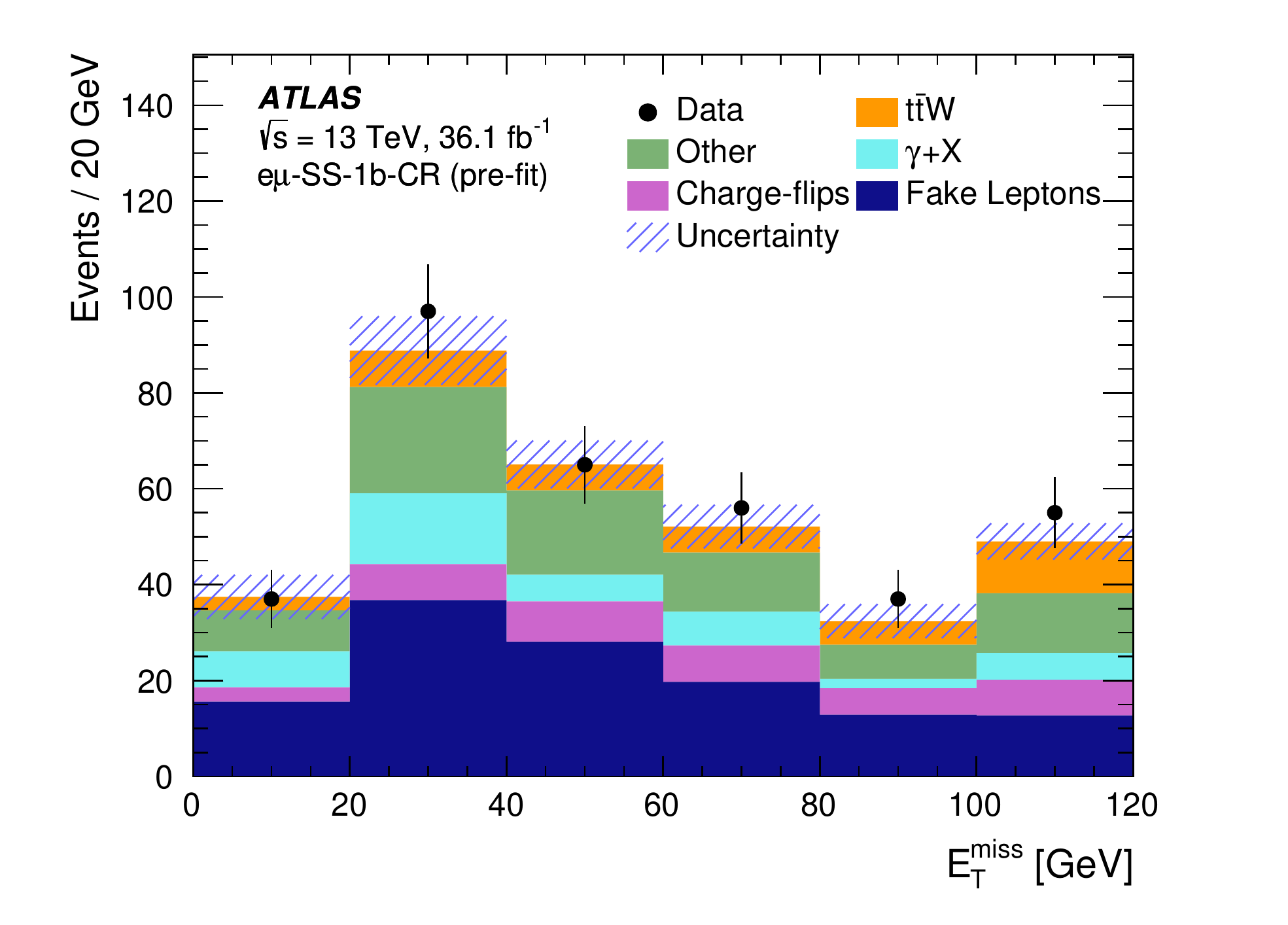}
}\\
\subfloat[\label{fig:ssmm_val_b3}]{
  \includegraphics[width=\twofigwidth]{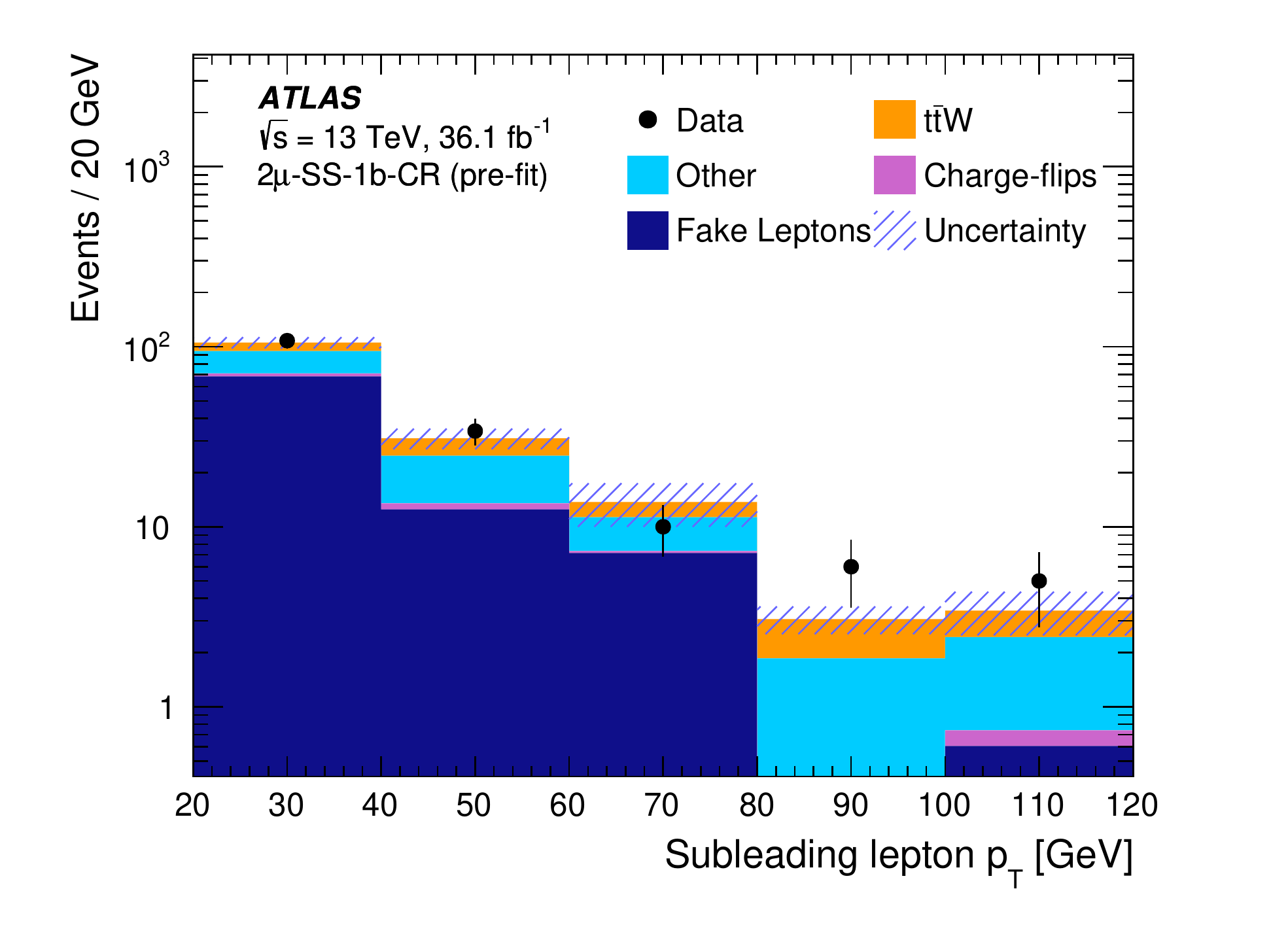}
}
\caption{\label{fig:ssmm_val_b} Distributions in fake-lepton control regions:
\protect\subref{fig:ssmm_val_b1} scalar sum of transverse momenta of leptons
and jets, \HT, in the $2e$-SS-$2b$-CR region, \protect\subref{fig:ssmm_val_b2}
missing transverse momentum, \met, in the $e\mu$-SS-$1b$-CR region and
\protect\subref{fig:ssmm_val_b3} subleading lepton transverse momentum, \pt, in
the $2\mu$-SS-$1b$-CR region. \hatch.  The `Other' background contains SM
processes with small cross sections producing two same-sign prompt leptons.
\oflo.}
\end{figure}

Figure~\ref{fig:ssmm_val_b} shows the distributions of \HT, \met and the
subleading lepton \pT, for the control regions $2e$-SS-$2b$-CR,
$e\mu$-SS-$1b$-CR and $2\mu$-SS-$1b$-CR. The data and the expectation agree
well, demonstrating the validity of the description of the fake-lepton
background determined by the matrix method.

\begin{figure}[htbp]
\centering
\subfloat[\label{fig:2lss_sr1}]{\includegraphics[width=\twofigwidth]{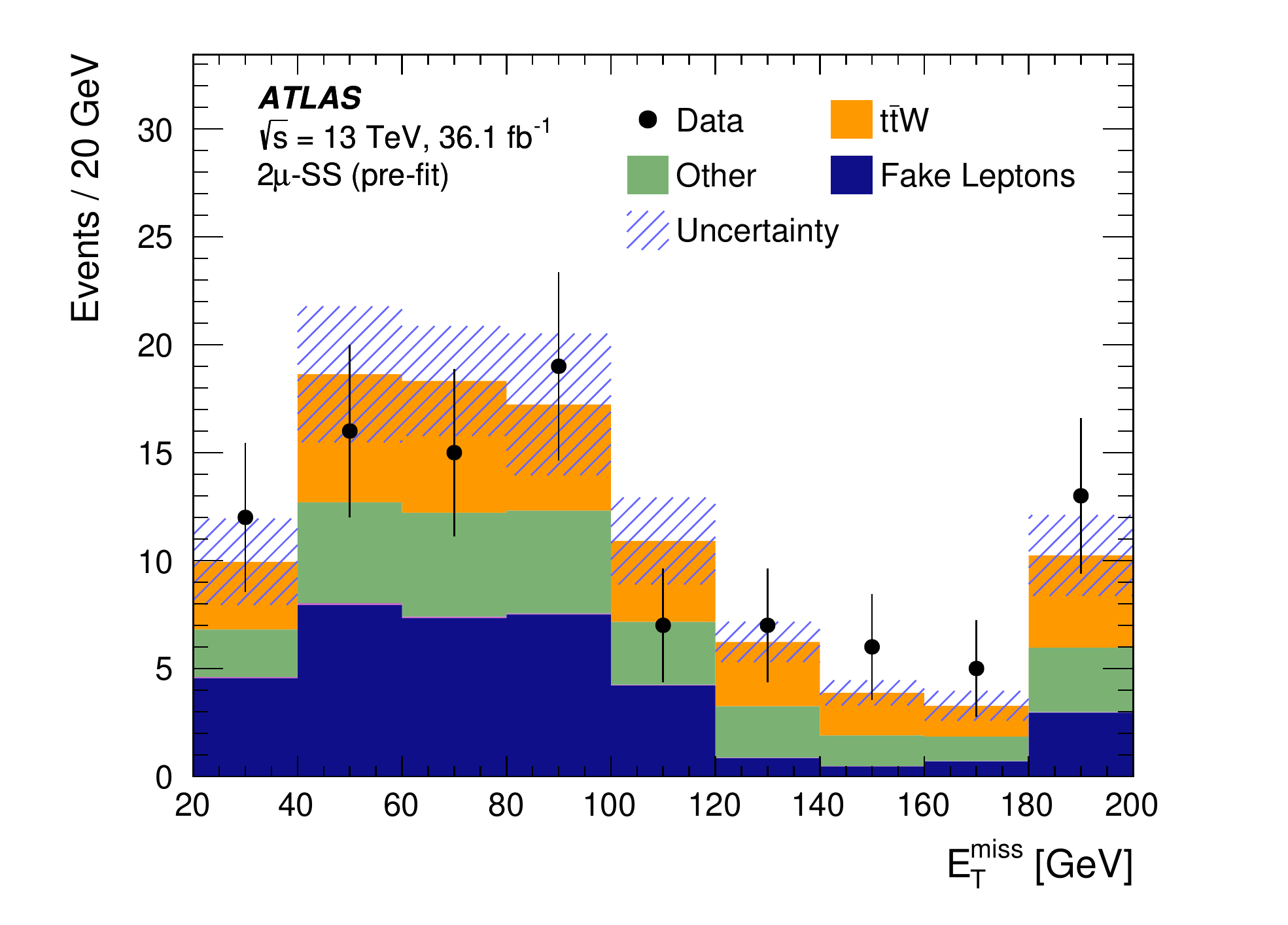}}
\subfloat[\label{fig:2lss_sr2}]{\includegraphics[width=\twofigwidth]{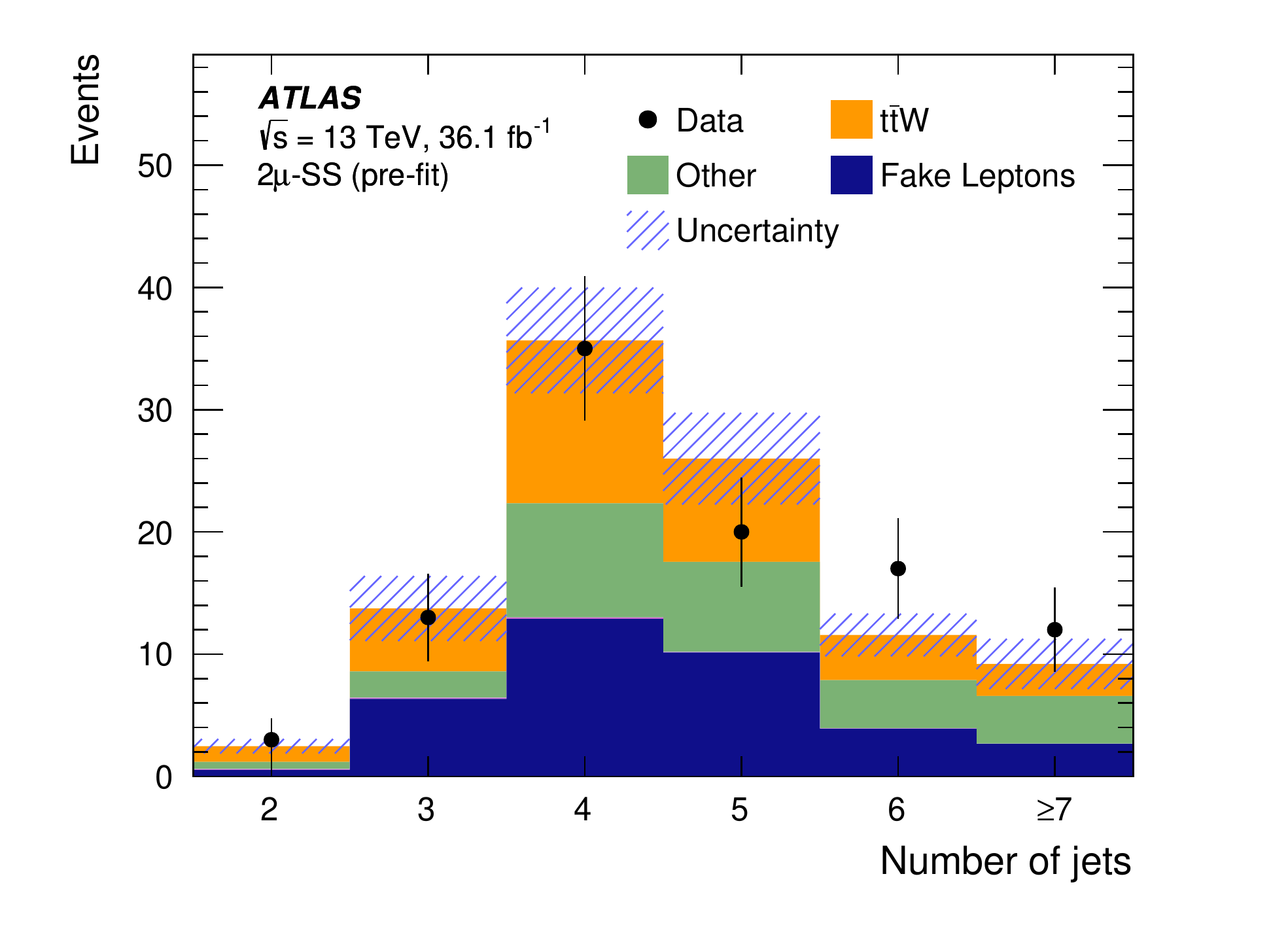}}\\
\subfloat[\label{fig:2lss_sr3}]{\includegraphics[width=\twofigwidth]{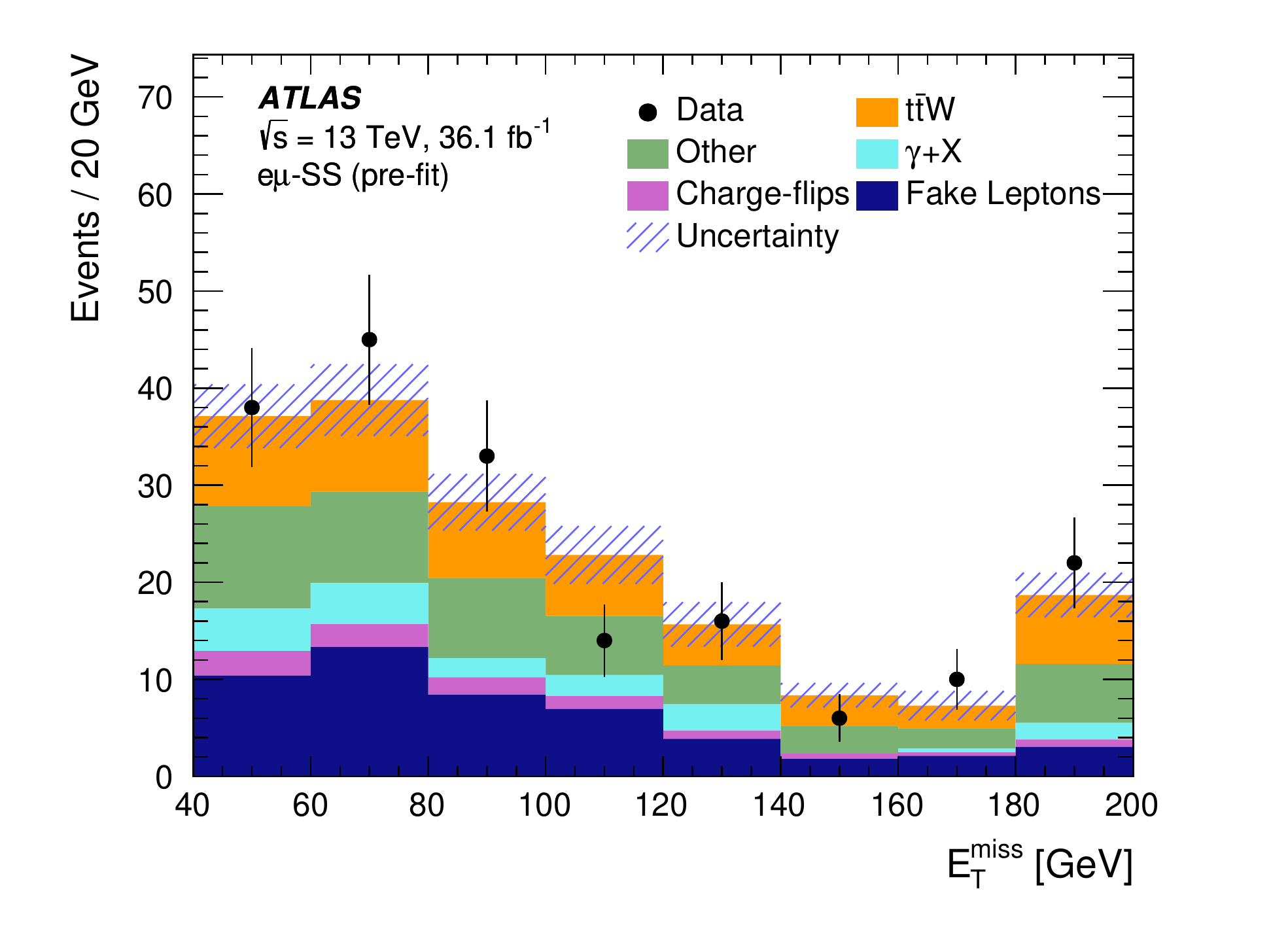}}
\subfloat[\label{fig:2lss_sr4}]{\includegraphics[width=\twofigwidth]{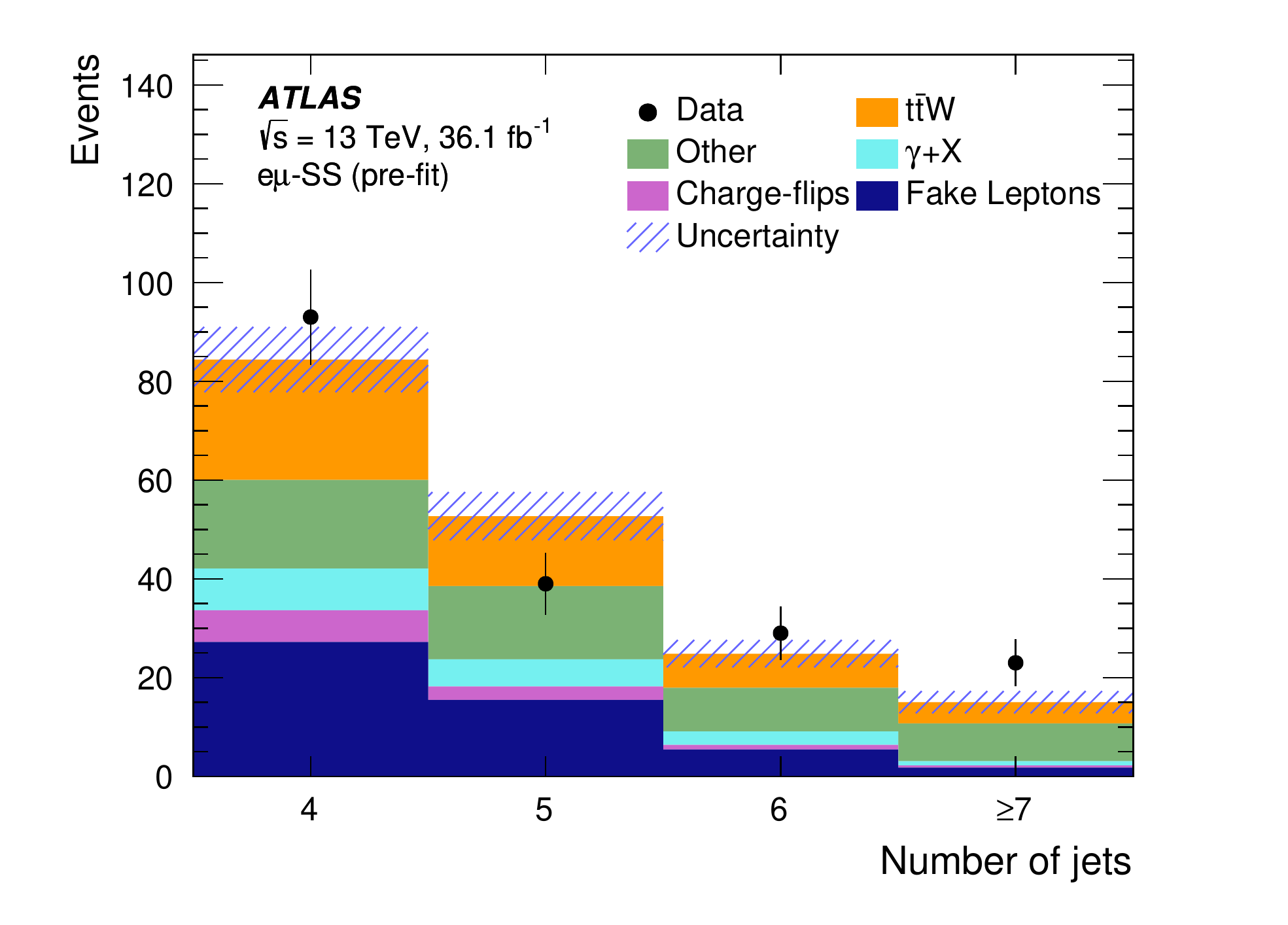}}\\
\subfloat[\label{fig:2lss_sr5}]{\includegraphics[width=\twofigwidth]{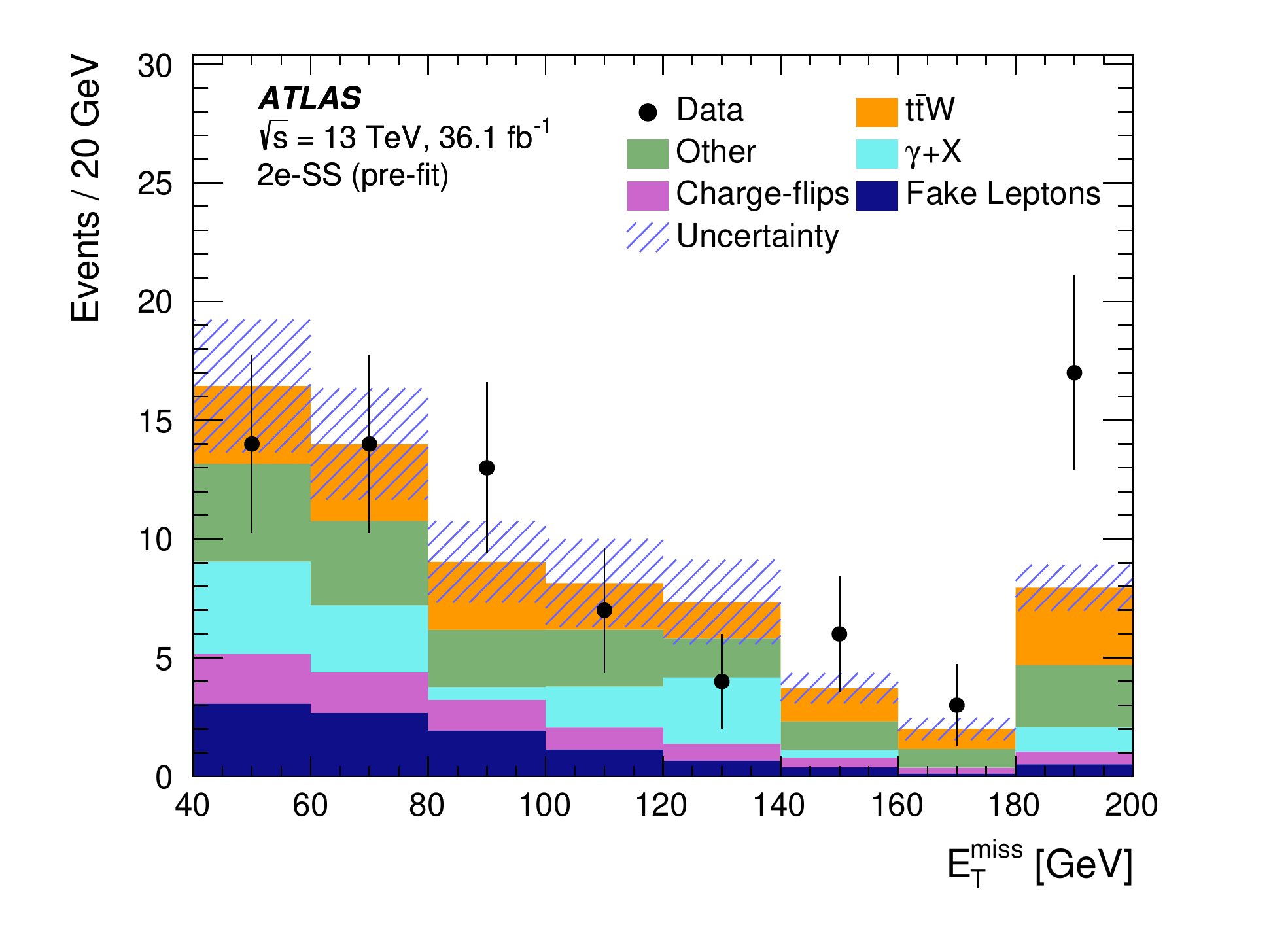}}
\subfloat[\label{fig:2lss_sr6}]{\includegraphics[width=\twofigwidth]{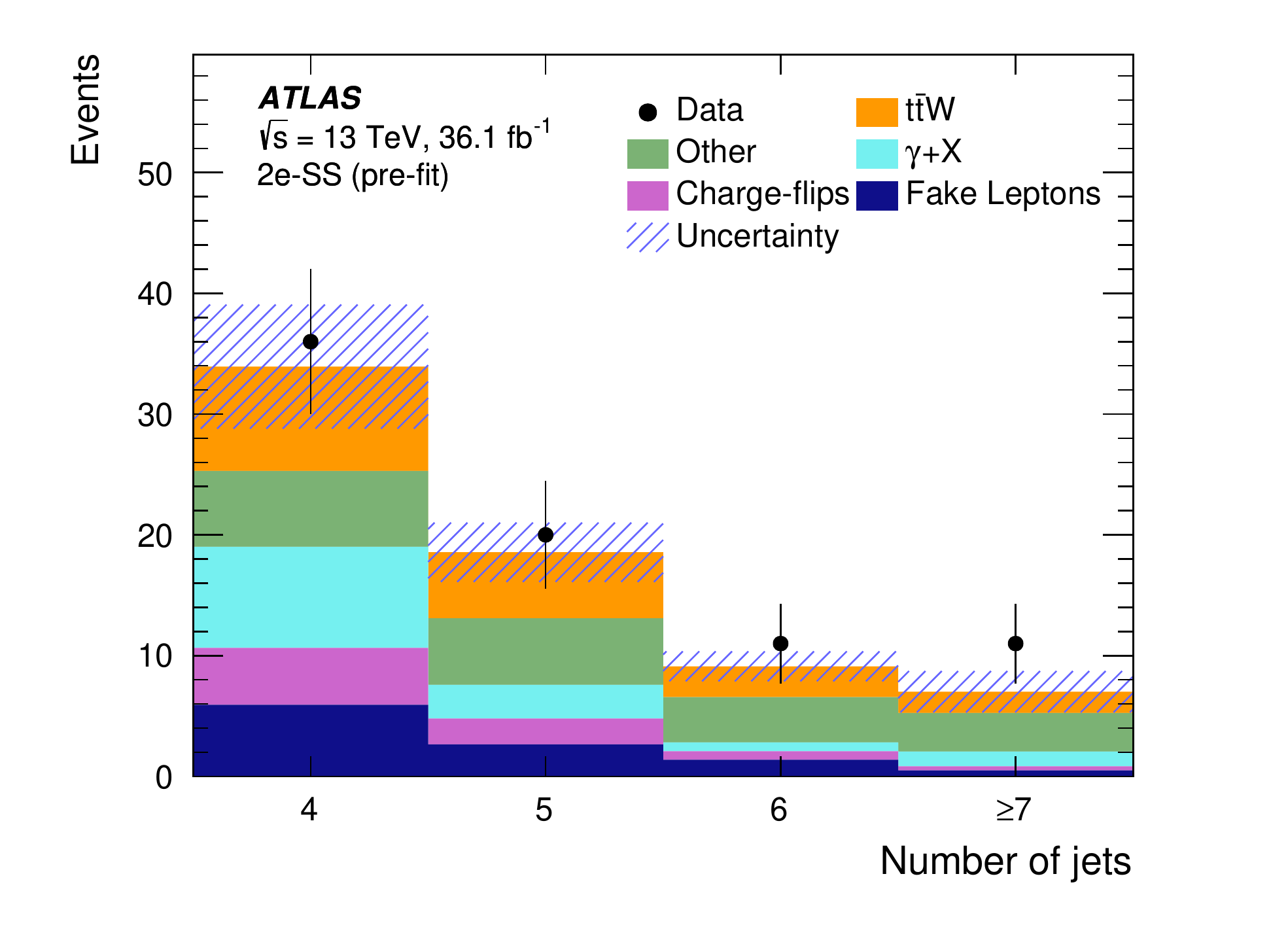}}\\
\caption{\label{fig:2lss_sr} Distributions in the $2\mu$-SS, $e\mu$-SS and
$2e$-SS regions of \protect\subref{fig:2lss_sr1} \protect\subref{fig:2lss_sr3}
\protect\subref{fig:2lss_sr5} the missing transverse momentum \met and
\protect\subref{fig:2lss_sr2} \protect\subref{fig:2lss_sr4}
\protect\subref{fig:2lss_sr6} the jet multiplicity. \hatch.  The `Other'
background contains SM processes with small cross sections producing two
same-sign prompt leptons.  \oflo.}
\end{figure}

To facilitate comparisons of data with the expectation, three regions $2e$-SS,
$e\mu$-SS and $2\mu$-SS are formed by combining all the same-sign signal
regions corresponding to a given lepton flavor combination. The distributions
of \met and the number of jets for these three regions are shown in
Figure~\ref{fig:2lss_sr}.

\subsection{Trilepton analysis}

Eight signal regions with exactly three leptons are considered, four of them
targeting the \ttZ and four targeting the \ttW process, as 
defined in Tables~\ref{tab:SRs3l_ttZ} and~\ref{tab:SRs3l_ttW}, respectively. The regions
are divided into two groups depending on whether or not a pair of OSSF leptons
with invariant mass within $10\,\GeV$ of the $Z$ boson mass is present. The
signal regions are further categorized according to jet and $b$-tagged jet
multiplicities. 

The four signal regions in the first group are sensitive to \ttZ. In the
\TLSRA\ region, at least four jets are required, exactly one of which is
$b$-tagged. In the \TLSRB\ region, exactly three jets with at least two
$b$-tagged jets are required.  In the \TLSRC\ region, at least four jets are
required, of which at least two are $b$-tagged.  In the \TLSRD\ region,
targeting events with an off-shell $Z^*$ or $\gamma^*$, at least four jets are
required, of which at least two are $b$-tagged; no OSSF lepton pair is allowed
in the $Z$ boson mass window. The sum of the lepton charges must be $\pm$1. 

The remaining four trilepton signal regions target the \ttW process. These
regions require two or three jets and veto events that contain an OSSF pair of
leptons whose invariant mass is within $10\,\GeV$ of the $Z$ boson mass. In the
first two regions, \TLnoZA\ and \TLnoZB, at least two jets are required to be
$b$-tagged. In the other two regions, \TLnoZC\ and \TLnoZD, exactly one jet is
required to be $b$-tagged. The sum of lepton charges is required to be +1 (--1)
in \TLnoZA\ and \TLnoZC\ (\TLnoZB\ and \TLnoZD). In regions \TLnoZC\ and
\TLnoZD, $\HT>240\,\GeV$ is also required. The signal region definitions for
the \TLC\ are summarized in  Tables~\ref{tab:SRs3l_ttZ} and \ref{tab:SRs3l_ttW}
for the signal regions targeting \ttZ and \ttW, respectively. 

The dominant backgrounds in the \TLSRA, \TLSRB\ and \TLSRC\ signal regions
arise from diboson production, the production of a single top quark in
association with a $Z$ boson ($tZ$ and $tWZ$) and $Z$+jets production with a
fake lepton. 

\begin{table}[htbp]
\centering
\caption{Summary of event selection requirements in the trilepton signal
regions targeting the \ttZ process.\spaceit} \label{tab:SRs3l_ttZ}
\begin{tabular}{@{}lrrrc@{}}
\toprule
Variable  & \TLSRA & \TLSRB & \TLSRC & \TLSRD\\
\midrule
Leading lepton & \multicolumn{4}{c}{$\pT> \SI{27}{\GeV}$} \\
Other leptons & \multicolumn{4}{c}{$\pT> \SI{20}{\GeV}$} \\
Sum of lepton charges &  \multicolumn{4}{c}{$\pm1$} \\
$Z$ requirement (OSSF pair) & \multicolumn{3}{c}{$|m_{\ell\ell} - m_Z| < \SI{10}{\GeV}$} & $|m_{\ell\ell} - m_Z| > \SI{10}{\GeV}$\\
$n_{\text{jets}}$ & $\ge 4$ & $3$ &  $\ge 4$ & $\ge4$\\
\nbtags & $1$ & $\ge2$ & $\ge2$ & $\ge2$\\
\bottomrule
\end{tabular}
\end{table}

\begin{table}[htbp]
\centering
\caption{Summary of event selection requirements in the trilepton signal
regions targeting the \ttW process.\spaceit} \label{tab:SRs3l_ttW}
\begin{tabular}{@{}lrrrr@{}}
\toprule
Variable  & \TLnoZA & \TLnoZB & \TLnoZC & \TLnoZD\\
\midrule
All leptons & \multicolumn{4}{c}{$\pT> \SI{27}{\GeV}$} \\
$Z$ veto (OSSF pair) & \multicolumn{4}{c}{$|m_{\ell\ell} - m_Z| > \SI{10}{\GeV}$} \\ 
$n_{\mathrm{jets}}$ & \multicolumn{4}{c}{2 or 3} \\
$\HT$  & \multicolumn{2}{c}{--} & \multicolumn{2}{c}{$>240$ \GeV} \\
Sum of lepton charges &  +1 & --1 & +1 & --1 \\
\nbtags & $\ge 2$ & $\ge 2$ & $1$ & $1$ \\
\bottomrule
\end{tabular}
\end{table}

\begin{figure}[htbp]
\centering
\subfloat[\label{fig:3l_wzcr1}]{\includegraphics[width=\twofigwidth]{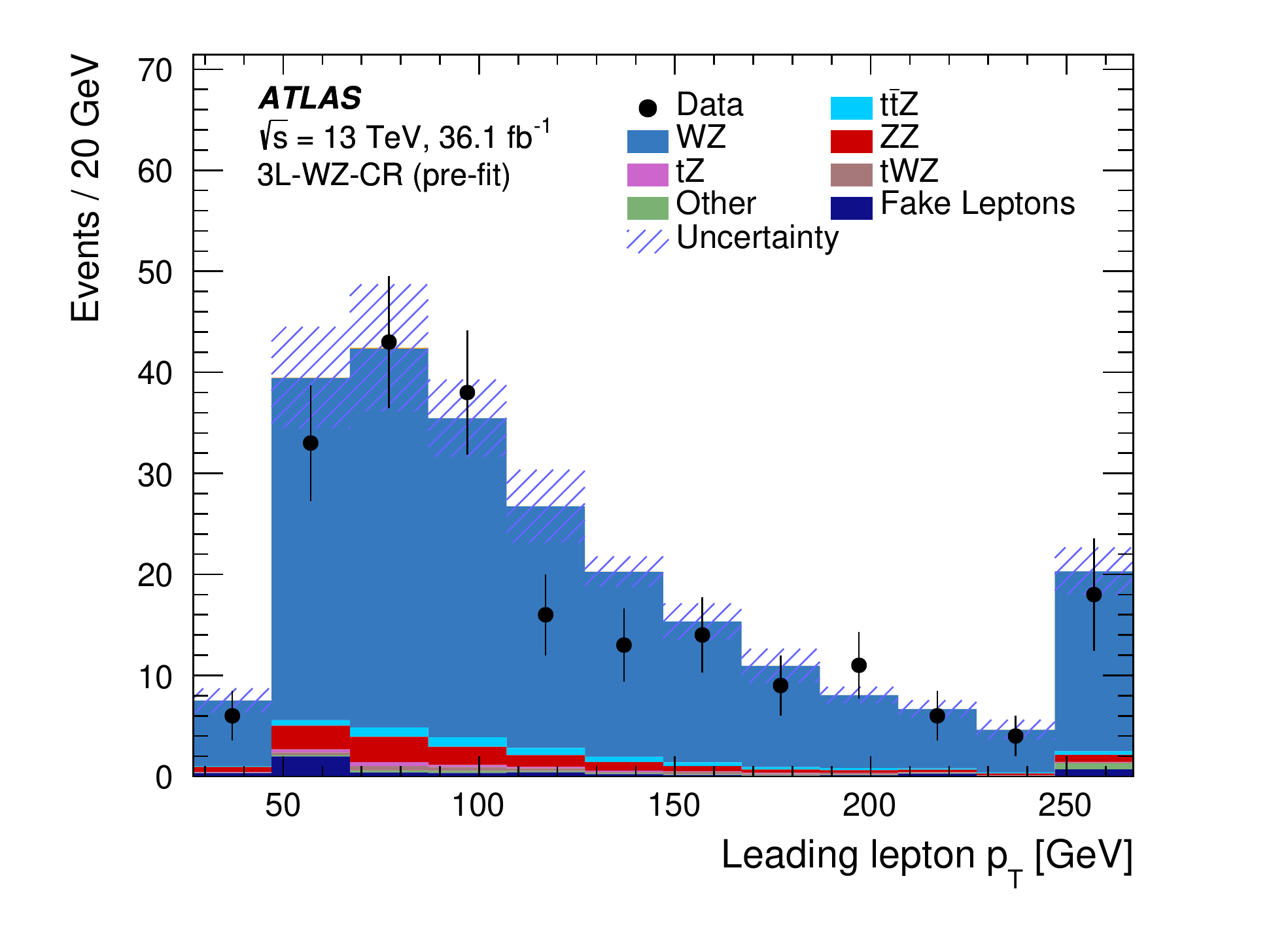}}
\subfloat[\label{fig:3l_wzcr2}]{\includegraphics[width=\twofigwidth]{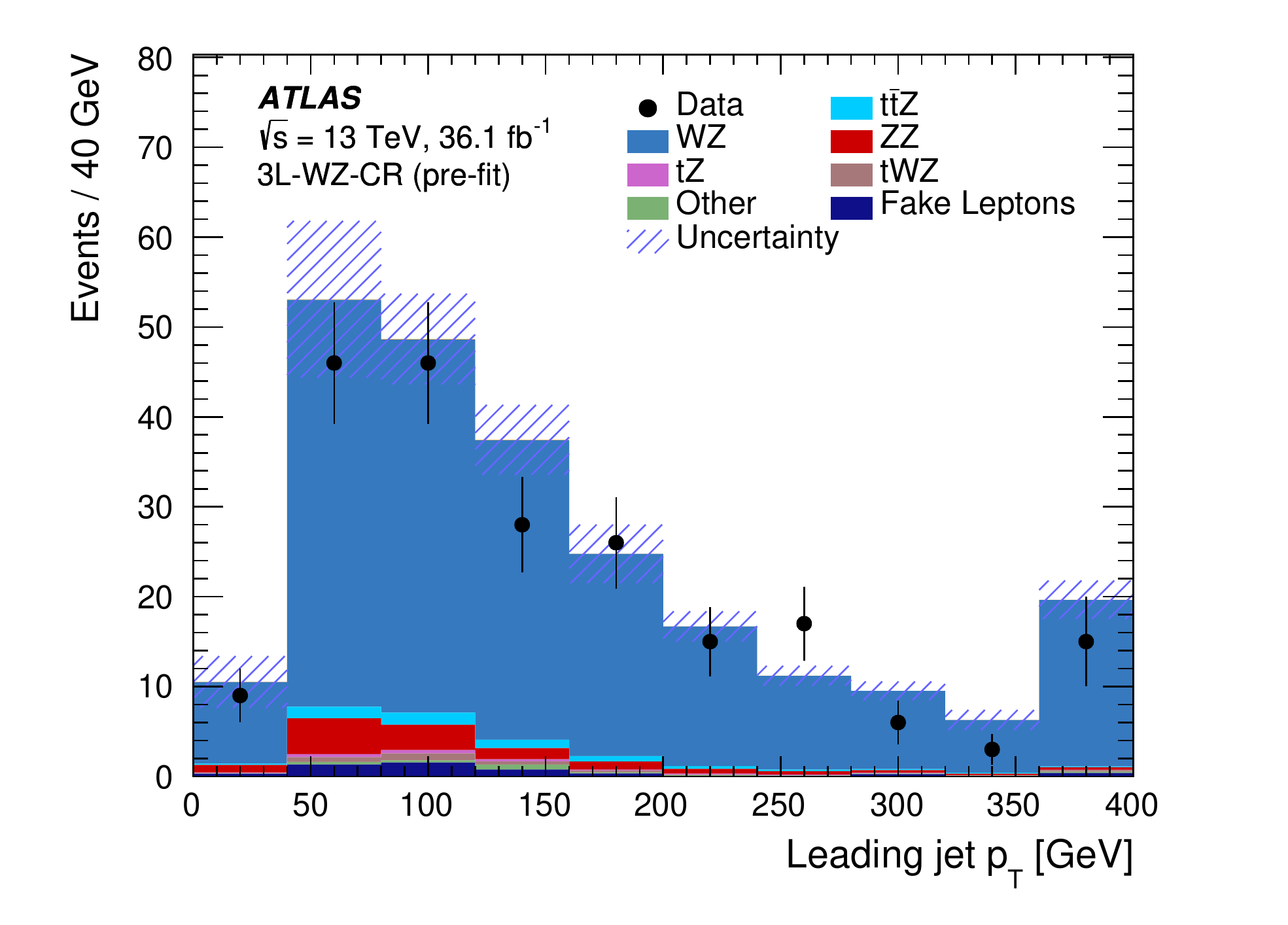}}
\caption{\label{fig:3l_wzcr} Distributions of \protect\subref{fig:3l_wzcr1} the
leading lepton transverse momentum \pt and \protect\subref{fig:3l_wzcr2} the
leading jet \pt in the \TLCR\ control region before the fit.  The `Other'
background contains SM processes with small cross sections producing three
prompt leptons.  \hatch.  \oflo.}
\end{figure}

A control region is used to determine the normalization of the $WZ$+jets
background in data. Exactly three leptons are required, at least one pair of
which must be an OSSF pair with an invariant mass within $10\,\GeV$ of the $Z$
boson mass.  There must be exactly three jets, none of which pass the
$b$-tagging requirement. This region is referred to as \TLCR\ and it is
included in the fit. Distributions comparing data with SM predictions in \TLCR\
are shown in Figure~\ref{fig:3l_wzcr}, demonstrating good modeling of the $WZ$
background.

\begin{figure}[htbp]
\centering
\subfloat[\label{fig:3l_srttW1}]{\includegraphics[width=\twofigwidth]{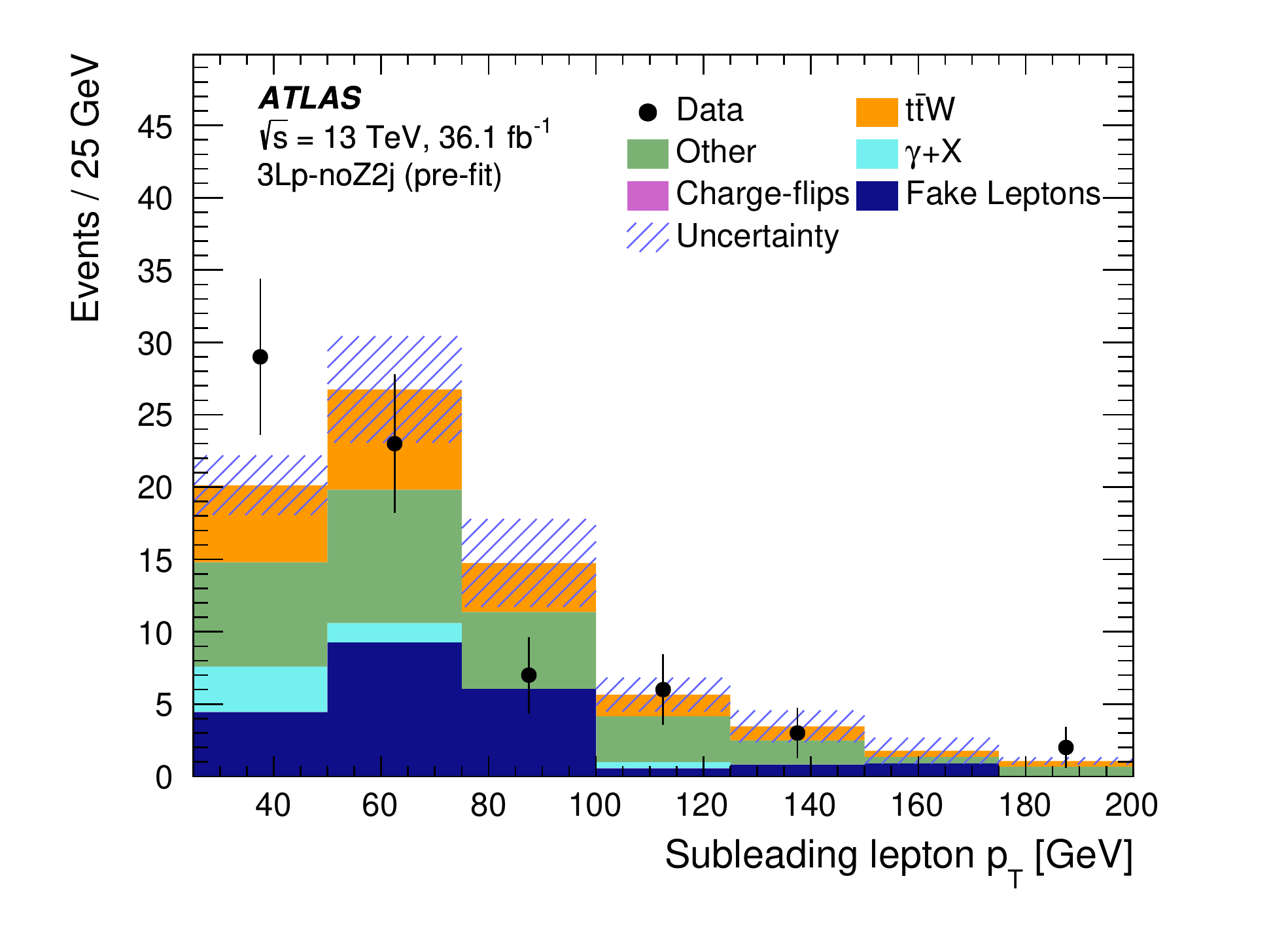}}
\subfloat[\label{fig:3l_srttW2}]{\includegraphics[width=\twofigwidth]{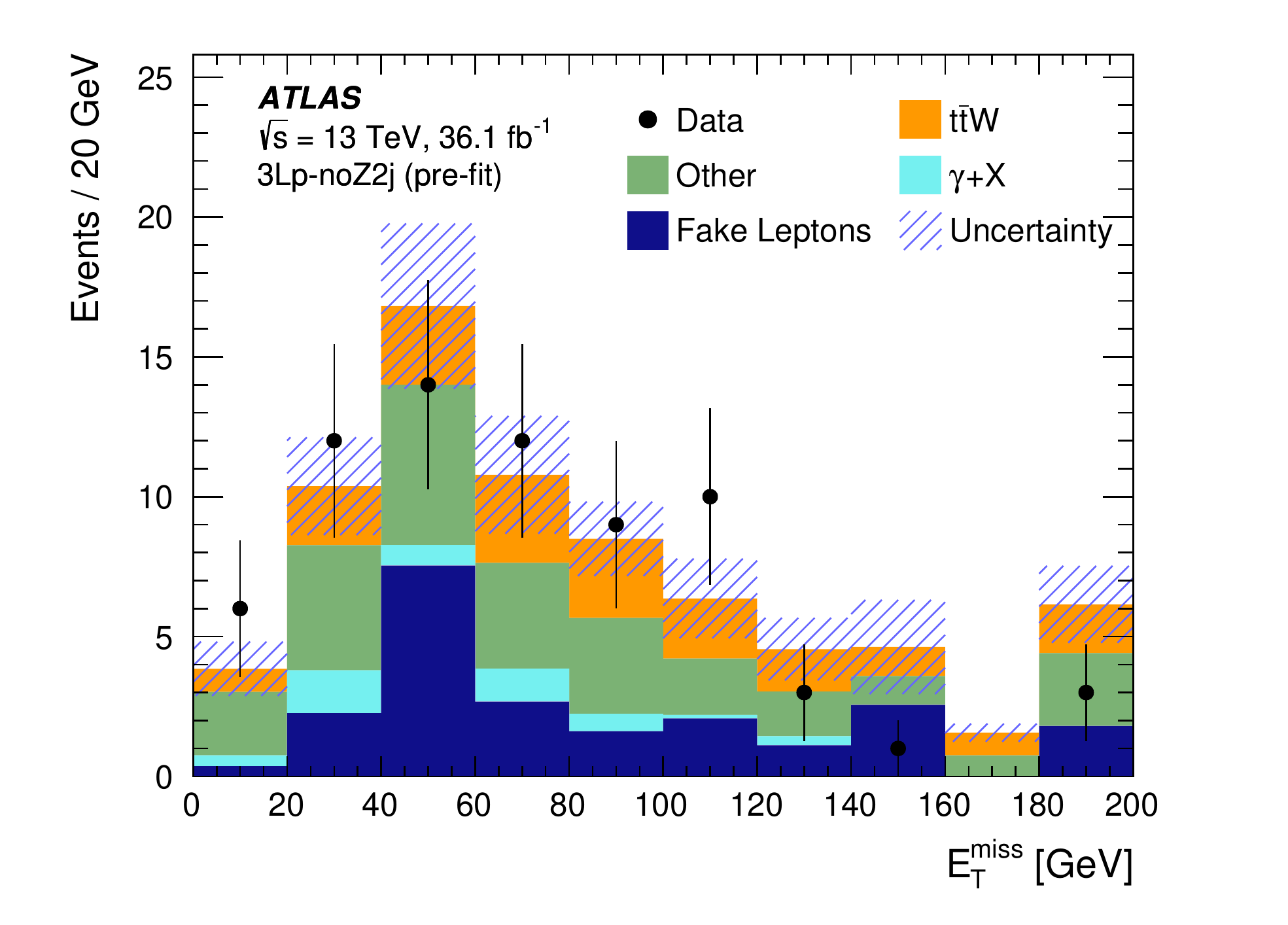}}
\caption{\label{fig:3l_srttW} Distributions of \protect\subref{fig:3l_srttW1}
the subleading lepton transverse momentum \pt and
\protect\subref{fig:3l_srttW2} missing transverse momentum \met for events
belonging to any of the four trilepton regions targeting the \ttW process. The
distributions are shown before the fit.  The `Other' background contains SM
processes with small cross sections producing three prompt leptons.  \hatch.
\oflo.} 
\end{figure}

\begin{figure}[htbp]
\centering
\subfloat[\label{fig:3l_sr1}]{\includegraphics[width=\twofigwidth]{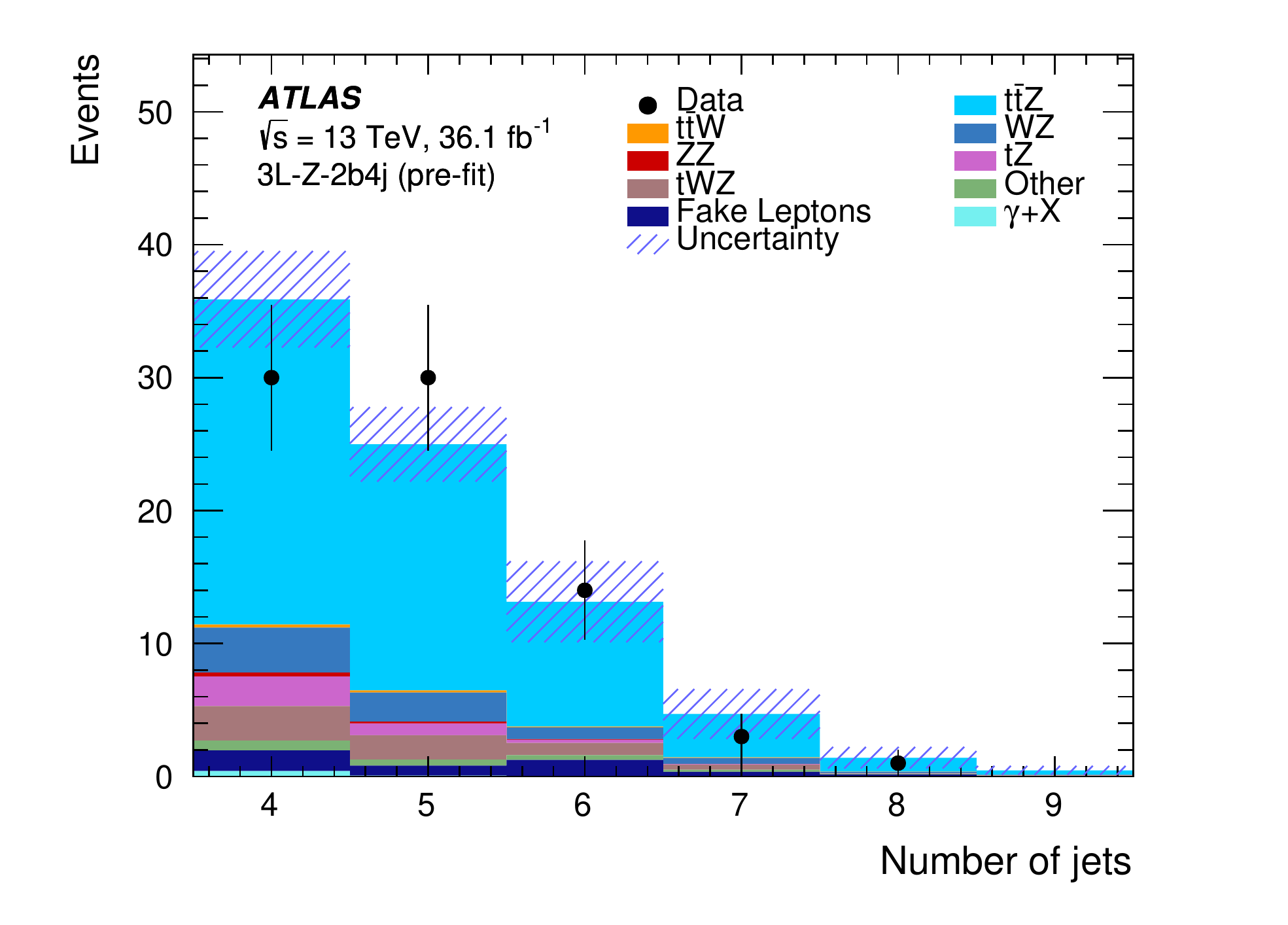}}
\subfloat[\label{fig:3l_sr2}]{\includegraphics[width=\twofigwidth]{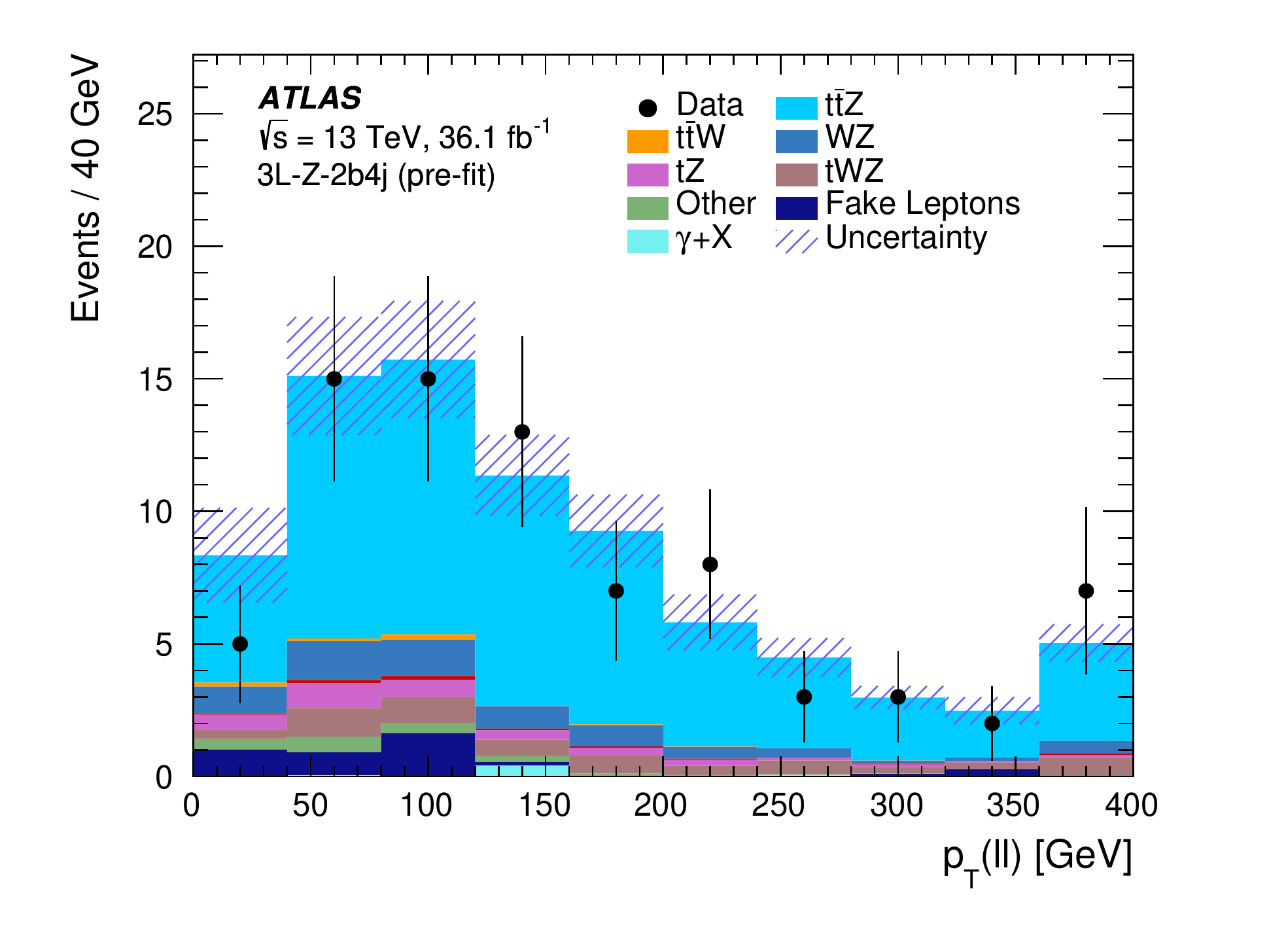}}\\
\subfloat[\label{fig:3l_sr3}]{\includegraphics[width=\twofigwidth]{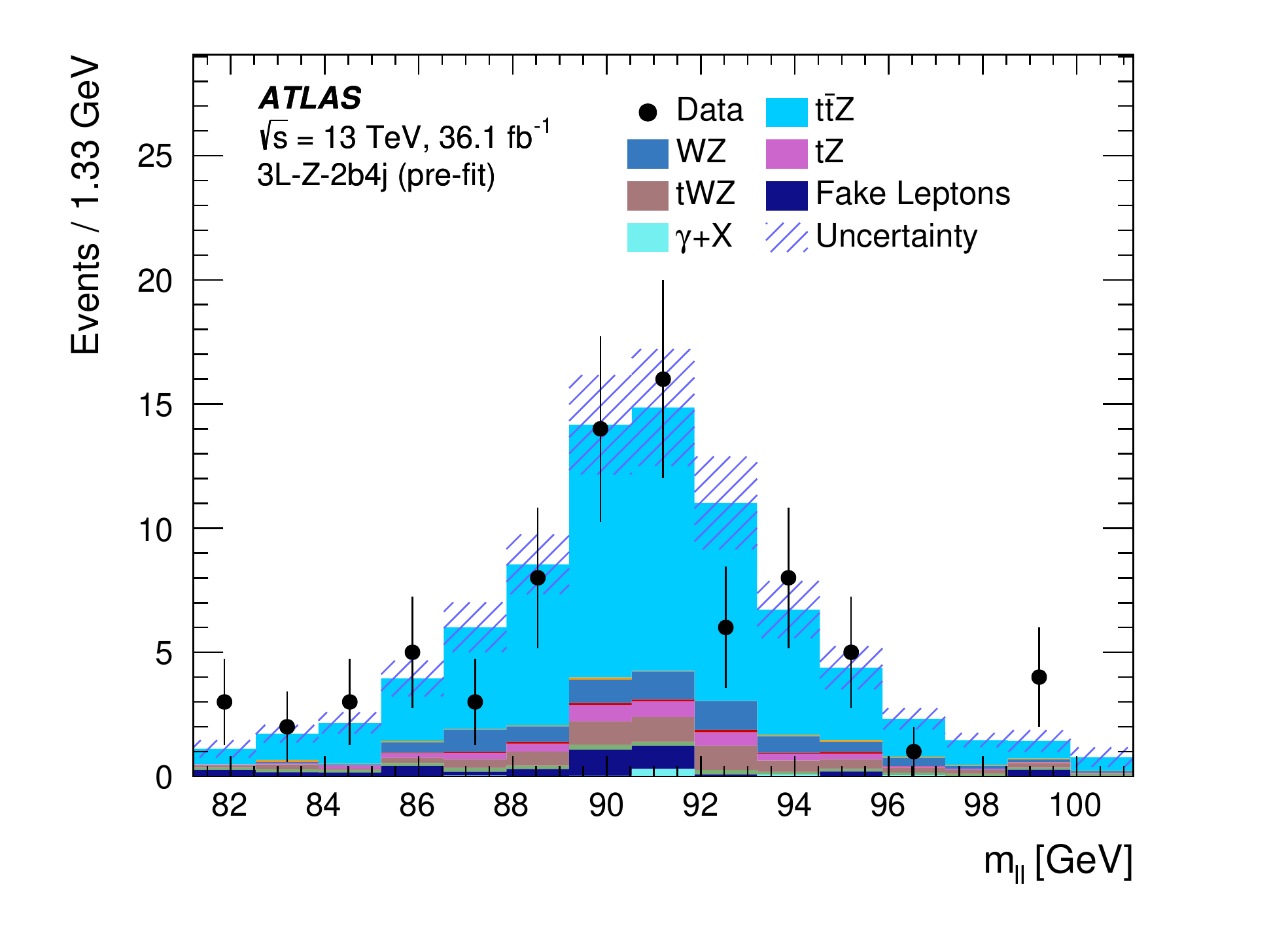}}
\caption{\label{fig:3l_sr} Distributions of \protect\subref{fig:3l_sr1} the
number of jets, \protect\subref{fig:3l_sr2} the transverse momentum
$\pt(\ell\ell)$ and \protect\subref{fig:3l_sr3} the mass $m_{\ell\ell}$ of the
reconstructed $Z$ boson candidate for events in \TLSRC. The distributions are
shown before the fit.  The `Other' background contains SM processes with small
cross sections producing three prompt leptons.  \hatch. \oflo.} 
\end{figure}

Figure~\ref{fig:3l_srttW} shows the leading lepton \pt and \met for events
belonging to any of the four trilepton regions targeting \ttW. Distributions of
the number of jets, the \pt and mass of the reconstructed $Z$ boson candidate
in the signal region most sensitive to \ttZ, \TLSRC, are shown in
Figure~\ref{fig:3l_sr}. 

\subsection{Tetralepton analysis}

The \FLC\ targets the \ttZ process for the case where both $W$ bosons,
resulting from top-quark decays, and the $Z$ boson decay leptonically.  Events
with two pairs of opposite-sign leptons are selected, and at least one pair
must have the same flavor.  The OSSF lepton pair with reconstructed invariant
mass closest to $m_Z$ is attributed to the \Zboson boson decay and denoted in
the following by $Z_1$.  The two remaining leptons are used to define $Z_2$.
The signal region definitions for the \FLC\ are summarized in
Table~\ref{tab:SRs4l}. Four signal regions are defined according to the
relative flavor of the two $Z_2$ leptons, different flavor (DF) or same flavor
(SF), and the number of $b$-tagged jets: one, or at least two ($1b$, $2b$).
The signal regions are thus \FLSRB, \FLSRC, \FLSRD\ and \FLSRE.

In the same-flavor regions, requirements on \met are applied to suppress the
$ZZ$ background. In the \FLSRD signal region, the \met is required to be
greater than 80~\GeV\ (40~\GeV) for events with $|m_{Z_{2}} - m_Z|<10$~\GeV\
($|m_{Z_{2}} - m_Z|>10$~\GeV). In the \FLSRE signal region, a requirement of
$\met>40$~\GeV\ is applied for events with $|m_{Z_{2}} - m_Z|<10$~\GeV.

To suppress events with fake leptons in the 1-$b$-tag multiplicity regions,
additional requirements on the scalar sum of the transverse momenta of the
third and fourth leptons ($\pttf$) are imposed.  In the \FLSRD\ and \FLSRB\
regions, events are required to satisfy $\pttf > \SI{25}{\gev}$ and $\pttf >
\SI{35}{\gev}$, respectively, while in the other regions all leptons are
required to satisfy $\pt > \SI{10}{\GeV}$.  

\begin{table}[htbp]
\centering \renewcommand{\arraystretch}{1.2}
\caption{\label{tab:SRs4l} Definitions of the four signal regions in the \FLC.
\spaceit}
\begin{tabular}{lcccr@{}cc@{}lr}
\toprule  
Region & $Z_2$ leptons & $\ptfour$ & \pttf && $|m_{Z_{2}} - m_Z| $ & \met && \nbtags\\
\midrule
\FLSRB & $e^{\pm}\mu^{\mp}$ & -- & $>\SI{35}{\gev}$ &&--&--&& 1\\
\FLSRC & $e^{\pm}\mu^{\mp}$ & $>\SI{10}{\gev}$&--&&--&--&& $\ge2$\\
\FLSRD & $e^{\pm}e^{\mp},\mu^{\pm}\mu^{\mp}$ & -- & $>\SI{25}{\gev}$ & \begin{tabular}{@{}r@{}} \ldelim\{{2}{2ex} \\ \\ \end{tabular} &
\begin{tabular}{c}$>\SI{10}{\gev}$\\ $<\SI{10}{\gev}$\end{tabular} & \begin{tabular}{c}$>\SI{40}{\gev}$\\ $>\SI{80}{\gev}$\end{tabular} &\begin{tabular}{@{}l@{}} \rdelim\}{2}{2ex} \\ \\ \end{tabular} & 1\\
\FLSRE & $e^{\pm}e^{\mp},\mu^{\pm}\mu^{\mp}$ & $>\SI{10}{\gev}$ & --&\begin{tabular}{@{}r@{}} \ldelim\{{2}{2ex} \\ \\ \end{tabular} &
\begin{tabular}{c}$>\SI{10}{\gev}$\\ $<\SI{10}{\gev}$\end{tabular} & \begin{tabular}{c}-\\ $>\SI{40}{\gev}$\end{tabular} &\begin{tabular}{@{}l@{}} \rdelim\}{2}{2ex} \\ \\ \end{tabular} & $\ge2$ \\
\bottomrule 
\end{tabular}
\end{table}

\begin{figure}[htbp]
\centering
\subfloat[\label{fig:zz_val1}]{\includegraphics[width=\twofigwidth]{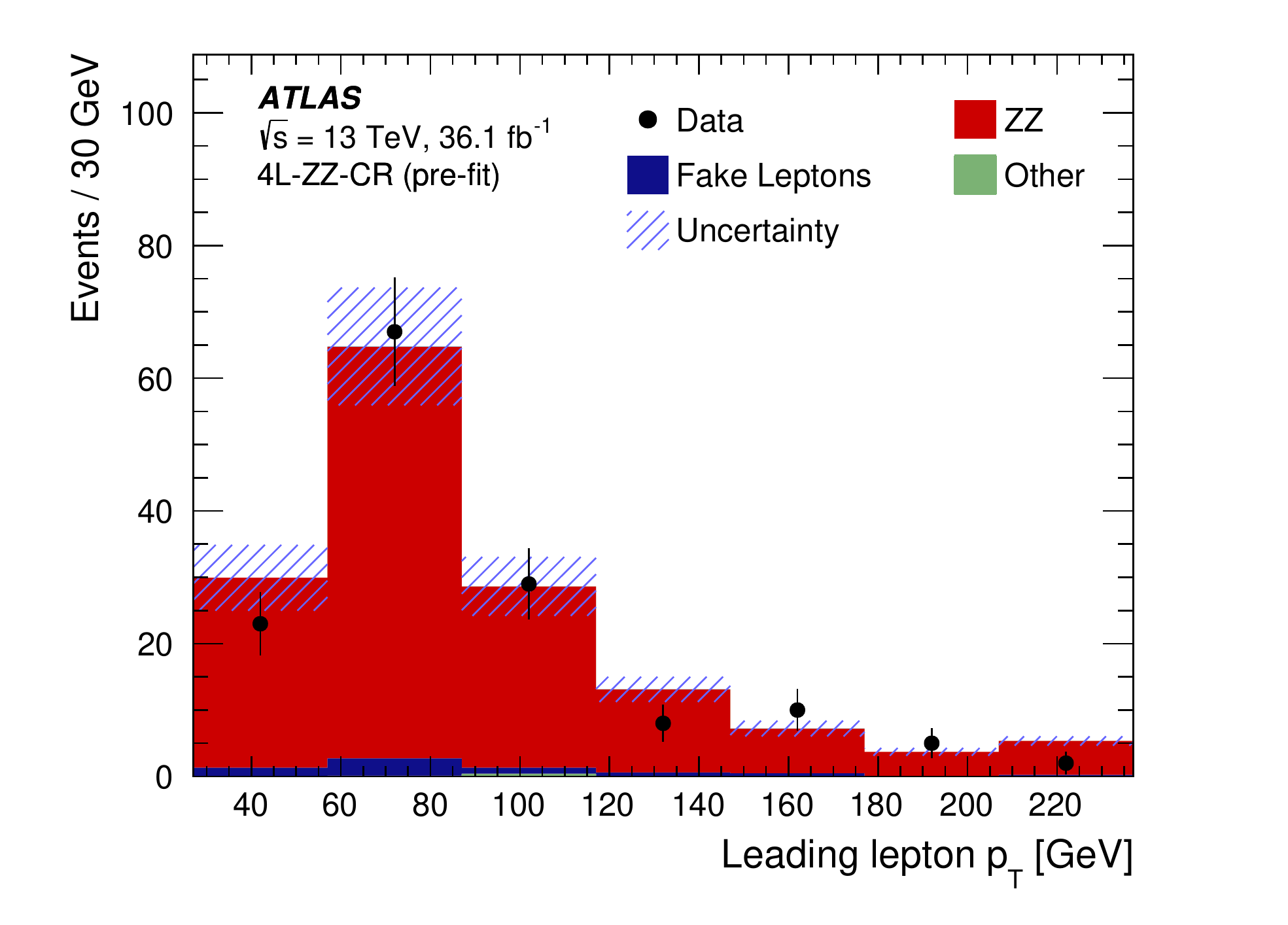}}
\subfloat[\label{fig:zz_val2}]{\includegraphics[width=\twofigwidth]{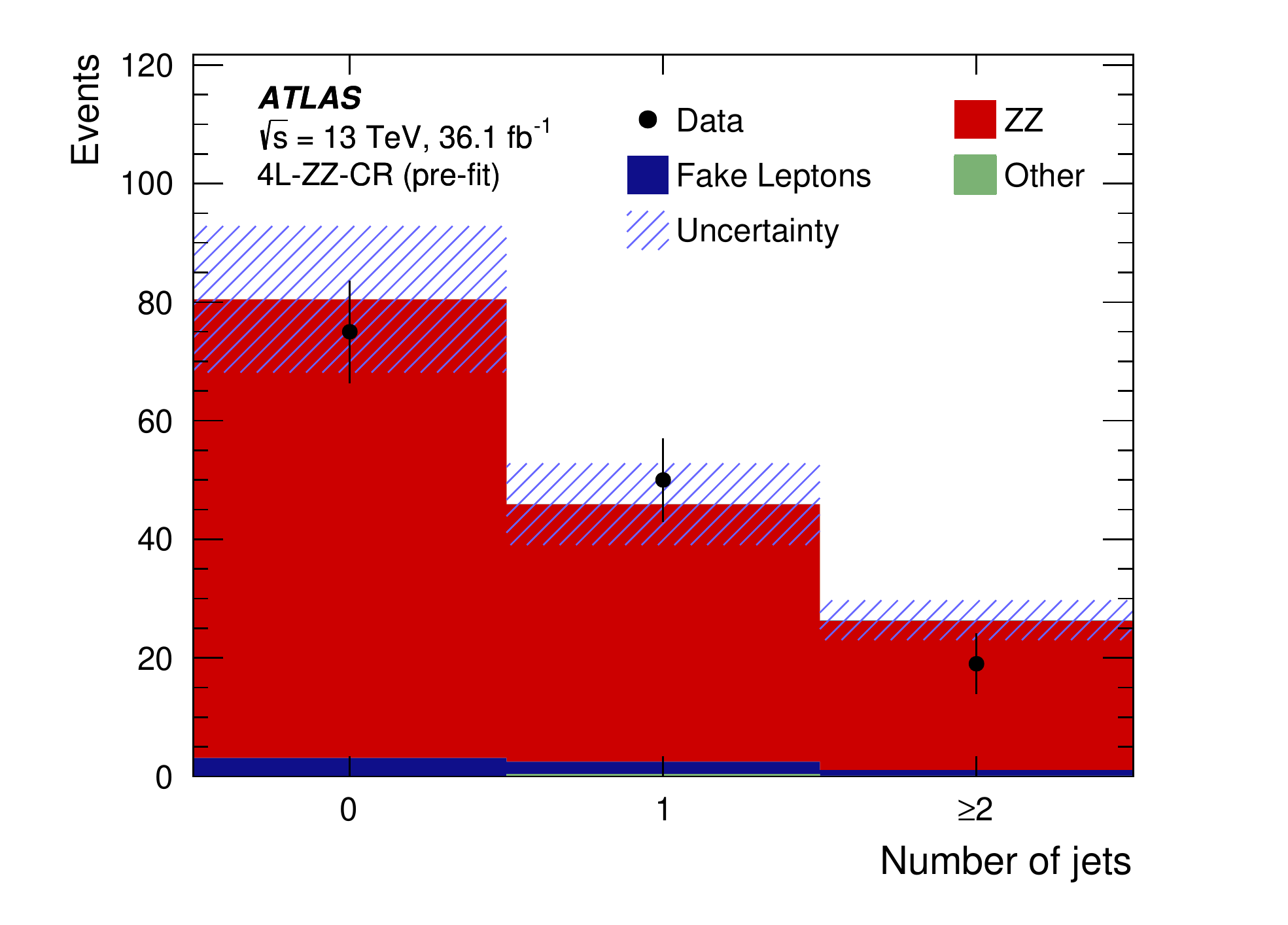}}
\caption{\label{fig:zz_val} Distribution of \protect\subref{fig:zz_val1} the
leading lepton transverse momentum \pt and \protect\subref{fig:zz_val2} jet
multiplicity in the \FLCR\ control region. The distributions are shown before
the fit. \hatch.  \oflo.} 
\end{figure}

A control region used to determine the $ZZ$ normalization, referred to as
\FLCR, is included in the fit and is defined to have exactly four reconstructed
leptons, a \SecondZ\ pair with OSSF leptons, the value of both \FirstZM\ and
\SecondZM\ within $10\,\GeV$ of the mass of the $Z$ boson, and $20\,\GeV< \met
<40\,\GeV$. The leading lepton \pT and the jet multiplicity in this control
region are shown in Figure~\ref{fig:zz_val}, and good agreement is seen between
data and prediction.

The contribution from backgrounds containing fake leptons is estimated from
simulation and corrected with scale factors determined in two control regions:
one region enriched in \ttbar events and one region enriched in $Z$+jets
events.  The scale factors are extracted and applied separately for electron
and muon fake-lepton candidates, and for leptons arising from heavy-flavor
hadrons and other sources. Therefore, a total of four scale factors are
determined.  The scale factors are applied to all MC simulation events with
fewer than four prompt leptons according to the number, flavor and origin of
the fake leptons.  It is verified that the scale factors for different
generators used in the simulation are consistent with each other.

\begin{figure}[htbp]
\centering
\subfloat[\label{fig:SR4l1}]{\includegraphics[width=\twofigwidth]{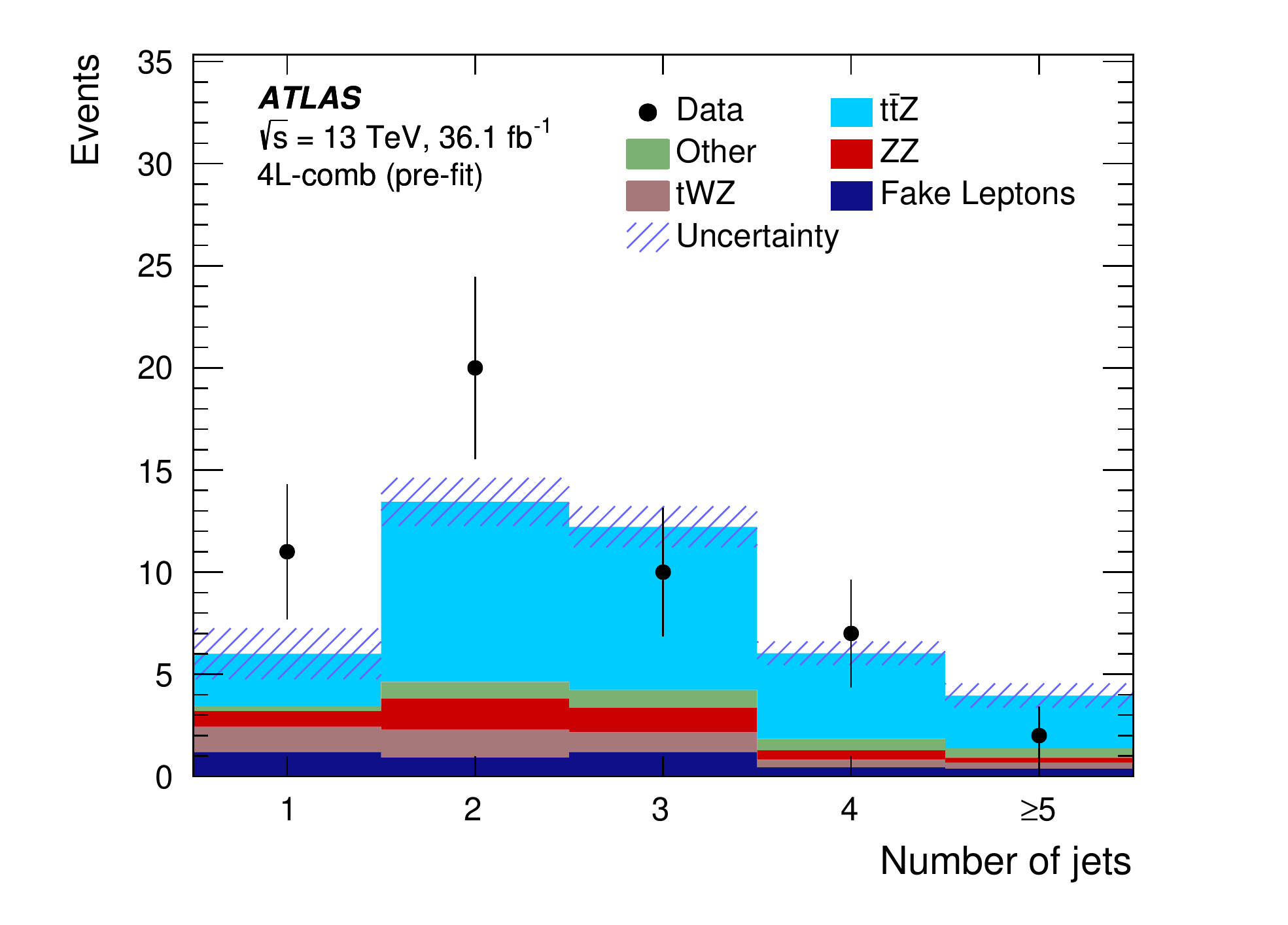}}
\subfloat[\label{fig:SR4l2}]{\includegraphics[width=\twofigwidth]{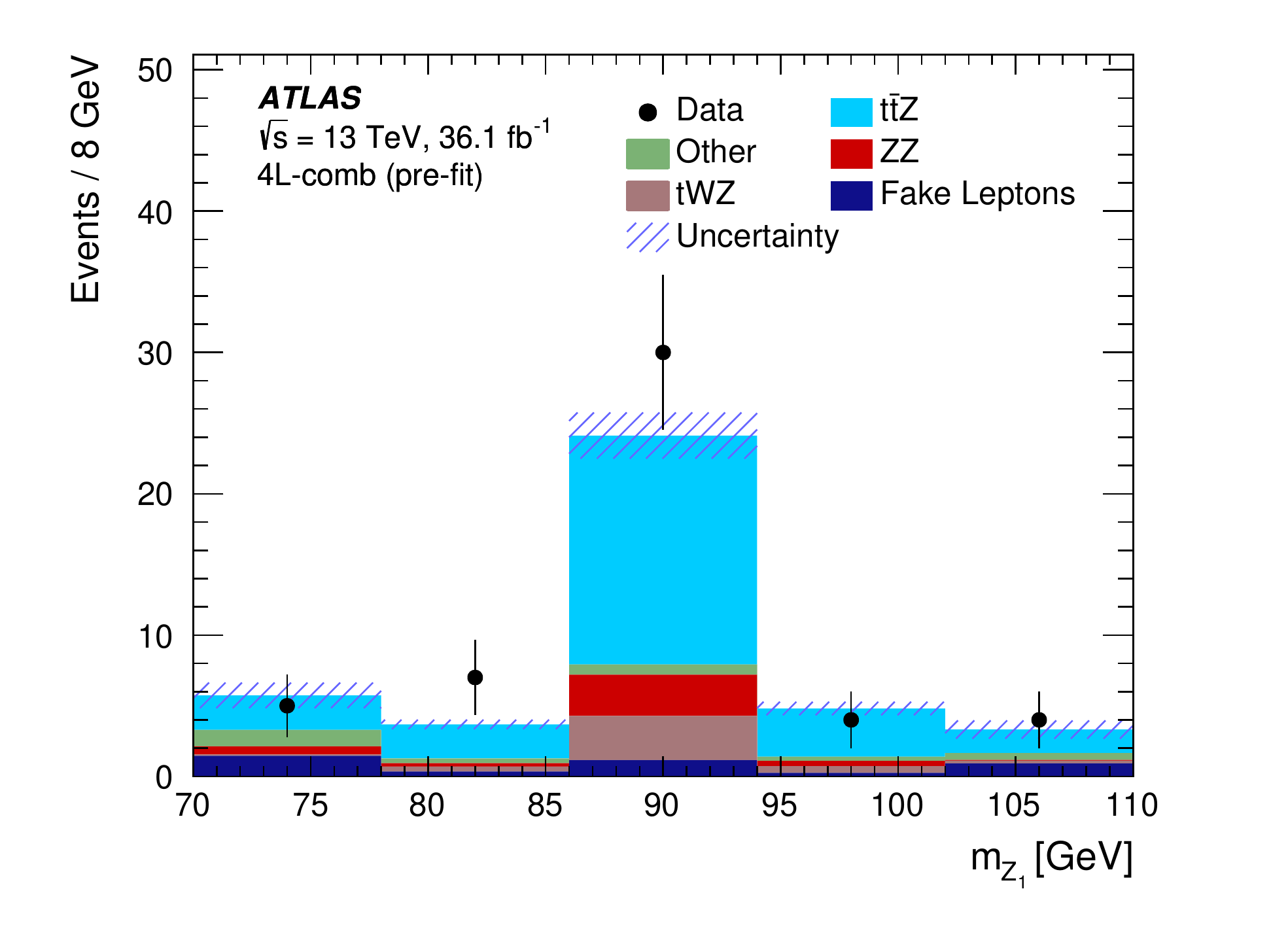}}\\
\subfloat[\label{fig:SR4l3}]{\includegraphics[width=\twofigwidth]{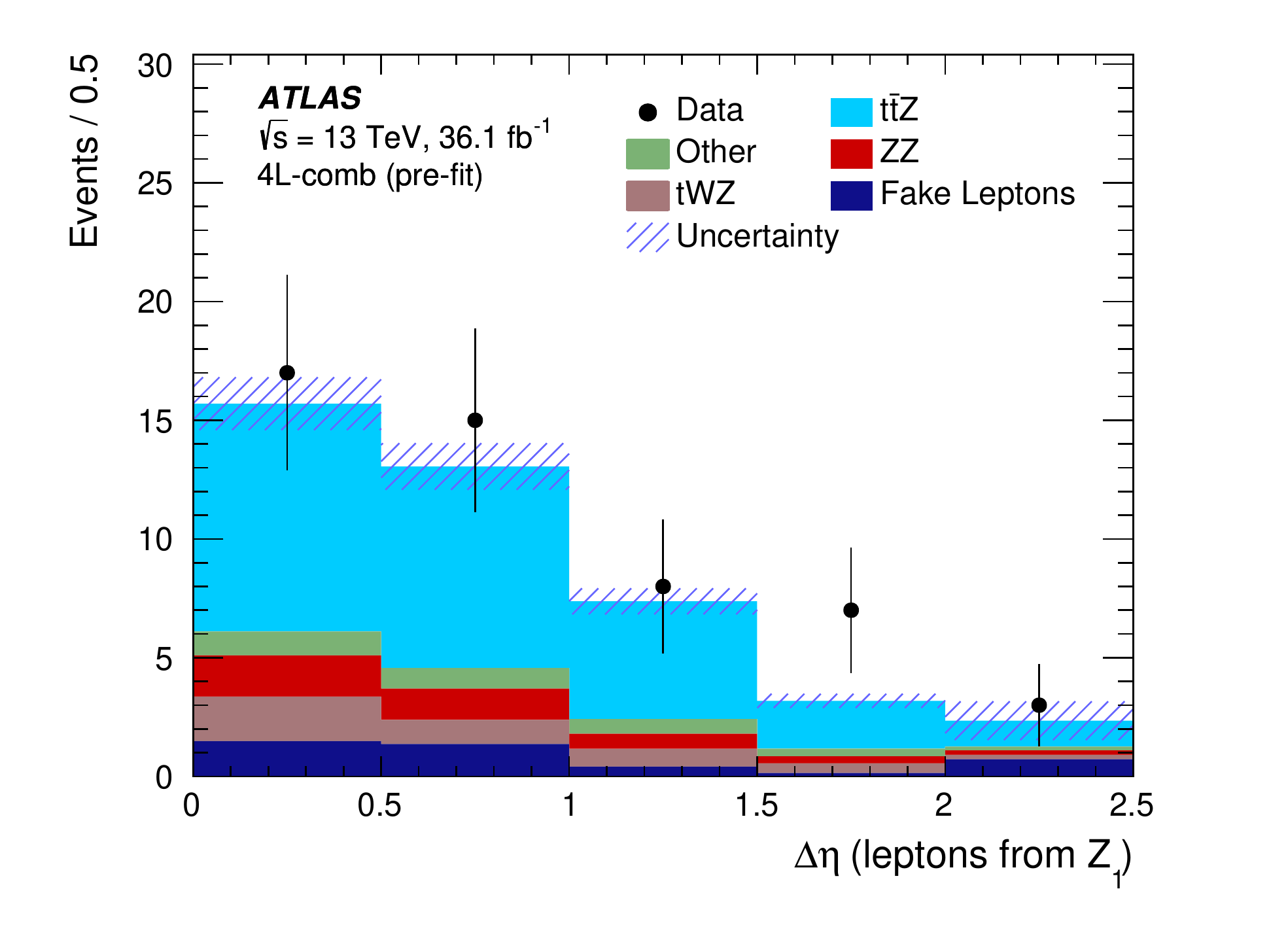}}
\subfloat[\label{fig:SR4l4}]{\includegraphics[width=\twofigwidth]{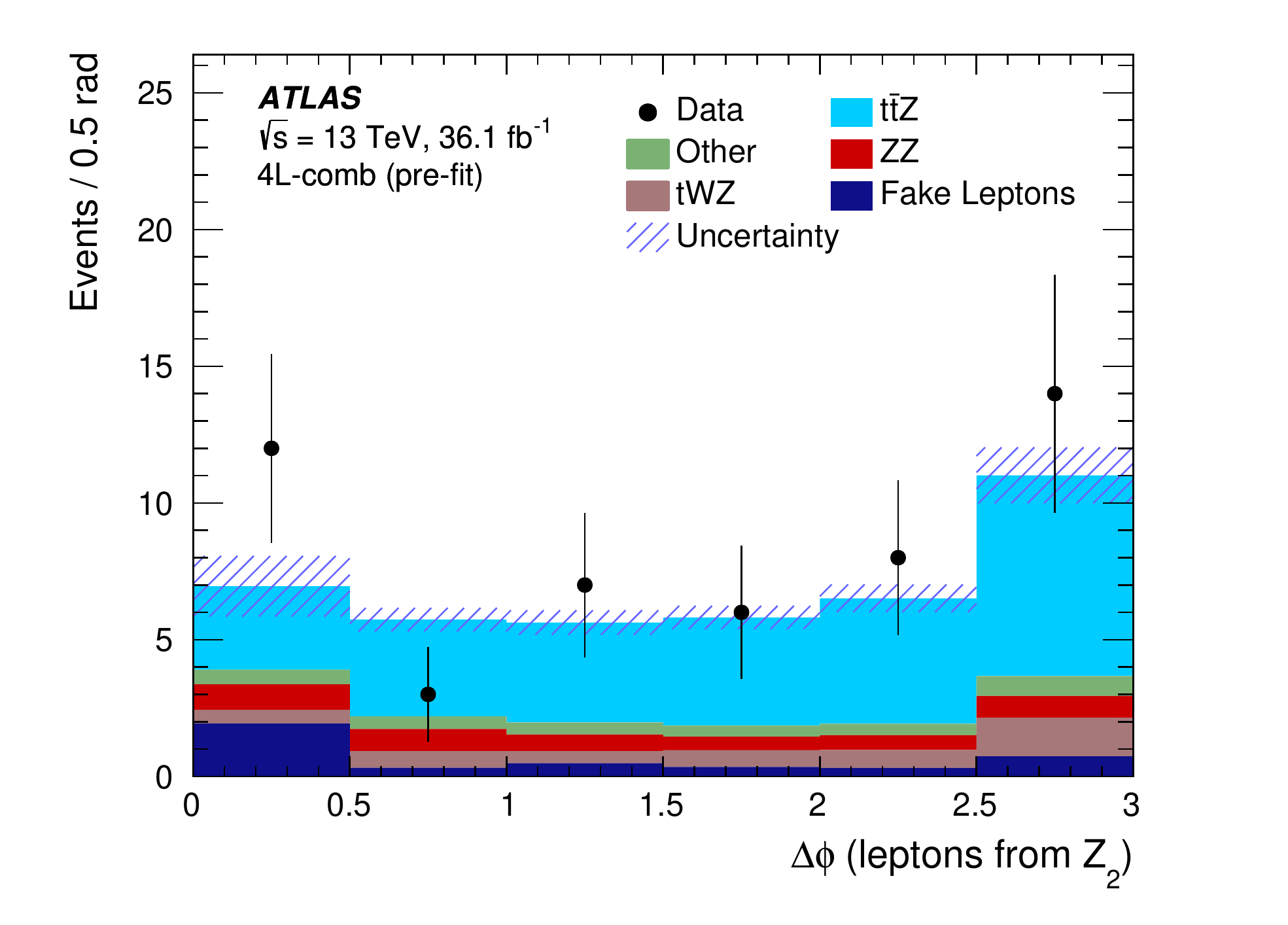}}
\caption{\label{fig:SR4l} Distributions, for all tetralepton signal regions
combined, of \protect\subref{fig:SR4l1} the number of jets,
\protect\subref{fig:SR4l2} the invariant mass of the OSSF lepton pair closest
to the $Z$ boson mass, $m_{Z_1}$, \protect\subref{fig:SR4l3}  the
pseudorapidity separation $\Delta\eta$ for that pair of leptons and
\protect\subref{fig:SR4l4}  the azimuthal angle $\Delta\phi$ between the
remaining two leptons.  The `Other' background contains SM processes with small
cross sections producing four prompt leptons.  \prefit. \hatch.  The first and
last bin include the underflow and overflow, respectively.} 
\end{figure}

Figure~\ref{fig:SR4l} compares the data with the expected distributions for all
four signal regions combined, showing good agreement between data and
expectation.

\section{Systematic uncertainties}
\label{s:systematics}

The signal and background yields in each signal and control region may be
affected by several sources of systematic uncertainty. These are implemented as
nuisance parameters in the fit, explained in Section~\ref{s:results}, are
constrained by Gaussian probability density functions and are described in the
following subsections.

\subsection{Luminosity}

The uncertainty in the integrated luminosity of the dataset is 2.1\%. It is
derived, following a methodology similar to that detailed in
Ref.~\cite{DAPR-2013-01}, and using the LUCID-2 detector for the baseline
luminosity measurements \cite{LUCID2}, from a calibration of the luminosity
scale using $x$--$y$ beam-separation scans.  This systematic uncertainty
affects all processes modeled using \MC\ simulations, apart from $Z$+1HF,
$Z$+2HF, $WZ$ and $ZZ$, whose normalizations are taken from data control
regions.

\subsection{Uncertainties associated with reconstructed objects}
 
Uncertainties associated with the lepton selection arise from the trigger,
reconstruction, identification and isolation efficiencies, and lepton momentum
scale and resolution \cite{PERF-2013-03, ATLAS-CONF-2014-032,
ATL-PHYS-PUB-2015-041, PERF-2015-10}. 

Uncertainties associated with the jet selection arise from the jet energy scale
(JES), the JVT requirement and the jet energy resolution (JER).  The JES and
its uncertainty are derived by combining information from test-beam data,
collision data and simulation~\cite{PERF-2016-04}. The uncertainties in the JER
and JVT have a significant effect at low jet \pt. 

The efficiency of the flavor-tagging algorithm is measured for each jet flavor
using control samples in data and in simulation. From these measurements,
correction factors are derived to correct the tagging rates in the simulation.
In the case of $b$-jets, correction factors and their uncertainties are
estimated from data using dileptonic \ttbar events~\cite{PERF-2016-05}.  In the
case of $c$-jets, they are derived using jets from $W$ boson decays in \ttbar
events~\cite{ATLAS-CONF-2018-001}.  In the case of light-flavor jets,
correction factors are derived using dijet events~\cite{ATLAS-CONF-2018-006}.
Sources of uncertainty affecting the $b$- and $c$-tagging efficiencies are
considered as a function of jet \pt, including bin-to-bin
correlations~\cite{ATLAS-CONF-2014-004}.  The uncertainty in the efficiency for
tagging light-flavor jets depends on the jet \pt and on $\eta$.  These
systematic uncertainties are taken as uncorrelated between $b$-jets, $c$-jets,
and light-flavor jets.  An additional uncertainty is assigned to account for
the extrapolation of the $b$-tagging efficiency measurement from the \pT region
used to determine the correction factors to regions with higher transverse momentum.

The treatment of the uncertainties associated with reconstructed objects is
common to all analysis channels, and thus these are considered as fully
correlated among different analysis regions. 

\subsection{Uncertainties in the signal modeling} 

Four sources of systematic uncertainties in the theoretical predictions of the
\ttZ and \ttW processes are considered. These signal modeling uncertainties are
treated as uncorrelated between the two processes, and correlated among
channels. Taking instead the uncertainties as correlated between the two
processes has a negligible impact on the results.  Acceptance effects due to
the choice of scale and PDF in the nominal \MGAMC+\PYTHIA~8 (\tune{A14} tune)
sample are considered. The renormalization and factorization scales $\muR =
\muF$ are varied simultaneously by factors $2.0$ and $0.5$. In addition, the
effects of a set of variations in the tune parameters (\tune{A14} eigentune
variations), sensitive to initial- and final-state radiation, multiple parton
interactions and color reconnection, are
evaluated~\cite{ATL-PHYS-PUB-2014-021}.  Studies performed at particle level
show that the largest impact comes from variations in initial-state
radiation~\cite{ATL-PHYS-PUB-2016-005}.  The systematic uncertainty due to the
choice of generator for the \ttZ and \ttW acceptance is estimated by comparing
the nominal sample with one generated with \SHERPA v2.2.  The \SHERPA sample
uses the LO matrix element with up to one (two) additional parton(s) included
in the matrix element calculation for \ttZ  (\ttW) and merged with the \SHERPA
parton shower~\cite{Schumann:2007mg}  using the \textsc{ME+PS@LO} prescription.
The \pdf{NNPDF3\!.\!0NLO} PDF set is used in conjunction with a dedicated
parton-shower tune developed by the \SHERPA authors.

\subsection{Uncertainties in the background modeling}

The $Z$+jets process is, together with \ttbar production, the dominant
background in the OS dilepton channel. Its normalization is extracted from data
as described in Section~\ref{s:selection2lOS}, but the shape of the BDT
distribution is obtained from simulation. To assess the systematic uncertainty
in the shape, the renormalization, factorization and resummation scales used in
the MC generation are varied by a factor of two with respect to the nominal
values.

The normalization and shape of the \ttbar background in the OS dilepton channel
is obtained using the data-driven method detailed in
Section~\ref{s:selection2lOS}. A systematic uncertainty arises from the factor
used to obtain \ttbar background yields in the same-flavor signal regions from
corresponding opposite-flavor dilepton control regions. The uncertainty is due
to the finite size of the samples of simulated events used, and the difference
between the values of the factor obtained with the nominal \POWHEGBOX\!+\PYTHIA
8 sample and an alternative sample generated using \MGAMC+\PYTHIA 8. The total
uncertainty in the factor is found to be 3\%.

In the \TL\ regions sensitive to \ttZ, the normalization of the $WZ$ background
is treated as a free parameter in the fit used to extract the \ttZ and \ttW
signals. The uncertainty in the extrapolation of the $WZ$ background estimate
from the control region to signal regions with specific jet and $b$-tag
multiplicities is evaluated by comparing predictions obtained by varying the
renormalization, factorization and resummation scales used in MC generation.
The uncertainties vary from 30\% to 50\%, depending on the signal region.

The normalization of the $ZZ$ background is treated as a free parameter in the
fit. An additional uncertainty arises from the extrapolation from the \FLCR\
control region to the signal regions. It is assessed by varying the
renormalization, factorization and resummation scales used in MC generation,
and found to be in the range 20--40\%.

The uncertainty in the \ttH background is evaluated by varying the
factorization and renormalization scales up and down by a factor of two with
respect to the nominal values. It is found to be around 10\%.

An overall normalization uncertainty of 30\% is assigned to the \tZ background,
motivated by the measurements of this process presented in
Refs.~\cite{TOPQ-2016-14, CMS-TOP-16-020}.  An additional uncertainty affecting
the distribution of this background as a function of jet and $b$-tagged jet
multiplicities is evaluated by varying the factorization and renormalization
scales, as well as the amount of radiation in the \tune{Perugia2012}
parton-shower tune.

An uncertainty of $10\%$ is assigned to the \WtZ background cross section,
resulting from different prescriptions for removing the interference with the
\ttZ process.  The shape uncertainty is evaluated by varying the factorization
and renormalization scales up and down by a factor of two with respect to the
nominal value.

For other prompt-lepton backgrounds, uncertainties of 20\% are assigned to the
normalizations of the $WH$ and $ZH$ processes, based on calculations from
Ref.~\cite{lhcxs}.  An uncertainty of 50\% is considered for triboson and
same-sign $WW$ processes.

A 10\% uncertainty is applied to the charge-flip background, resulting from
uncertainties in the charge-flip rates extracted from a control sample as
described in Section~\ref{s:selection}.

A 30\% uncertainty is assigned to the contribution from events with two prompt
leptons or one prompt lepton and a photon conversion in the control regions used
to measure the fake-lepton efficiency. In the SS dilepton channel regions and
trilepton regions targeting \ttW, there are 22 nuisance parameters
corresponding to the statistical uncertainty in the measurement of the
fake-lepton efficiencies. One nuisance parameter is used for each \pt bin used
in the measurement of the fake-lepton efficiencies. For fake-lepton
efficiencies in events with one (at least two) $b$-tagged jet(s), seven (four)
bins are used, and there is one nuisance parameter for each of the two lepton
flavors.  In the trilepton signal regions targeting \ttZ, where the fake-lepton
background is less important, a simplified description of the fake-lepton
uncertainties is used, with one nuisance parameter for each of the two lepton
flavors. These nuisance parameters correspond to the maximum of the up and down
shifts of the fake-lepton efficiencies resulting from statistical uncertainties
and the prompt lepton background subtraction in the control regions used to
measure the fake-lepton efficiency. The uncertainties in the fake-lepton
background in the \ttZ and \ttW analysis regions are considered to be
uncorrelated, due to the different lepton selection requirements used in the
two sets of regions.

Uncertainties in scale factors applied to the fake-lepton background are taken
into account in the fake-lepton background yield in the \FL\ channel and the
$t\bar{t}\gamma$ background contribution in the \TL\ and \SSLSR\ channels.
These uncertainties are associated with reconstructed objects and the limited
sizes of control regions in which the scale factors are obtained. The scale
factors have uncertainties between 10\% and 50\%, depending on the fake-lepton
flavor and source. The \SSLSR\ and \TL\ fake-lepton systematic uncertainties
from the matrix method are assumed to be uncorrelated with the systematic
uncertainties in the fake-lepton scale factors.

\section{Results}
\label{s:results}

The signal strengths $\mu_{\ttZ}$ and $\mu_{\ttW}$, defined as the ratios of
the measured values of the inclusive production cross sections to the
corresponding SM predictions discussed in Section~\ref{s:samples}, are
extracted simultaneously using a binned maximum-likelihood fit to the numbers
of events in the dilepton, trilepton and tetralepton signal and control
regions. In the OS dilepton channel signal regions \OSLSRA, \OSLSRB and
\OSLSRC, the BDT output distribution is fitted. In the SS dilepton channel, the
twelve signal regions \eloneeltwo-SSp-$1b$, \eloneeltwo-SSm-$1b$,
\eloneeltwo-SSp-$2b$ and \eloneeltwo-SSm-$2b$ are fitted together with the
twelve control regions \eloneeltwo-SSp-$1b$-CR, \eloneeltwo-SSm-$1b$-CR,
\eloneeltwo-SSp-$2b$-CR and \eloneeltwo-SSm-$2b$-CR defined in
Section~\ref{s:selection}. The contribution from the \ttW signal in the SS
dilepton control regions is taken into account in the fit.  The dependence of
the fake-lepton background in these regions on the \ttW signal strength is also
taken into account.  In the trilepton channel, the eight signal regions
described in Section~\ref{s:selection} are included in the fit, as is the
\TLCR\ control region. Finally, in the tetralepton channel, the four signal
regions \FLSRB, \FLSRC, \FLSRD\ and \FLSRE\ and the control region \FLCR\ are
included in the fit.

The fit is based on the profile-likelihood technique, where systematic
uncertainties are included in the fit as nuisance parameters constrained by
Gaussian functions.  None of the uncertainty parameters are found to be
significantly constrained or pulled in the fit.  The calculation of confidence
intervals and hypothesis testing is performed using a modified frequentist
method as implemented in RooStats~\cite{RooFit, RooFitManual}.

\begin{figure}[htbp]
\centering
\subfloat[\label{fig:BDTOS1}]{\includegraphics[width=\twofigwidth]{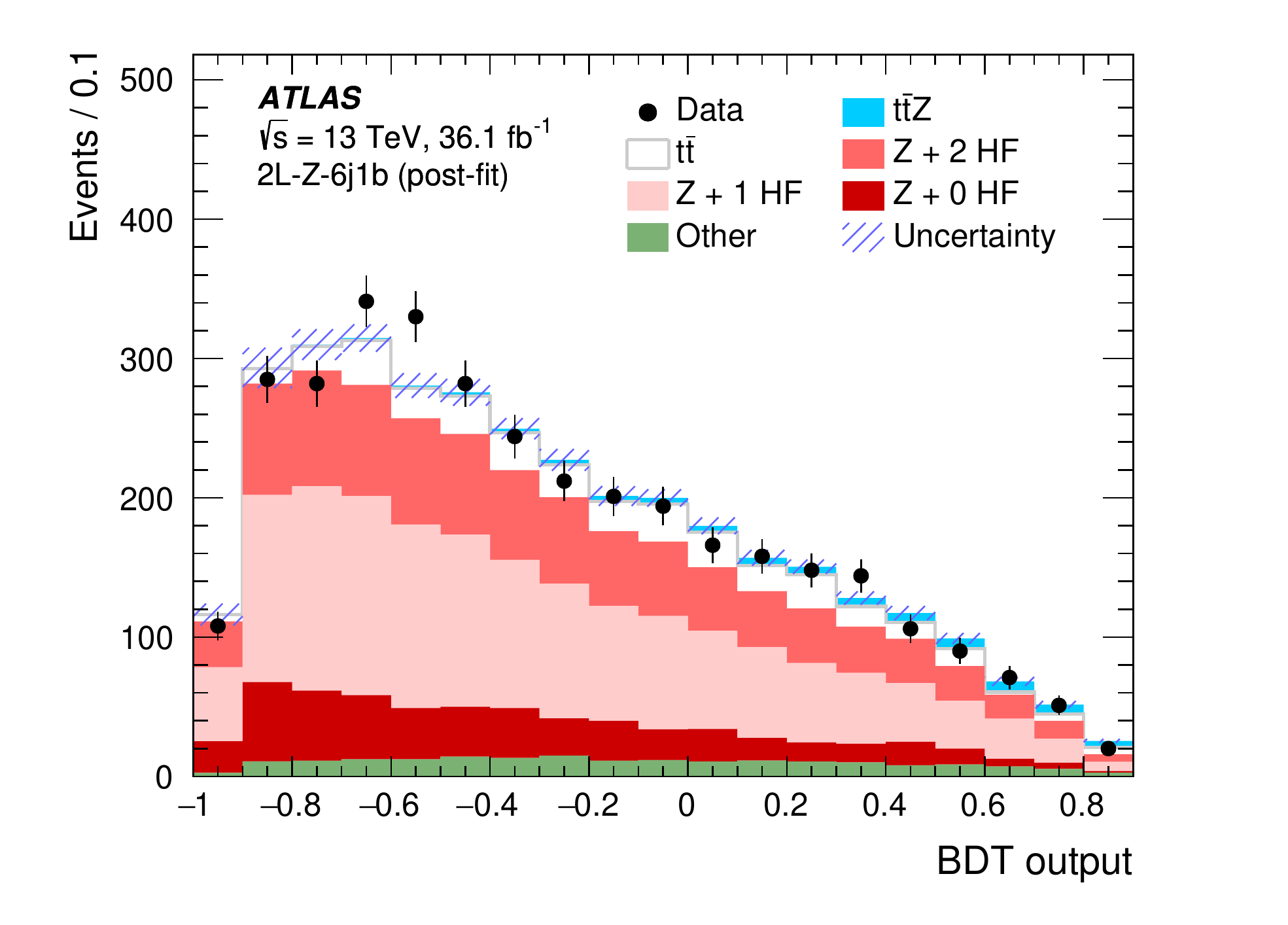}}
\subfloat[\label{fig:BDTOS2}]{\includegraphics[width=\twofigwidth]{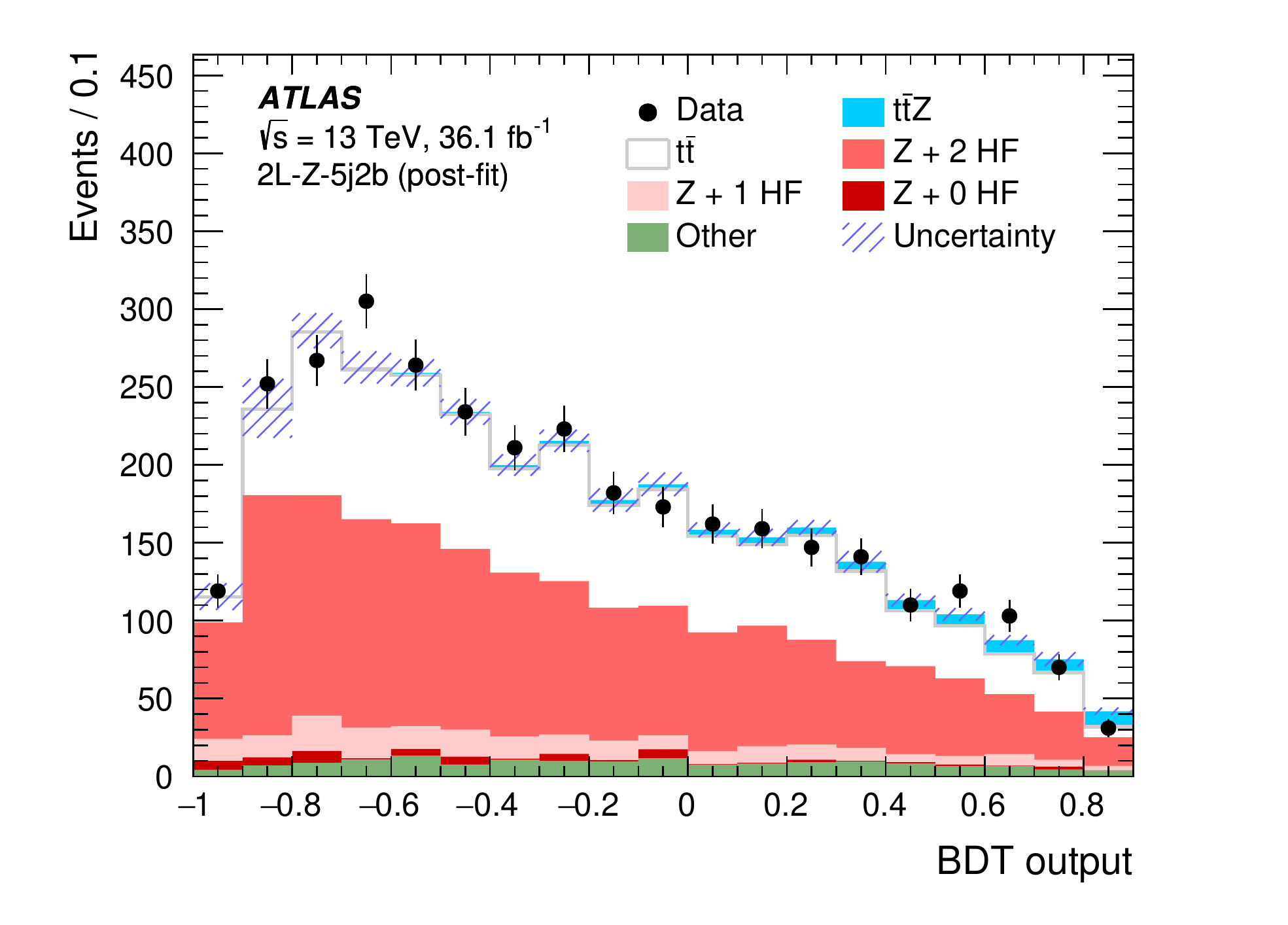}}\\
\subfloat[\label{fig:BDTOS3}]{\includegraphics[width=\twofigwidth]{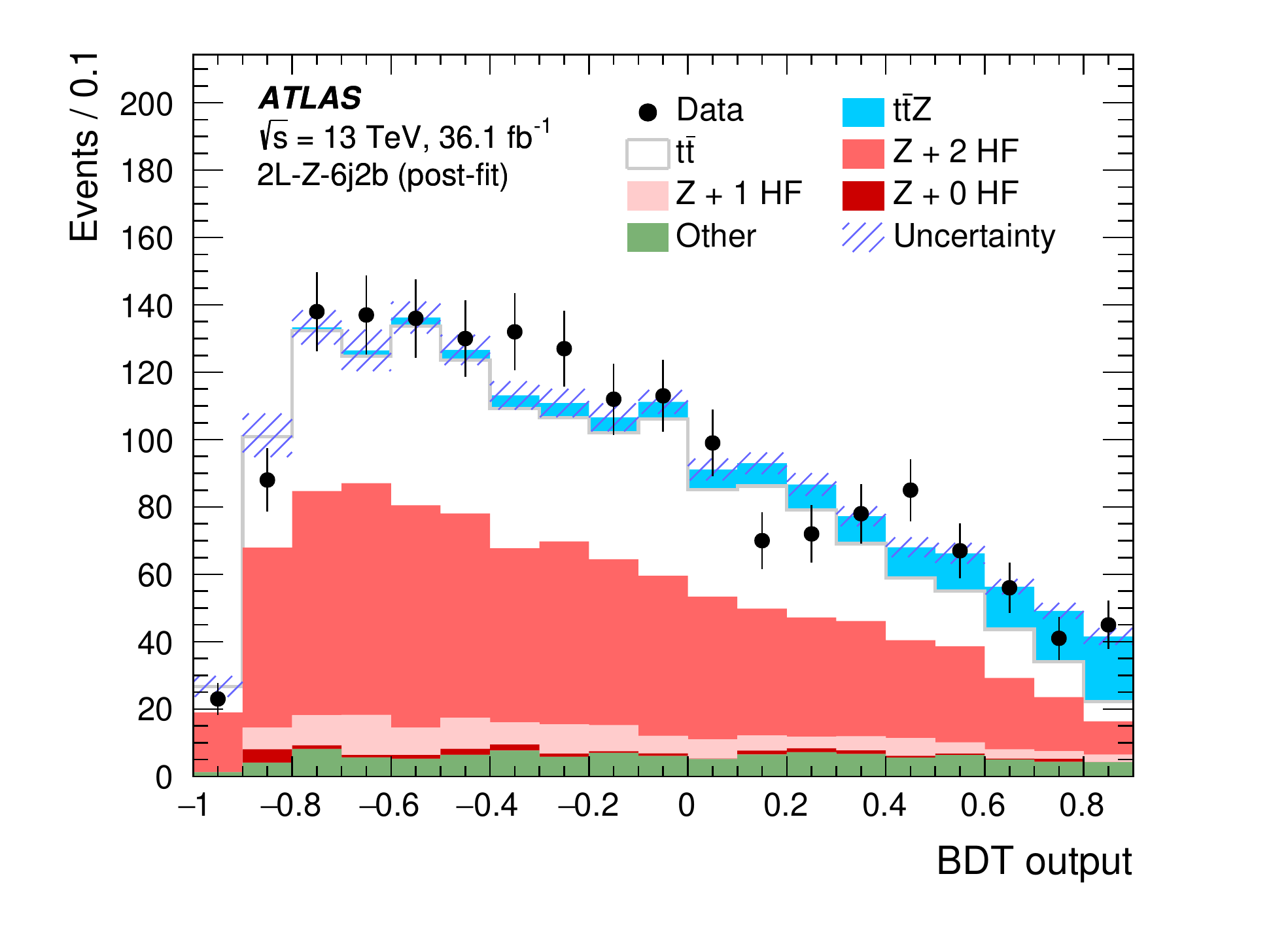}}
\caption{The BDT distributions for the OS dilepton signal regions,
\protect\subref{fig:BDTOS1} \OSLSRA, \protect\subref{fig:BDTOS2} \OSLSRB,
\protect\subref{fig:BDTOS3} \OSLSRC.  The distributions are shown after the
fit. The `Other' background contains SM processes with small cross sections
producing two opposite-sign prompt leptons. \hatch. The last bin of each
distribution contains the overflow.\label{fig:BDTOS}}
\end{figure}

\begin{figure}[htbp]
\centering
\subfloat[\label{fig:summaryttZ1}]{\includegraphics[width=0.52\textwidth]{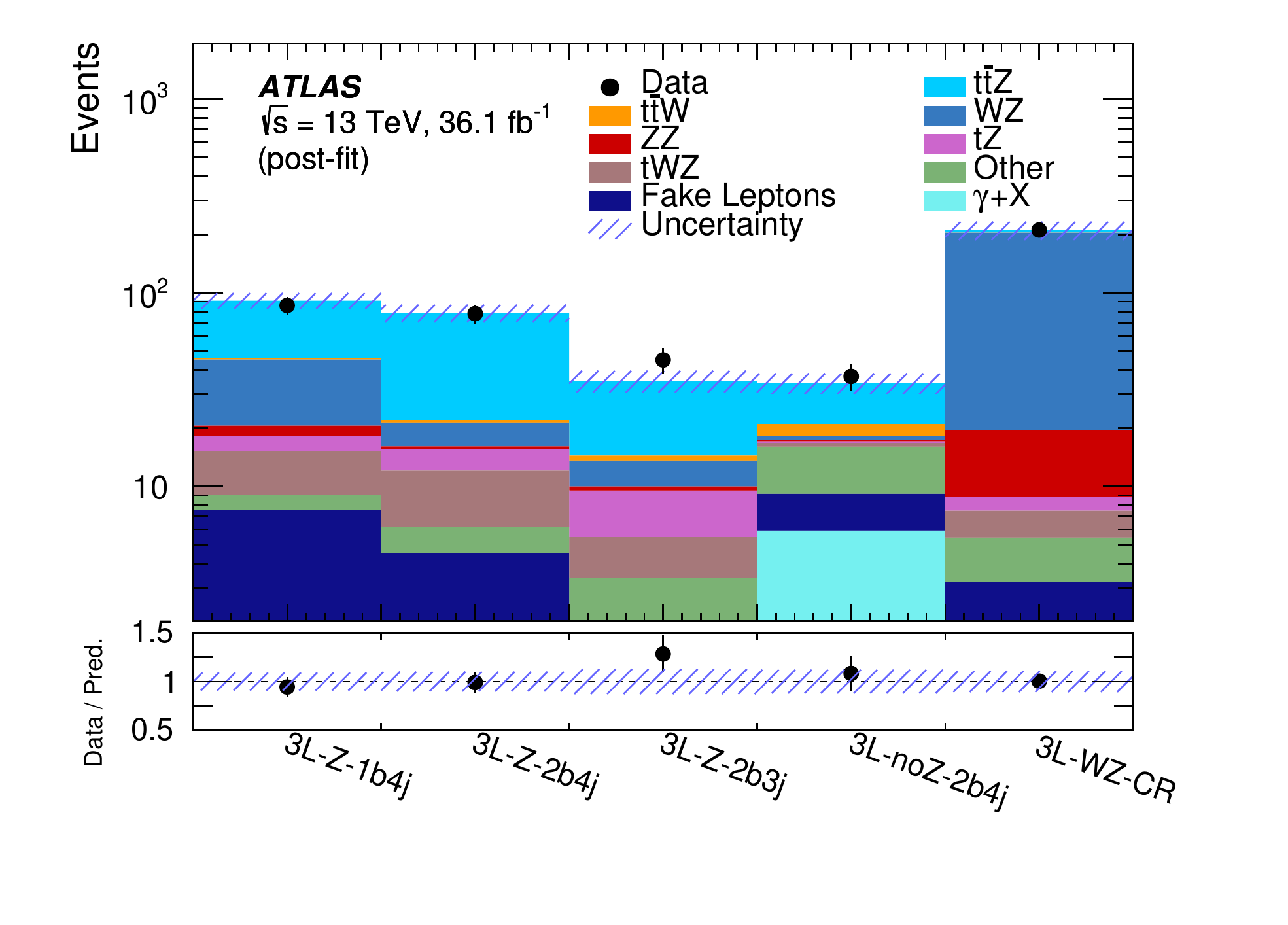}}
\subfloat[\label{fig:summaryttZ2}]{\includegraphics[width=0.52\textwidth]{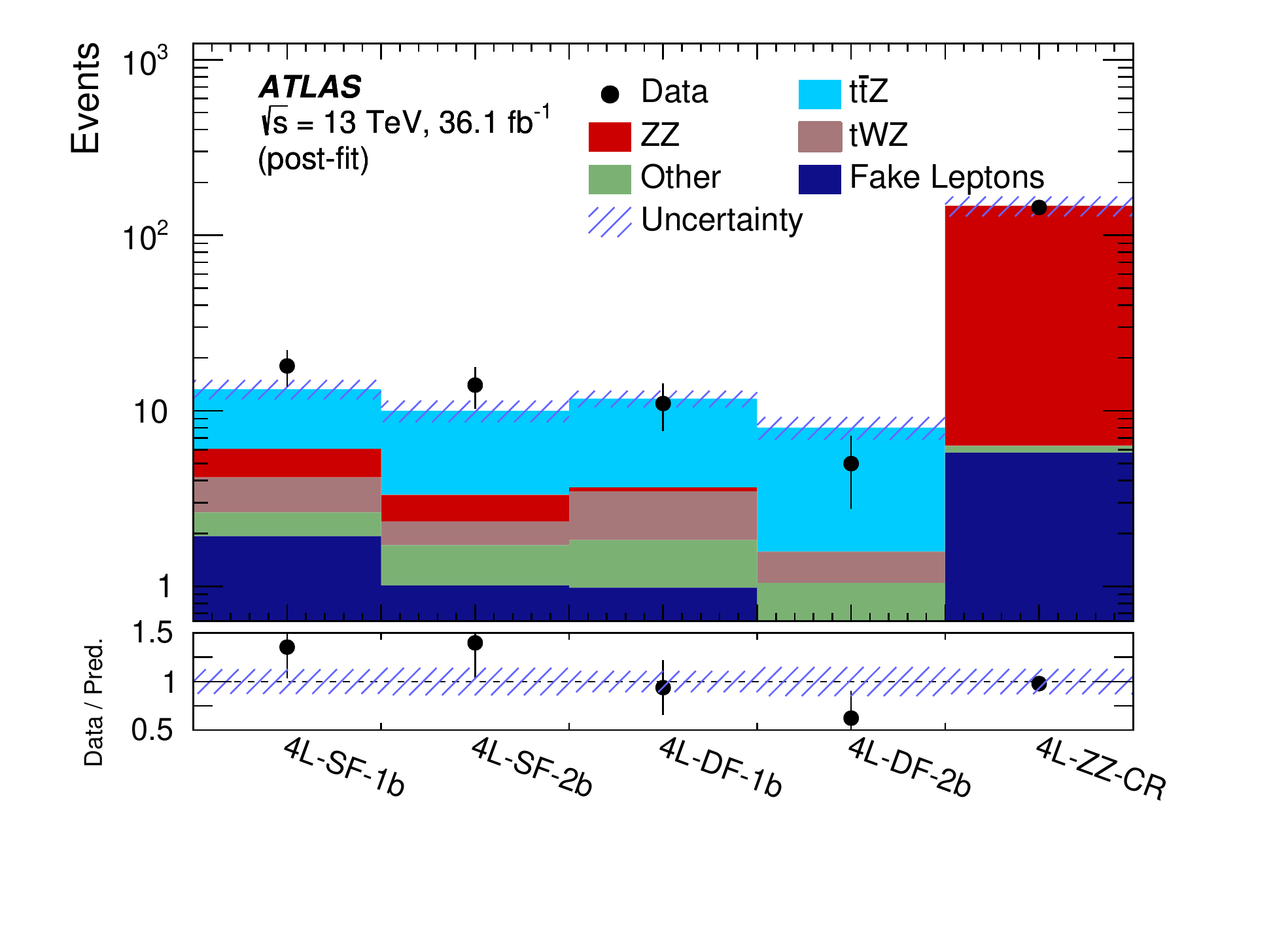}}
\caption{Event yields in data compared with the results of the fit that
extracts \sttZ and \sttW simultaneously in the \protect\subref{fig:summaryttZ1}
trilepton and \protect\subref{fig:summaryttZ2} tetralepton signal regions
targeting the \ttZ process. Yields for the control regions used to extract the
normalization of the $WZ$ and $ZZ$ backgrounds are also shown.  The `Other'
background summarizes all small SM backgrounds described in
Section~\ref{s:samples}.  \hatch.\label{fig:summaryttZ}}
\end{figure}

\begin{figure}[htbp]
\centering
\includegraphics[width=0.9\textwidth]{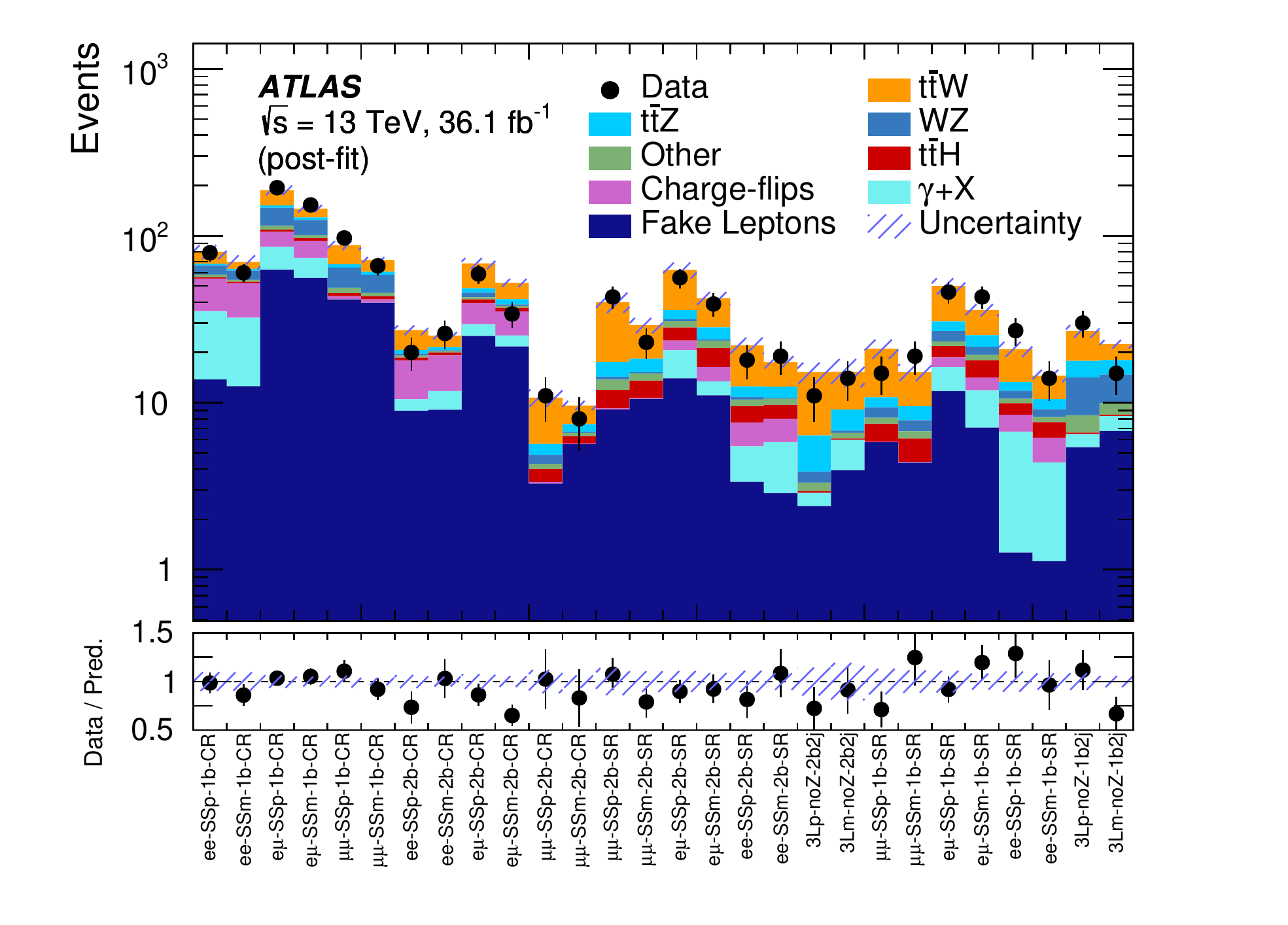}
\caption{Event yields in data compared with the results of the fit that
extracts \sttZ and \sttW simultaneously in the regions targeting the \ttW
process.  The `Other' background summarizes all small SM backgrounds described
in Section~\ref{s:samples}.  \hatch.\label{fig:summaryttW}}
\end{figure}

Figure~\ref{fig:BDTOS} shows the BDT output distribution in signal regions
\OSLSRA, \OSLSRB and \OSLSRC after performing the fit.
Figures~\ref{fig:summaryttZ} and \ref{fig:summaryttW} summarize the comparison
between data and the post-fit signal and background yields for regions
sensitive to \ttZ and \ttW production, together with the relevant control
regions.  In all cases, good agreement between observed values and the
expectation is seen.  The normalization corrections for the $WZ$ and $ZZ$
backgrounds with respect to the predictions are obtained from the fits as
described in Section~\ref{s:selection} and found to be compatible with unity:
$0.91 \pm 0.10$ for the $WZ$ background and $1.11 \pm 0.17$ for the $ZZ$
background.  The normalizations of the $Z$+1HF and $Z$+2HF backgrounds are
mainly constrained in the low BDT output bins of the OS dilepton channel signal
regions, where the signal contamination is low. Their values are found to be
$1.19\pm 0.25$ and $1.09\pm 0.13$, respectively.  

In addition to the combined fit described above, fits in individual channels
are performed.  The \ttZ signal strength is extracted through fits to the \OSL\
regions alone, to the \TLC\ regions alone and to the \FLC\ signal regions
alone. The \ttW signal strength is extracted using the four trilepton signal
regions targeting \ttW and the \SSL\ regions considered in the combined fit.
The measured values of the signal strengths $\mu_{\ttZ}$ and $\mu_{\ttW}$ are
reported in Table~\ref{tab:results} for each channel separately and for the
combined fit.  Agreement is observed for the measured values between all the
different fit configurations.

\begin{table}[htbp]
\centering \renewcommand{\arraystretch}{1.2}
\caption{Measured signal strengths of \ttZ and \ttW for different fit
configurations and the combined fit. The uncertainties include statistical and
systematic components.\spaceit} \label{tab:results}
\begin{tabular}{lcc}
\toprule
Fit configuration &   $\mu_{\ttZ}$ & $\mu_{\ttW}$ \\
\midrule
Combined                    &  1.08 $\pm$ 0.14 &  1.44 $\pm$ 0.32\\
\OSLSR         & 0.73 $\pm$ 0.28 & --\\
$3\ell$ \ttZ   & 1.08 $\pm$ 0.18 & --\\
\SSLSR and $3\ell$ \ttW & -- &  1.41 $\pm$ 0.33\\
$4\ell$ & 1.21 $\pm$ 0.29  & -- \\
\bottomrule
\end{tabular}
\end{table}

\begin{figure}[htbp]
\centering
\includegraphics[width=0.85\textwidth]{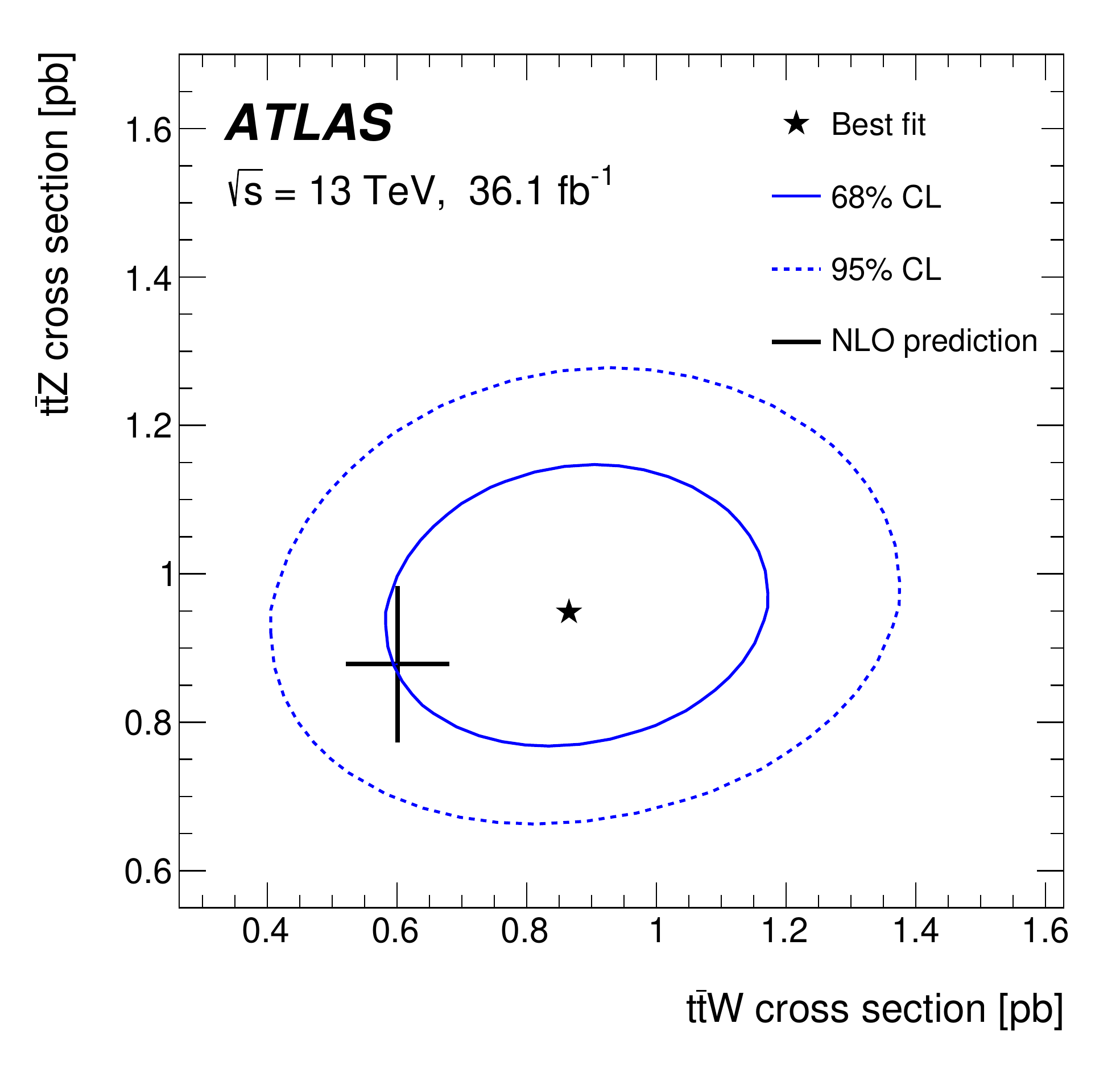}
\caption{The result of the simultaneous fit to the \ttZ and \ttW cross sections
along with the 68\% and 95\% confidence level (CL) contours. The cross shows
the SM calculations and their uncertainties, including
renormalization and factorization scale uncertainties as well as
uncertainties including $\alphas$ variations.}
\label{fig:simfit}
\end{figure}

The measured signal strengths from the combined fit and their uncertainties are
converted to inclusive cross-section measurements using the signal simulation
described in Section~\ref{s:samples} and the central values of the theoretical
predictions. The results are: $\sttZ =
0.95\;\!\pm\;\!0.08_{\text{stat.}}\;\!\pm\;\!0.10_{\text{syst.}}\,\text{pb} =
0.95\;\!\pm\;\!0.13\,\text{pb}$ and $\sttW =
0.87\;\!\pm\;\!0.13_{\text{stat.}}\;\!\pm\;\!0.14_{\text{syst.}}\,\text{pb} =
0.87\;\!\pm\;\!0.19\,\text{pb}$.  Figure~\ref{fig:simfit} shows a comparison of
the fit results with theoretical predictions, $\sigma^{\textrm{th}}_{\ttZ} =
0.88^{+0.09}_{-0.11}$ pb and $\sigma^{\textrm{th}}_{\ttW} =
0.60^{+0.08}_{-0.07}$ pb, demonstrating good agreement between the measured and
predicted cross sections.

For the \ttZ process, both the observed and the expected significances are
found to be much larger than five standard deviations. For the \ttW process, an
excess of events over the expected background-only hypothesis is found with an
observed (expected) significance of $4.3$ ($3.4$) standard deviations. The
significance values are computed using the asymptotic approximation described
in Ref.~\cite{cls_3}.

Table~\ref{tab:syst} shows the uncertainties in the measured \ttZ and \ttW
cross sections, grouped in categories, along with the total uncertainties.  For
both processes, the precision of the measurement is affected by statistical and
systematic uncertainties in similar proportions.  For the \ttZ determination,
the dominant systematic uncertainty sources are the modeling of the backgrounds
and of the signal.  For the \ttW determination, the dominant systematic
uncertainty sources are the modeling of the signal and the limited amount of
data available in the control regions and simulated event samples.

\begin{table}[htbp]
\centering \renewcommand{\arraystretch}{1.2}
\caption{List of relative uncertainties in the measured cross sections of the
\ttZ and \ttW processes from the fit, grouped in categories.  All uncertainties
are symmetrized. The sum in quadrature may not be equal to the total due to
correlations between uncertainties introduced by the fit.  \spaceit}
\label{tab:syst}
\begin{tabular}{lcc}
\toprule
Uncertainty                 &   $\sigma_{\ttZ}$ & $\sigma_{\ttW}$ \\
\midrule
Luminosity                                 &  2.9$\%$&   4.5$\%$\\
Simulated sample statistics         &  2.0$\%$&   5.3$\%$\\
Data-driven background statistics         &  2.5$\%$&   6.3$\%$\\
JES/JER                                    &  1.9$\%$&   4.1$\%$\\
Flavor tagging                            &  4.2$\%$&   3.7$\%$\\
Other object-related                       &  3.7$\%$&   2.5$\%$\\
Data-driven background normalization       &  3.2$\%$&   3.9$\%$\\
Modeling of backgrounds from simulation   &  5.3$\%$&   2.6$\%$\\
Background cross sections                  &  2.3$\%$&   4.9$\%$\\
Fake leptons and charge misID              &  1.8$\%$&   5.7$\%$\\
\ttZ modeling                             &  4.9$\%$&   0.7$\%$\\
\ttW modeling                             &  0.3$\%$&   8.5$\%$\\
\midrule
Total systematic            & 10$\%$ & 16$\%$\\
Statistical                 & 8.4$\%$& 15$\%$\\
\midrule
Total                       &  13$\%$& 22$\%$ \\
\bottomrule
\end{tabular}
\end{table}

\section{Interpretation}
\label{s:interpretation}
The effective field theory (EFT) framework provides a model-independent
approach to the parameterization of possible deviations from the SM
predictions. In this framework, effects due to BSM physics are described by
adding additional operators of dimension six or higher to the SM Lagrangian.
Each EFT operator $\mathcal O_i$ is associated with a Wilson coefficient $C_i$,
and the operators enter the modified Lagrangian in the form $(C_i/\Lambda^2)
\mathcal O_i$, where $\Lambda$ is the characteristic energy scale of the BSM physics.

The complete set of independent, gauge-invariant and baryon-number conserving
EFT operators at dimension six contains 59 different
operators~\cite{AguilarSaavedra:2008zc,Grzadkowski:2010es}. In the present
analysis, five of these operators are considered, all of which modify the $tt
Z$ vertex: $\ofqt$, $\mathcal O_{\phi Q}^{(1)}$, $\oft$, $\otw$, $\otb$. The
operators are defined in Table~\ref{tab:EFToperators}, following
Ref.~\cite{AguilarSaavedra:2018nen}. The first two operators enter the $ttZ$
vertex as a linear combination, such that the measurement is sensitive to the
difference $\cfqt - \cfqo$. For this paper, the effect of this combination is
evaluated by varying $\cfqt$ with $\cfqo$ set to zero.

\begin{table}[htbp]
\centering \renewcommand{\arraystretch}{1.2}
\caption{\label{tab:EFToperators} Effective field theory operators considered
and their form in terms of SM fields. The notation of
Ref.~\cite{AguilarSaavedra:2018nen}
is used.\spaceit}
\begin{tabular}{cc}
\toprule
Operator & Expression \\
\midrule
$\ofqt$ & $( \phi^\dagger i \covariantI \phi) (\bar{Q}\gamma^\mu \tau^I Q)$ \\
$\mathcal O_{\phi Q}^{(1)}$ & $ ( \phi^\dagger i \covariant \phi) (\bar{Q}\gamma^\mu Q) $\\
$\oft$ & $ ( \phi^\dagger i \covariant \phi) (\bar{t}\gamma^\mu t) $ \\
$\otw$ & $ (\bar{Q} \sigma^{\mu\nu} \tau^I t)\tilde{\phi}W^I_{\mu\nu}$\\
$\otb$ & $ (\bar{Q} \sigma^{\mu\nu} t)\tilde{\phi} B_{\mu\nu}$\\
\bottomrule
\end{tabular}
\end{table}

Considering only one EFT operator at a time, any observable, such as the \ttZ
event rate in a certain signal region, can be expressed as a quadratic function
of the coefficient $C_i$:
\begin{eqnarray}
\sigma_{\text{tot}, i} = \sigma_{\text{SM}} + \frac{\mathcal
C_i}{(\Lambda/1{\TeV})^2} \sone + \frac{\mathcal C_i^2}{(\Lambda/1{\TeV})^4}
\stwo \label{eq:quad_form}.
\end{eqnarray}
The term linear in $C_i$ on the right-hand side of Eq.~(\ref{eq:quad_form})
results from the interference of the BSM operators with the SM. For $\mathcal
C_i/\Lambda^2$ of order $1~\TeV^{-2}$, the interference term dominates
in Eq.~(\ref{eq:quad_form}) for $\ofqt$ and $\oft$, while the quadratic term
dominates for $\otw$ and $\otb$. 

The values of $\sone$ and $\stwo$ are computed using simulated event samples
generated with \MGAMC\ interfaced to \PYTHIA 8~\cite{Christensen:2008py,
Degrande:2011ua, Bylund:2016phk}. The computation is performed at NLO,
separately for all trilepton and tetralepton signal regions. The detector
reconstruction efficiency is verified to be compatible between SM \ttZ samples
and samples with non-zero values of $C_i$, for ranges of $C_i$ considered here.

A fit is then performed to extract $\mathcal C_i/\Lambda^2$. The fit is similar
to the one described in Section~\ref{s:results}, except that only the four
trilepton and four tetralepton signal regions targeting \ttZ are used and a
normalization uncertainty of 12\%, corresponding to the uncertainty in the NLO
cross-section computation, is applied to the SM \ttZ prediction.  Uncertainties
resulting from the limited sizes of MC samples used to derive the values of
$\sone$ and $\stwo$ are propagated to the measured values of $\mathcal
C_i/\Lambda^2$. 

The profile-likelihood test statistic is defined as $-\Delta \log (L) = \log
({\cal{L}} (\hat{C}_i) / {\cal{L}}(C_i))$, where ${\cal{L}}$ is the profile
likelihood as a function of the Wilson coefficient $C_i$, and $\hat{C}_i$ is
the best-fit value of $C_i$.  Approximate confidence intervals for the Wilson
coefficients are computed using the formula $ -\Delta \log (L) = \varepsilon$,
where the threshold $\varepsilon$ is set to 0.5 and $1.92$ for the 68\%
($1\sigma$) and 95\% confidence levels (CL), respectively. 

The confidence intervals for $C_i$ are computed considering only the minimum of
$\lambda (C_i)$ near $C_i=0$. For coefficient $\cfqt$ ($\cft$), another, deeper
minimum exists for negative values of $C_i\sim30~(20)$, which is excluded by
previous constraints.  The 68\% and 95\% confidence intervals are shown in
Table~\ref{tab:EFTresults}, together with previous constraints on the EFT
coefficients obtained from Refs.~\cite{Zhang:2012cd, Buckley:2015lku,
CMS-TOP-12-038, TOPQ-2016-02}.  The lower boundary of the 95\% confidence
interval for $\cft$ is at large negative values, which are excluded by indirect
constraints. The \ttZ measurement provides competitive constraints for positive
$\cft$ values.  The full likelihood scans are shown in
Figure~\ref{fig:likelihoodComData} in the Appendix.

The fits are repeated while assuming that the quadratic terms are zero, and the
results of these fits are also reported in Table~\ref{tab:EFTresults}.  For
$\ctw$ and $\ctb$, where the quadratic terms dominate, the fits do not
converge.  Compared with the nominal fits for $\cfqt$ and $\cft$, the limits
shift to larger values, consistent with removing a positive term from the
prediction. The most notable change is the improvement in the lower limit for
$\cft$ at 95\% CL, as the second minimum disappears when a linear expression is
assumed.

\begin{table}[htbp]
\centering 
\caption{\label{tab:EFTresults}The expected and observed 68\% and 95\%
confidence intervals, which include the value $0$, for $\mathcal C_i/\Lambda^2$
for the EFT coefficients $\cfqt$, $\cft$, $\ctb$ and $\ctw$. The intervals for
$\cfqt$ are derived setting $\cfqo$ to zero; the measurement is sensitive to
the difference $\cfqt-\cfqo$. All results are obtained by varying one
coefficient at the time and are given in units of $1/{\TeV}^2.$ Previous
indirect 68\% CL constraints \cite{Zhang:2012cd} are also quoted. The previous direct
constraints at 95\% CL are obtained from Ref.~\cite{Buckley:2015lku} for $\cft$
and $\ctb$ and from Refs.~\cite{CMS-TOP-12-038, TOPQ-2016-02} for $\cfqt$ and
$\ctw$, using Ref.~\cite{AguilarSaavedra:2008zc} to translate the measurements
into limits on the coefficients.  Limits from fits for the EFT coefficients
with only the linear term are also shown. \spaceit }
\begin{tabular}{lcccc}
\toprule
Coefficients  & $\cfqt/\Lambda^2$ & $\cft/\Lambda^2$ & $\ctb/\Lambda^2$ & $\ctw/\Lambda^2$\\
\midrule
Previous indirect constraints at 68\% CL & [--4.7, 0.7] & [--0.1, 3.7] & [--0.5, 10] & [--1.6, 0.8]\\
Previous direct constraints at 95\% CL & [--1.3, 1.3] & [--9.7, 8.3] & [--6.9, 4.6] & [--0.2, 0.7]\\
\midrule
Expected limit at 68\% CL & [--2.1, 1.9] & [--3.8, 2.7] & [--2.9, 3.0] & [--1.8, 1.9]\\
Expected limit at 95\% CL & [--4.5, 3.6] & [--23, 4.9] & [--4.2, 4.3] & [--2.6, 2.6]\\
Observed limit at 68\% CL & [--1.0, 2.7] & [--2.0, 3.5] & [--3.7, 3.5] & [--2.2, 2.1]\\
Observed limit at 95\% CL & [--3.3, 4.2] & [--25, 5.5] & [--5.0, 5.0] & [--2.9, 2.9]\\
\midrule
Expected limit at 68\% CL (linear) & [--1.9, 2.0] & [--3.0, 3.2] & -- & --\\
Expected limit at 95\% CL (linear) & [--3.7, 4.0] & [--5.8, 6.3] & -- & --\\
Observed limit at 68\% CL (linear) & [--1.0, 2.9] & [--1.8, 4.4] & -- & --\\
Observed limit at 95\% CL (linear) & [--2.9, 4.9] & [--4.8, 7.5] & -- & --\\
\bottomrule
\end{tabular}
\end{table}

\FloatBarrier

\section{Conclusion}
\label{s:conclusion}

Measurements of the production cross sections of a top-quark pair in
association with a $Z$ or $W$ boson using \lumi of data collected by the ATLAS
detector in $\sqrt{s} = 13\,\TeV$ $pp$ collisions at the LHC are presented.
Final states with two same- or opposite-sign leptons, three leptons or four
leptons are analyzed.  The \ttZ and \ttW production cross sections are
determined to be  $\sttZ = 0.95 \pm 0.08_{\text{stat.}} \pm
0.10_{\text{syst.}}\,\text{pb} = 0.95 \pm 0.13 \,\text{pb}$ and $\sttW = 0.87
\pm 0.13_{\text{stat.}} \pm 0.14_{\text{syst.}}\,\text{pb} = 0.87 \pm 0.19
\,\text{pb}$.  The measured values are consistent with the SM predictions. The
measurements are used to derive confidence intervals for the Wilson
coefficients of dimension-6 effective field theory operators involving the top
quark and the $Z$ boson.

\section*{Acknowledgments}

% Acknowledgements for papers with collision data
% Version 24-Oct-2018

% Standard acknowledgements start here
%----------------------------------------------
We thank CERN for the very successful operation of the LHC, as well as the
support staff from our institutions without whom ATLAS could not be operated
efficiently.

We acknowledge the support of ANPCyT, Argentina; YerPhI, Armenia; ARC,
Australia; BMWFW and FWF, Austria; ANAS, Azerbaijan; SSTC, Belarus; CNPq and
FAPESP, Brazil; NSERC, NRC and CFI, Canada; CERN; CONICYT, Chile; CAS, MOST and
NSFC, China; COLCIENCIAS, Colombia; MSMT CR, MPO CR and VSC CR, Czech Republic;
DNRF and DNSRC, Denmark; IN2P3-CNRS, CEA-DRF/IRFU, France; SRNSFG, Georgia;
BMBF, HGF, and MPG, Germany; GSRT, Greece; RGC, Hong Kong SAR, China; ISF and
Benoziyo Center, Israel; INFN, Italy; MEXT and JSPS, Japan; CNRST, Morocco;
NWO, Netherlands; RCN, Norway; MNiSW and NCN, Poland; FCT, Portugal; MNE/IFA,
Romania; MES of Russia and NRC KI, Russian Federation; JINR; MESTD, Serbia;
MSSR, Slovakia; ARRS and MIZ\v{S}, Slovenia; DST/NRF, South Africa; MINECO,
Spain; SRC and Wallenberg Foundation, Sweden; SERI, SNSF and Cantons of Bern
and Geneva, Switzerland; MOST, Taiwan; TAEK, Turkey; STFC, United Kingdom; DOE
and NSF, United States of America. In addition, individual groups and members
have received support from BCKDF, CANARIE, CRC and Compute Canada, Canada;
COST, ERC, ERDF, Horizon 2020, and Marie Sk{\l}odowska-Curie Actions, European
Union; Investissements d' Avenir Labex and Idex, ANR, France; DFG and AvH
Foundation, Germany; Herakleitos, Thales and Aristeia programmes co-financed by
EU-ESF and the Greek NSRF, Greece; BSF-NSF and GIF, Israel; CERCA Programme
Generalitat de Catalunya, Spain; The Royal Society and Leverhulme Trust, United
Kingdom. 

The crucial computing support from all WLCG partners is acknowledged
gratefully, in particular from CERN, the ATLAS Tier-1 facilities at TRIUMF
(Canada), NDGF (Denmark, Norway, Sweden), CC-IN2P3 (France), KIT/GridKA
(Germany), INFN-CNAF (Italy), NL-T1 (Netherlands), PIC (Spain), ASGC (Taiwan),
RAL (UK) and BNL (USA), the Tier-2 facilities worldwide and large non-WLCG
resource providers. Major contributors of computing resources are listed in
Ref.~\cite{ATL-GEN-PUB-2016-002}.

\clearpage
\printbibliography
\clearpage
\part*{Appendix}
\addcontentsline{toc}{part}{Appendix}

\begin{figure}[htbp]
\begin{center}
\subfloat[\label{fig:cfq3}]{\includegraphics[width=\rtwofigwidth]{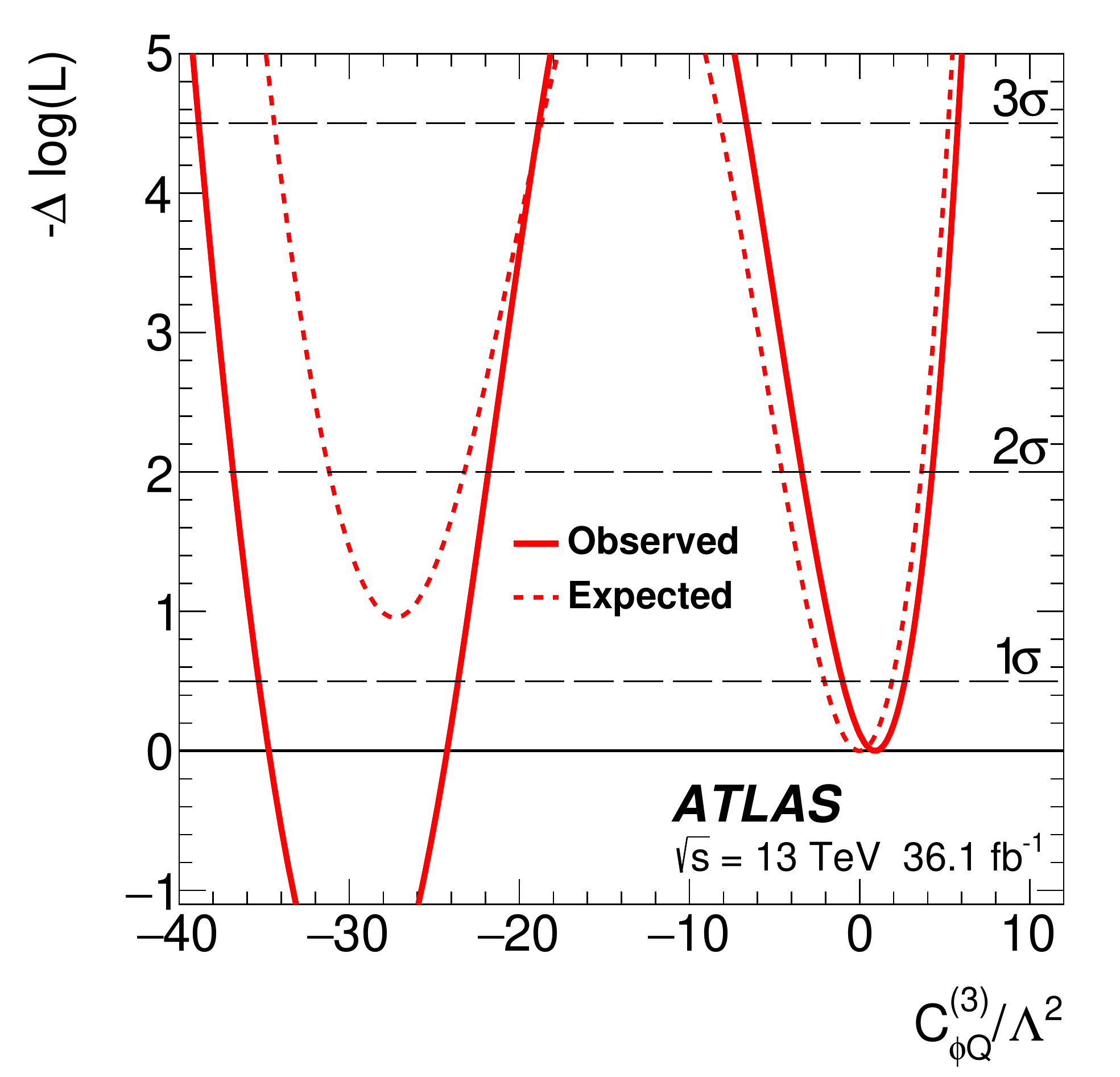}}
\subfloat[\label{fig:cft}]{\includegraphics[width=\rtwofigwidth]{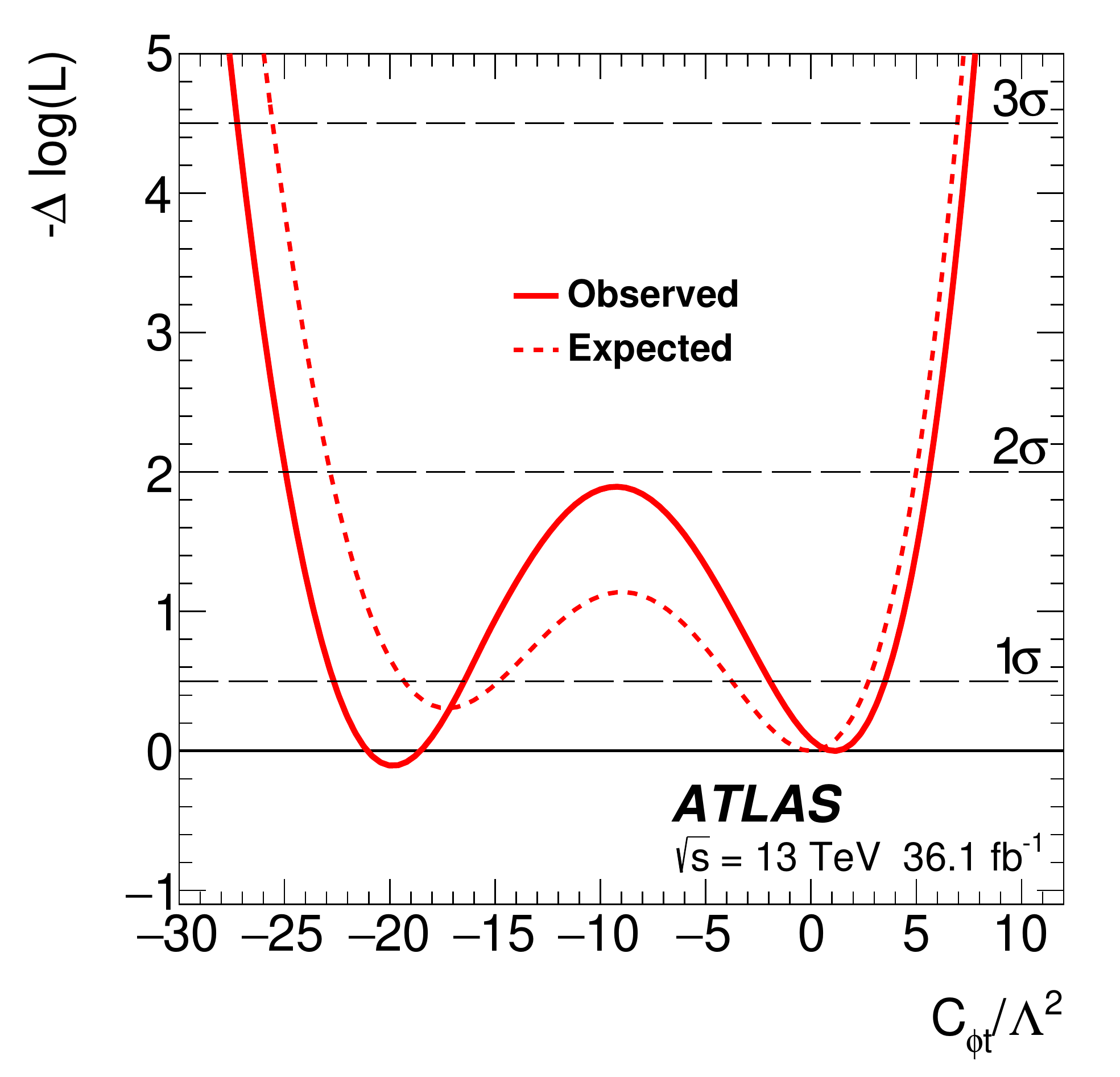}}\\
\subfloat[\label{fig:ctb}]{\includegraphics[width=\rtwofigwidth]{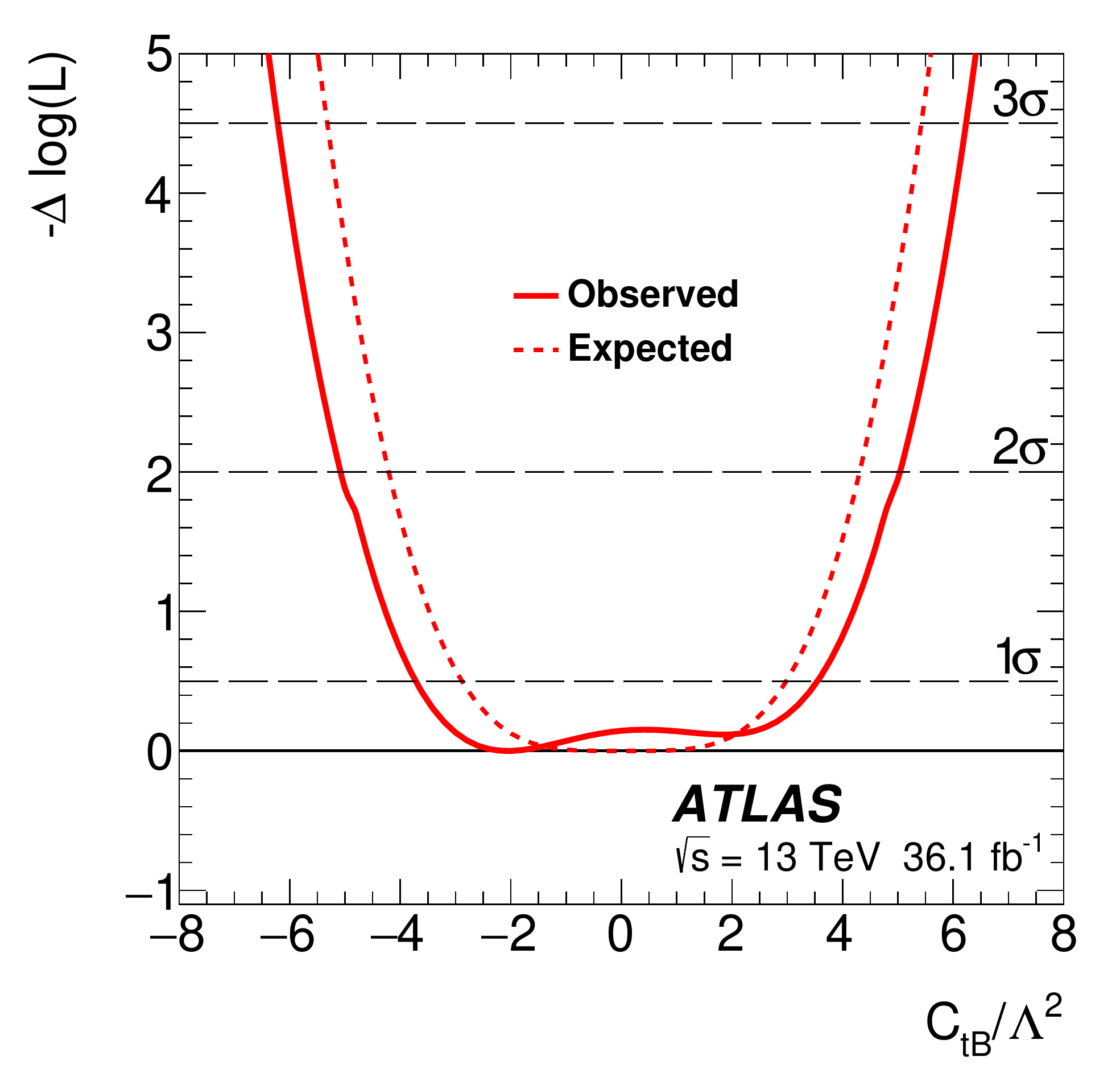}}
\subfloat[\label{fig:ctw}]{\includegraphics[width=\rtwofigwidth]{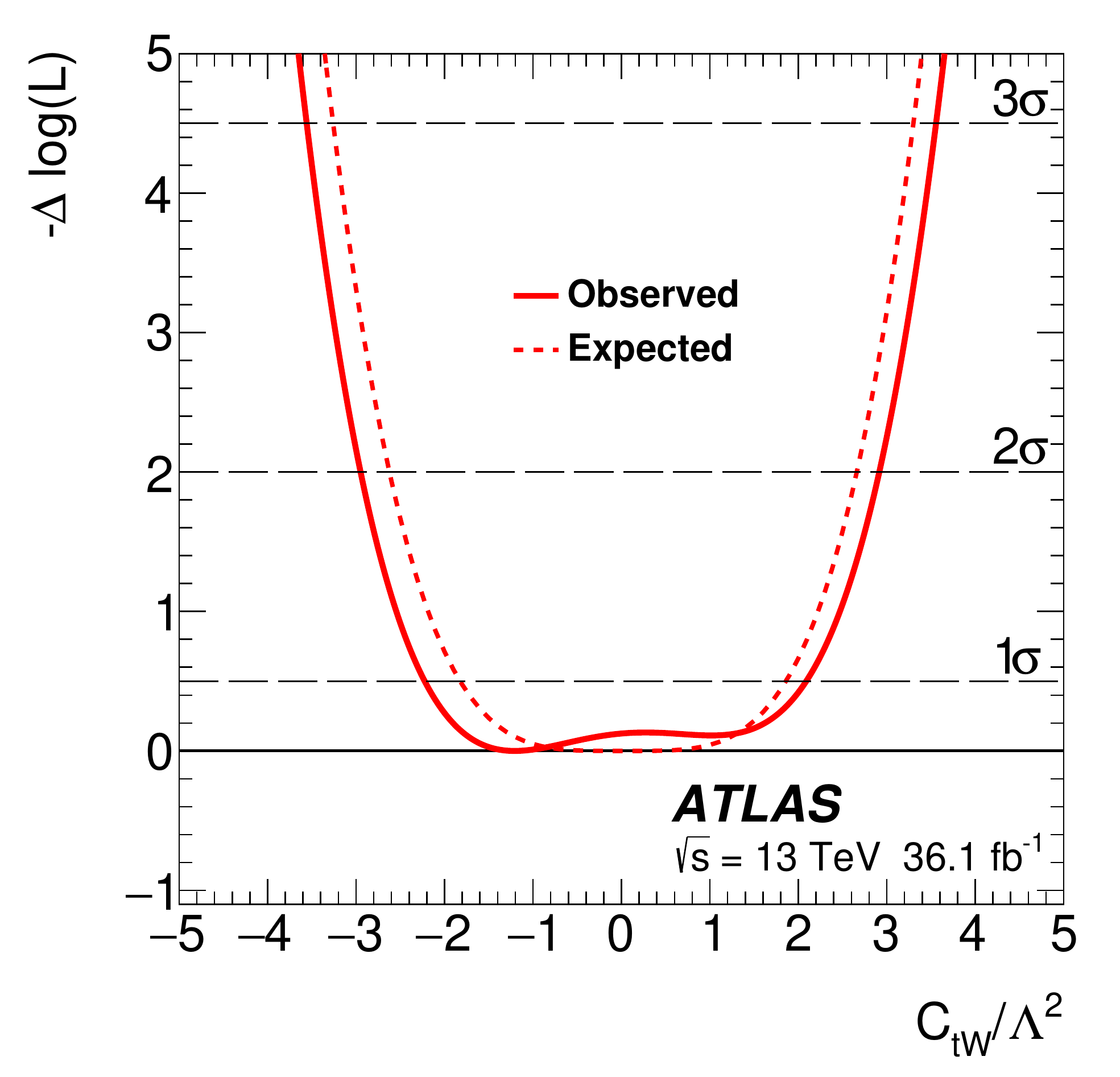}}
\caption{\label{fig:likelihoodComData} The value of the profile-likelihood test
statistic as a function of $c/\Lambda^2$, for \protect\subref{fig:cfq3} \cfqt,
\protect\subref{fig:cft} \cft, \protect\subref{fig:ctb} \ctb, and
\protect\subref{fig:ctw} \ctw.  
In the \cfqt and \cft distributions, another, deeper minimum exists for large
negative values of $C_i$, which is excluded by indirect measurements. There, the vertical
axis is chosen such that the value of the likelihood at the minimum near
$C_i=0$ is zero.}
\end{center}
\end{figure}

\begin{table}[htbp]
\renewcommand{\arraystretch}{1.3}
\begin{center}
\caption{The definitions and ranking of input variables for the BDT in the
OS dilepton analysis. Jets and leptons are ordered in descending
order of \pt. Only the first eight jets are considered when calculating the input
variables.\spaceit} \label{tab:NN_inputv_vars}
\begin{tabular}{p{0.60\textwidth}@{\hskip 10ex}rrr}
\toprule
 & \multicolumn{3}{c}{Ranking} \\
 Definition & 6j1b & 5j2b & 6j2b \\
\midrule
 \pt of the lepton pair & 8 & 11 & 8 \\ 
 \pt of the fourth jet & 6 & 12 & 6 \\ 
 \pt of the fifth jet & -- & 14 & -- \\ 
 \pt of the sixth jet & 9 & -- & 11 \\
 $\Delta R_{\eta}$ between the two leptons & 7 & 8 & 12 \\
 Number of jet pairs with mass within a window of 30~\GeV\ around 85~\GeV\ & 4 & 6 & 4 \\
 Number of three-jet combinations (containing exactly one $b$-tagged jet) compatible with a top quark ($|m_{bjj}-m_t| <15\,\GeV$) and  ($|m_{jj}-m_W| < 15\,\GeV$) & -- & -- & 17 \\
 Invariant mass of the two jets with the smallest $\Delta R_{\eta}$ & 13 & 7 & 14 \\
 Invariant mass of the two untagged jets with the highest \pt & 15 & 13 & -- \\
 Invariant mass of the two jets with the highest value of the $b$-tagging discriminant & -- & 10 & 9   \\ 
 Scalar sum of \pt divided by the sum of energy of all jets & 2 & 1 & 2 \\ 
 Average $\Delta R_{\eta}$ of all jet pairs & 5 & 4 & 5 \\
 Maximum invariant mass of a lepton and the $b$-tagged jet with the smallest $\Delta R_{\eta}$ & 14 & -- & 13 \\
 First Fox--Wolfram moment built from jets and leptons & 3 & 2 & 1 \\
 Sum of jet \pt, using up to six jets & 12 & 5 & 10 \\
 $\eta$ of dilepton system & 1 & 3 & 3 \\
 Sum of the two closest two-jet invariant masses from $\mathrm{jjj}_1$ and $\mathrm{jjj}_2$ divided by two & 10 & -- & 15 \\
 $\Delta R_{\eta}$ between two jets with the highest value of the $b$-tagging discriminant in the event & -- & 9 & 7 \\
 \pt of the $b$-tagged jet with the highest \pt & 11 & -- & 16 \\
\bottomrule
\end{tabular}
\end{center}
\end{table}

\clearpage
% ATLAS Collaboration author list
% Reference date of TOPQ-2016-11 is 2018-08-29
% Author list last updated on date 07-DEC-18
% Data extracted on 07-Dec-2018 for paper reference TOPQ-2016-11
% at 9:05am
 
\begin{flushleft}
{\Large The ATLAS Collaboration}

\bigskip

M.~Aaboud$^\textrm{\scriptsize 34d}$,    
G.~Aad$^\textrm{\scriptsize 99}$,    
B.~Abbott$^\textrm{\scriptsize 125}$,    
D.C.~Abbott$^\textrm{\scriptsize 100}$,    
O.~Abdinov$^\textrm{\scriptsize 13,*}$,    
B.~Abeloos$^\textrm{\scriptsize 129}$,    
D.K.~Abhayasinghe$^\textrm{\scriptsize 91}$,    
S.H.~Abidi$^\textrm{\scriptsize 164}$,    
O.S.~AbouZeid$^\textrm{\scriptsize 39}$,    
N.L.~Abraham$^\textrm{\scriptsize 153}$
% ...
\bigskip
\\

% ...
$^{13}$Institute of Physics, Azerbaijan Academy of Sciences, Baku; Azerbaijan.\\
% ...
$^{34}$$^{(a)}$Facult\'e des Sciences Ain Chock, R\'eseau Universitaire de Physique des Hautes Energies - Universit\'e Hassan II, Casablanca;$^{(b)}$Centre National de l'Energie des Sciences Techniques Nucleaires (CNESTEN), Rabat;$^{(c)}$Facult\'e des Sciences Semlalia, Universit\'e Cadi Ayyad, LPHEA-Marrakech;$^{(d)}$Facult\'e des Sciences, Universit\'e Mohamed Premier and LPTPM, Oujda;$^{(e)}$Facult\'e des sciences, Universit\'e Mohammed V, Rabat; Morocco.\\
% ...
$^{39}$Niels Bohr Institute, University of Copenhagen, Copenhagen; Denmark.\\
% ...
$^{91}$Department of Physics, Royal Holloway University of London, Egham; United Kingdom.\\
% ...
$^{99}$CPPM, Aix-Marseille Universit\'e, CNRS/IN2P3, Marseille; France.\\
$^{100}$Department of Physics, University of Massachusetts, Amherst MA; United States of America.\\
% ...
$^{125}$Homer L. Dodge Department of Physics and Astronomy, University of Oklahoma, Norman OK; United States of America.\\
% ...
$^{129}$LAL, Universit\'e Paris-Sud, CNRS/IN2P3, Universit\'e Paris-Saclay, Orsay; France.\\
% ...
$^{153}$Department of Physics and Astronomy, University of Sussex, Brighton; United Kingdom.\\
% ...
$^{164}$Department of Physics, University of Toronto, Toronto ON; Canada.\\
% ...
$^{a}$ Also at Borough of Manhattan Community College, City University of New York, NY; United States of America.\\
$^{b}$ Also at California State University, East Bay; United States of America.\\
$^{c}$ Also at Centre for High Performance Computing, CSIR Campus, Rosebank, Cape Town; South Africa.\\
$^{d}$ Also at CERN, Geneva; Switzerland.\\
$^{e}$ Also at CPPM, Aix-Marseille Universit\'e, CNRS/IN2P3, Marseille; France.\\
% ...
$^{*}$ Deceased

\end{flushleft}

% Created with Glance <Atlas.Glance@cern.ch>

\clearpage
\end{document}